\documentclass{aa}

\renewcommand{\linenumbers}{}
\renewcommand{\nolinenumbers}{}
\renewcommand{\runninglinenumbers}{}

\usepackage{graphicx}
\usepackage{txfonts}
\usepackage{xcolor}
\usepackage{booktabs}
\usepackage{orcidlink}
\usepackage{xparse}
\usepackage{enumitem}
\usepackage{amsmath}
\usepackage{makecell}
\usepackage{pifont}
\usepackage{xspace}
\usepackage{siunitx}

\newcommand{\ppxf}{{\sc pPXF}\xspace}
\newcommand{\ifscube}{{\sc ifscube}\xspace}
\newcommand{\sinopsis}{{\sc sinopsis}\xspace}

\begin{document} 
\nolinenumbers

\title{With arms wide open: a VLT/MUSE view of the mechanisms driving unwinding spiral arms in cluster galaxies}

\author{
    Augusto E. Lassen\inst{1}\thanks{\email{augusto.lassen@inaf.it}}
    \orcidlink{0000-0003-3575-8316}
    \and
    Benedetta Vulcani\inst{1}\orcidlink{0000-0003-0980-1499}
    \and
    Jacopo Fritz\inst{2}\orcidlink{0000-0002-7042-1965}
    \and
    Bianca M. Poggianti\inst{1}\orcidlink{0000-0001-8751-8360}
    \and
    Antonino Marasco\inst{1}\orcidlink{0000-0002-5655-6054}
    \and
    Yara Jaffé\inst{3,4}\orcidlink{0000-0003-2150-1130}
    \and
    Marco Gullieuszik\inst{1}\orcidlink{0000-0002-7296-9780}
    \and
    Alessia Moretti\inst{1}\orcidlink{0000-0002-1688-482X}
    \and
    Mario Radovich\inst{1}\orcidlink{0000-0002-3585-866X}
    \and
    Rory Smith\inst{3,4}\orcidlink{0000-0001-5303-6830}
    \and
    Stephanie Tonnesen\inst{5}\orcidlink{0000-0002-8710-9206}
    \and
    Neven Tomi\v{c}i\'{c}\inst{6}\orcidlink{0000-0002-8238-9210}
    \and
    Koshy George\inst{7}\orcidlink{0000-0002-1734-8455}
    \and
    Alessandro Ignesti\inst{1}\orcidlink{0000-0003-1581-0092}
    \and
    Luka Matijevi\'{c}\inst{6,1}\orcidlink{0009-0004-2049-7701}
    \and
    Eric Giunchi\inst{8}\orcidlink{0000-0002-3818-1746}
}

\institute{
  INAF- Osservatorio astronomico di Padova, Vicolo Osservatorio 5, I-35122 Padova, Italy
  \and
  Instituto de Radioastronom\'{\i}a y Astrof\'{\i}sica, UNAM, Campus Morelia, A.P.~3--72, C.P.~58089, Mexico
  \and
  Departamento de Física, Universidad Técnica Federico Santa María, Avenida España 1680, Valparaíso, Chile
  \and
  Millennium Nucleus for Galaxies (MINGAL)
  \and
  Flatiron Institute, Center for Computational Astrophysics, 162 5th Avenue, New York, NY 10010, USA
  \and
  Department of Physics, Faculty of Science, University of Zagreb, Bijeni\v{c}ka Cesta 32, 10000 Zagreb, Croatia
  \and
  Universit\"ats-Sternwarte\,M\"unchen,\,\,Fakult\"at\,f\"ur\,Physik,\,\,Ludwig-Maximilians-Universit\"at\,M\"unchen,\,\,Scheinerstra{\ss}e\,1,\,81679\,\,M\"unchen, Germany
  \and
  Dipartimento di Fisica e Astronomia "Augusto Righi", Universit\`a di Bologna, via Piero Gobetti 93/2, I-40129 Bologna, Italy
}

\date{Received March 4, 2026; accepted April 25, 2026}

\titlerunning{A VLT/MUSE view of the mechanisms unwinding the spiral arms of cluster galaxies}

\abstract
      {The environmental mechanisms driving unwinding spiral arms in cluster galaxies remain debated. While earlier studies attributed this phenomenon primarily to gravitational interactions, more recent works suggest that ram-pressure stripping (RPS) alone can induce unwinding arms.}
   {Using VLT/MUSE observations, we present a spatially resolved analysis to investigate the external mechanisms responsible for spiral-arm unwinding. We focus on two galaxies, UG101 and UG103, selected from a larger sample of unwinding systems. They are selected as tidal- and RPS-driven candidates, respectively, based on the presence/absence of close neighbors.}
   {We estimate the galactocentric radius at which tidal forces, either from a companion or the cluster potential, become relevant ($R_{\mathrm{tid}}$). We examine gas and stellar kinematics, exploiting how these two components respond differently to gravitational and hydrodynamical perturbations. The spectrophotometric code \sinopsis is used to map stellar populations in different age bins and constrain the unwinding timescale for each galaxy.}
   {For UG101, we find $R_{\mathrm{tid}}$\,$\approx$\,$1.5\,R_e$, while the unwound features extend beyond this radius. Additionally, UG101 shows irregular stellar and gas kinematics; its rotation curve indicates similar stellar and gas motions, although the gas appears truncated on one side and extended on the other. For UG103, neither the closest companion ($r_{\rm proj}$$\sim$117\,kpc), nor the cluster appear capable of triggering the unwinding arms through tidal forces.
   UG103 displays regular stellar kinematics but disturbed gas kinematics, with truncation on the disk side that likely faces the intracluster medium wind, and gas extended toward the opposite direction. Stellar population maps show the emergence and subsequent unwinding of the spiral arms in UG103 on timescales consistent with its inferred cluster infall time ($\sim$\,1.6\,Gyr).}
   {We conclude that spiral-arm unwinding in UG101 and UG103 is primarily driven by tidal interactions and RPS, respectively, although a combined effect cannot be excluded in the case of UG101. Our methodology provides a robust framework to disentangle the mechanisms driving spiral-arm unwinding in cluster galaxies from their spatially resolved properties.}
   
\maketitle

\section{Introduction}
\label{sec:intro}

Understanding galaxy evolution requires a detailed view of the physical mechanisms that regulate and modify the cold gas reservoir of galaxies. Internal processes, such as the gravitational influence of spiral arms and bars as well as outflows driven by stellar feedback or active galactic nuclei (AGN), can remove gas from the interstellar medium (ISM) \citep{Tremonti2004, Finlator2008}. Conversely, inflows from the intergalactic and circumgalactic media (IGM and CGM, respectively) can replenish the gas supply required for sustained star formation \citep[SF,][]{Tumlinson2017}. In dense environments, however, additional external mechanisms can strongly influence the multiphase gas content of galaxies, in particular within galaxy clusters and their hot intracluster medium \citep[ICM,][]{Dressler1980, Kenney1989, Cortese2021, Boselli2022, Poggianti2025}.

In addition to regulating the gas content, environmental mechanisms can disturb the morphology of galaxies mainly through gravitational and/or hydrodynamical perturbations.
Gravitational perturbations---including tidal interactions with the cluster potential \citep{Byrd1990, Valluri1993} or with close companions---can distort both the stellar and gaseous components.
As interacting pairs approach each other, the influence of the gravitational field upon both their gas and stellar components increases, acting differently in one side of the galaxy compared to the other \citep{Bournaud2004, Duc2013}. After plunging into a tidal field, galaxy material can be deformed, often resulting in tidal features such as tails, bridges and plumes, although the observation and formation of these features depend on the evolutionary stage of the system and on the initial morphology of the perturbed galaxy \citep{Struck1999, Duc2013}. Tidal torques acting at large radii can push material outward, enhancing the formation of these features \citep{Bournaud2010}.
In the cluster environment, the frequency of galaxy–galaxy interactions and their efficiency in triggering significant morphological disturbances depend on the clustercentric distance. Although galaxy density increases toward the cluster center---consequently encounters become more frequent---the relative speeds of galaxies also increase in these regions. As a result, both the duration of the interactions and their capacity to induce prominent morphological features is decreased \citep{Boselli2006, Adams2012}. Nevertheless, these frequent high-speed encounters at small clustercentric radii can still induce substantial mass loss through a process widely known as ``harassment'' \citep{Moore1996, Moore1998}.

Among a variety of hydrodynamical processes experienced by cluster galaxies, ram-pressure stripping \citep[RPS;][]{Gunn1972} is one of the most extensively studied \citep[e.g.,][among many others]{Abadi1999, Poggianti2017, Cortese2021, Boselli2022}. As galaxies fall into a cluster, the ICM exerts pressure on the ISM cold gas, while the stellar component is less affected \citep{Kapferer2009, Smith2010}. Consequently, the gas disk is compressed along the leading edge \citep{Rasmussen2006, Poggianti2019b} and stripped away along the trailing edge, which creates extended gas tails in the wake of the galaxy \citep{Fumagalli2014, Poggianti2017, Poggianti2019a}. 
The morphological perturbation in the gas component of RPS galaxies undergoing the peak of the stripping phase is so intense that they are often referred to as ``jellyfish'' galaxies due to the peculiar appearance of their extended gas tails, resembling the tentacles of the sea animal.
Although RPS can enhance the star formation rate (SFR) during the peak of the stripping phase \citep{Vulcani2018b}, the long-term aftermath of removing the cold gas reservoir is the quenching of SF \citep{Vulcani2020a, Cortese2021}. Moreover, RPS has been shown to be sufficient to induce a morphological evolution from spirals to S0s \citep{Marasco2023, Marasco2026}.

Recently, \citet[][B21 hereafter]{Bellhouse2017, Bellhouse2021} provided observational evidence for another morphological consequence of RPS in spiral galaxies. Using 11 confirmed RPS galaxies from the GASP \citep[GAs Stripping Phenomena in galaxies;][]{Poggianti2017, Poggianti2025} sample, B21 showed that their spiral arms open up with increasing galactocentric distance, i.e. they appear ``unwound''. Spatially resolved star formation history (SFH) maps revealed that the unwound component hosts exclusively young stellar populations, whereas older stars remain confined to the undisturbed stellar disk, therefore indicating that the unwound component traces stars formed in the gas displaced by the ICM wind.
Observational evidence for RPS-induced unwinding spiral arms has also been reported for NGC~2276 \citep{Matijevic2025}\footnote{The relative contribution of RPS and tidal interactions to the morphological perturbations observed in NGC~2276 has long been debated \citep[e.g., see][]{Forbes1992, Wolter2015, Ilina2016, Tomicic2018}. For a comprehensive and updated discussion on this matter, we refer the reader to \citet{Matijevic2025}.} 
and for UGC 2665 \citep{George2025}. This observational picture is consistent with hydrodynamical simulations showing that RPS compresses and displaces the inner gas disk relative to the halo potential while stretching and shearing the outer arms before stripping them away \citep{Schulz2001, Machado2025}.

However, tidal interactions can produce qualitatively similar curved or open spiral features \citep[e.g.,][]{Struck1999, Dobbs2010}. Because these features have traditionally been associated to gravitational perturbations, unwinding galaxies without prominent optical tails have often been deliberately excluded from visually selected RPS galaxy samples \citep{Poggianti2016, Durret2021}. The exclusion of this subpopulation may lead to an underestimation of the role and efficiency of RPS in driving galaxy evolution in dense environments. Indeed, \citet{Vulcani2022} showed that including unwinding galaxies can double the fraction of ram pressure stripped candidates among blue late-type cluster galaxies.
Moreover, it is worth noting that the presence of stellar bars has been also reported to influence the geometry of spiral arms. For instance, observational studies have found a correlation between the presence of strong stellar bars and increased pitch angles \citep{Hart2017, Masters2019}, a relationship supported by the invariant manifold theoretical framework \citep{Romero-Gomez2006, Romero-Gomez2007}, which suggests that the orbits of stars in barred galaxies should produce more open spirals as bar strength increases. However, this remains a matter of debate, as other observational studies find no such correlation \citep{Font2019, Lingard2021}.

Identifying the physical mechanism responsible for the observed morphological perturbations is therefore crucial for improving the accuracy of RPS galaxy censuses. Integral field spectroscopy (IFS) is particularly suited for this task: it provides spatially resolved emission-line diagnostics, kinematics, and stellar population properties for each galaxy, enabling a direct comparison of how the gas and stellar components respond to the disturbing mechanisms \citep{Merluzzi2016, Vulcani2021}. As an example, regular stellar and gas rotation fields may be transformed into irregular distributions depending on the disturbing forces. For a comprehensive characterization of how galaxy-galaxy interactions, mergers and environmental mechanisms affect the spatially resolved properties of galaxies, with a wealth of observational examples from the GASP sample, we refer the reader to \citet{Vulcani2021}.

Motivated by the need to understand the nature and frequency of RPS-induced unwinding spiral arms, \citet{Vulcani2022} identified 143 blue ($B$\,<\,18.2) spiral galaxies displaying unwinding features across 52 clusters using WINGS and OMEGAWINGS imaging \citep{Fasano2006, Moretti2014, Gullieuszik2015, Moretti2017}, visually classifying them from UClass = 1 to 5 according to the strength of the unwinding signature. 
To investigate the underlying mechanisms at play, ESO program 109.23DA (PI: B. Vulcani) obtained IFS observations with the multi-unit spectroscopic explorer \citep[MUSE,][]{Bacon2010} for 13 of these galaxies. Combined with existing GASP data, these observations provide the first statistically robust IFS sample of unwinding galaxies, allowing an assessment of the relative incidence of RPS-driven spiral-arm unwinding in clusters.

In this work, we characterize the spatially resolved ionized gas, stellar, and morphological properties of two galaxies from this sample: UG103, a candidate for RPS-driven unwinding, and UG101, likely disturbed by tidal forces.
We use these two systems to illustrate how IFU observations can be used to reveal and distinguish the external mechanisms driving spiral-arm unwinding. This analysis establishes the methodological framework to be applied consistently to the full dataset, enabling the classification of galaxies into RPS- and tidal-driven cases and the assessment of the relative incidence of RPS within this particular population of cluster galaxies (Lassen et al. in prep.).
Section~\ref{sec:obs} describes the observations and data reduction, followed by the methodology in Sect.~\ref{sec:methods}. The selection of UG101 and UG103 as tidal- and RPS-driven candidates, respectively, is discussed in Sect.~\ref{sec:obs_justification}, with a quantification of the unwinding effect on each galaxy presented in Sect.~\ref{subsec:pitch_angles}. The results are shown in Sect.~\ref{sec:results} and discussed in Sect.~\ref{sec:discussion}.

Throughout this work, we adopt a \citet{Chabrier2003} Initial Mass Function (IMF) with a 0.1\,$M_{\odot}$\,--\,100$M_{\odot}$ stellar mass range, and standard cosmological parameters of $H_0 = 70\,$km\,s$^{-1}$\,Mpc$^{-1}$, $\Omega_{m} = 0.3$ and $\Omega_{\Lambda} = 0.7$. 

\begin{figure}[!t]
    \centering
    \includegraphics[width=0.9\columnwidth]{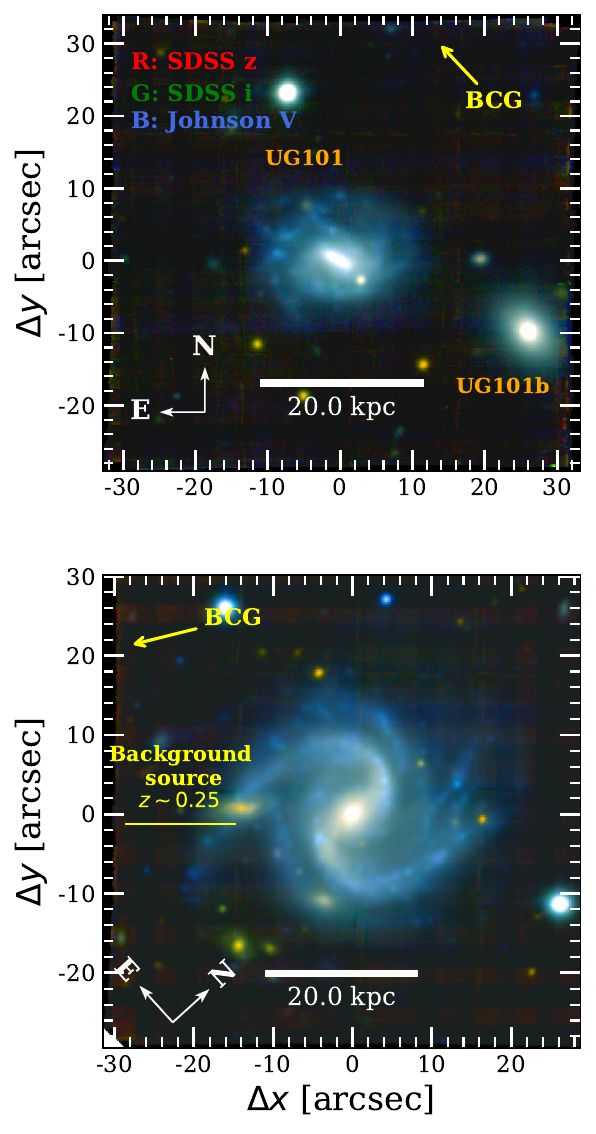}
    \caption{\textit{Top panel}: Color-composite image of UG101 (at the center of the image, labeled in orange) generated from the MUSE datacube. The filter transmission curves used to integrate the flux and their corresponding RGB channels are indicated at the top. A close companion, UG101b, lies approximately 28\arcsec\ to the southwest (also indicated in orange).
    \textit{Bottom panel}: Color-composite image of UG103, generated from the MUSE datacube. The extended yellow/green source located within one of UG103 spiral arms at southeast (indicated by the yellow annotation) is identified as a background source with $z_{\mathrm{spec}}$$\sim$\,0.25\,(see Appendix \ref{appendix:back_source}). Other point-like sources seen in the image correspond to foreground stars or background galaxies. A yellow arrow indicates the direction of the brightest cluster galaxy \citep{Biviano2017}.}
    \label{fig:rgb_figs}
\end{figure}

\section{Observations and data reduction}
\label{sec:obs}
We use IFU observations from VLT/MUSE in the wide-field mode (NOAO-WFM). MUSE covers the approximate nominal wavelength range of 4650\AA\,--\,9300\AA, with a spectral sampling of 1.25\,\AA/pixel. The spectrograph resolving power varies from $R$$\sim$2000 at 4650\,\AA\, to $R$$\sim$4000 at 9300\,\AA, with a median spectral resolution of $\sim$2.6\,\AA. The WFM covers a field-of-view (FoV) of approximately 1\arcmin\,$\times$\,1\arcmin with angular sampling of 0\farcs{2}/pixel. In this work, we use the available science-ready MUSE datacubes that are part of the observing program 109.23DA (P.I.: B. Vulcani). As mentioned in Sect.~\ref{sec:intro}, throughout this work we focus on two galaxies, UG101 and UG103 (see Fig.~\ref{fig:rgb_figs}). UG101 was observed on 2022 April 4, UG103 was observed on 2022 May 24; both with a total exposure time of 2700\,s.
Before detailing how these two galaxies were selected as the representative candidates of tidal and RPS-driven unwinding galaxies, we described in the next section the methodologies applied.

\section{Methods}
\label{sec:methods}
This section presents in detail the methods used to derive the spatially resolved properties of the unwinding galaxies. These techniques are applied to the entire sample and will serve as the methodological framework for forthcoming papers based on this dataset.

\subsection{Preliminary steps}
\label{subsec:prelim_methods}
As a first step, we use GAIA DR3 \citep{Gaia2023}, Pan-STARRS1 \citep{Chambers2016} and DES DR2 \citep{Abbott2021} to perform an astrometry correction of the datacubes. We identify non-saturated stars within the FoV covered by the MUSE observations and use their celestial coordinates as reference values, which are compared to the detector coordinates derived from 2D Moffat profiles fitted to the corresponding sources in the white-light MUSE images. Based on these matched positions, a corrected World Coordinate System (\verb|WCS|) is computed and applied to the original datacubes.
When a source appears in multiple catalogs, the reference value is selected according to the catalog providing the best precision.

Additionally, the datacubes are corrected for Galactic foreground extinction. We adopt the reddening law of \citet[CCM]{Cardelli1989} with a total-to-selective extinction ratio of $R_V$\,=~3.1. The color excess $E_{(B-V)}$ due to Milky Way dust along the line of sight is estimated from the dust maps of \citet{Schlafly2011}, at the position of each galaxy on the sky. After these corrections, we apply an average filter along the spatial direction of the datacubes, as described in \citet{Poggianti2017}, using a 5~$\times$~5 pixel kernel, corresponding to the seeing of our observations ($\sim$1\arcsec).

\subsection{Modeling stellar emission and kinematics with full-spectral fitting}
\label{subsec:ppxf}
A proper treatment of the stellar continuum emission is essential for retrieving stellar kinematics and subtracting the stellar component to accurately estimate nebular emission line fluxes, particularly for the Balmer recombination lines, whose line profiles are often superimposed on absorption features produced by young and intermediate-age stellar populations \citep{Gonzalez1999a,Gonzalez1999b}.

To ensure a sufficiently high signal-to-noise ratio (S/N) in the modeled spectra, we apply the Voronoi binning technique \citep{Cappellari2003}, using the weighted Voronoi tessellation adaptation \citep{Diehl2006}. The binning targets S/N~=~30 
\footnote{This S/N ensures high-quality spectral fitting and reliable stellar kinematic maps, which are essential to distinguish between gravitational and hydrodynamical perturbations.} across featureless continuum spectral regions.
These spectral windows are carefully selected via visual inspection for each galaxy to avoid contamination from emission-line features from potentially unmasked foreground/background sources within the FoV. For UG101 and UG103, these regions correspond to 
$5130 \lesssim \lambda_{\rm rest}$\,[\AA]\,$\lesssim 5300$ and
$5320 \lesssim \lambda_{\rm rest}$\,[\AA]\,$\lesssim 5490$, respectively. Stars and spaxels with S/N~$\leqslant 2$ over these spectral windows are masked prior to the binning.

We model the stellar continuum emission in each Voronoi bin using the full-spectral fitting code \textsc{pPXF} \citep[Penalized Pixel-Fitting;][]{Cappellari2004, Cappellari2017}. 
The input spectra are fitted by \textsc{pPXF} through a convolution of simple stellar population (SSP) templates with the line-of-sight velocity (LOS) distribution. To ensure that any deviations between the fitted templates and the observed spectra originate solely from the broadening caused by the galaxy LOS distribution, we match the spectral resolution of the stellar templates to the instrumental line spread function (LSF\footnote{Throughout this work we adopt the polynomial parameterization of MUSE LSF derived by \citet{Guerou2017}: FWHM$_{\rm{inst}}\,$[\AA] = $5.866 \times 10^{-8} \lambda^{2} - 9.187 \times 10^{-4}\lambda +6.04$.}) using a 1D wavelength-dependent Gaussian kernel. We adopt the MILES \citep{Vazdekis2010} SSP models, with a \citet{Chabrier2003} initial mass function (IMF) and the PADOVA2000 evolutionary tracks \citep{Girardi2000}, within the stellar mass range of 0.1\,$M_{\odot}$--100\,$M_{\odot}$. Due to the contamination from residual sky lines at NIR wavelengths, the fitting is restricted to $\lambda_{\mathrm{rest}} \leqslant 7000\,$\AA.
We derive stellar emission intensity, LOS velocity and velocity dispersion maps (moment 1 and 2, respectively) and the $h3$ and $h4$ moments using a 12th-order additive Legendre polynomial to correct the shape of the template continuum during the fit \citep[e.g., as in][]{Poggianti2017}.

\subsection{Modeling ionized gas emission}
\label{subsec:linefitting}
\subsubsection{Spaxel selection}
When the galaxies of interest occupy a relatively small portion of the observed FoV, fitting all the spaxels is unnecessary and  computationally inefficient, as many spaxels contain no relevant information. To pre-select the spaxels for the emission-line fitting, we first create a 2D boolean mask based on the S/N of H$\alpha$, which is typically the brightest emission feature. An accurate estimate of $(\mathrm{S/N})_{\mathrm{H}\alpha}$ is formally obtained from the fitting procedure itself; however an initial proxy is required at this stage. We therefore estimate $(\mathrm{S/N})_{\mathrm{H}\alpha}$ prior to fitting by applying a tophat filter centered on the H$\alpha$---computed using the stellar redshift derived from \ppxf---and adopt the following approximation: \citep{Rola1994}:

\begin{equation}
    (\mathrm{S/N})_{\mathrm{H}\alpha} \approx \frac{\sqrt{2\pi N}}{6} \times \frac{A_{\mathrm{H}\alpha}}{\sigma_{\mathrm{cont}}}
    \label{eq:sn_approx}
\end{equation}

\noindent where $A_{\mathrm{H}\alpha}$ is the peak flux value within the spectral window, $N$ is the number of spectral pixels within this window, and $\sigma_{\mathrm{cont}}$ is the standard deviation of the flux density in adjacent featureless continuum regions.
We note that contamination from nearby [\ion{N}{ii}] lines is possible, and in particular physical conditions the flux peak can correspond to [\ion{N}{ii}]~$\lambda6583$ instead of H$\alpha$. However, this does not impact the spaxel selection.
Since Eq.~\ref{eq:sn_approx} provides only an approximate estimate and to ensure faint ionized gas emission is not missed, we adopt a conservative selection threshold of $(\mathrm{S/N})_{\mathrm{H}\alpha} \geqslant 1$ to identify the spaxels to be included in the emission-line fitting.

\subsubsection{Emission line fitting}
To estimate the ionized gas physical properties, we fit the
following emission lines: H$\alpha$, H$\beta$,
[\ion{O}{iii}]~$\lambda\lambda$4959,5007,
[\ion{O}{i}]~$\lambda\lambda$6300,6364,
[\ion{N}{ii}]~$\lambda\lambda$6548,6583,
[\ion{S}{ii}]~$\lambda\lambda$6716,6731.
To model the emission-line profiles, we use the \ifscube\footnote{\label{foot:ifscube}\url{https://ifscube.readthedocs.io/en/latest/}} Python package \citep{ifscube2021}, adopting single Gaussian components, and using the same spectral range applied for the stellar continuum fitting described in Sect.~\ref{subsec:ppxf}.
The continuum emission is modeled using the best-fit stellar emission from \textsc{pPXF}. To model the pseudo-continuum after subtraction of the stellar emission, a supplementary 5th-order polynomial is adopted, with a 3$\sigma$ rejection threshold. 
As described in Sect.~\ref{subsec:ppxf}, the best-fit stellar emission is obtained for each Voronoi bin. To rescale these solutions to the native spaxel resolution, we adopt the simplifying assumption that the stellar emission at the spaxel resolution  differs from the corresponding binned solution only by a multiplicative scale factor. This factor is derived from the median difference between the observed spectrum of each spaxel and the corresponding bin spectrum, after masking all emission lines. For a previous application of this methodology, we refer the reader to \citet{Cid2013, DellaBruna2020}. 

A different strategy is adopted for the emission-line fitting of spaxels with $\mathrm{S/N}\leqslant 2$ over the featureless continuum. This is a relatively common situation, since H$\alpha$ emission is often detected at larger galactocentric distances than the stellar continuum. At these locations, we instead adopt a boxcar median filter to model the continuum, using a spectral width of $\Delta \lambda = 200\,$\AA
\footnote{Although this approach does not account for absorption features underlying the Balmer lines, it is more robust to variations in the continuum shape than high-order polynomials. The adopted spectral width is large enough to prevent strong emission lines or sky residuals from biasing the derived medians. Moreover, this method is typically employed in the galaxy outskirts, where the intensity of the detected emission lines is already weak.}.
As constraints, we assume the theoretical line ratios
[\ion{O}{iii}]\,($\lambda$5007/$\lambda$4959)~=~2.98 and
[\ion{N}{ii}]\,($\lambda$6583/$\lambda$6548)~=~3.07 \citep{Storey2000, Osterbrock2006}, as well as kinematic groups (i.e., transitions from the same ion species have bound kinematical properties).  
To estimate the uncertainties on the amplitudes and two first moments of the fitted Gaussian profiles, we perform 150 Monte Carlo Markov Chain (MCMC) realizations per spectrum, drawing realizations from the variance of the datacube. Emission-line flux uncertainties are computed by \ifscube following the empirical relation \citep{Lenz1992, Wesson2016}:

\begin{equation}
    \delta f = \frac{f}{0.67} \times \frac{\delta A}{A} \times \sqrt{\frac{\Delta \lambda}{\mathrm{FWHM}}}
    \label{eq:flux_errors}
\end{equation}

\noindent where $f$ is the integrated emission-line flux, $A$ and $\delta A$ are the amplitude of the line profile and its uncertainty, respectively, $\Delta \lambda$ is the wavelength sampling interval, and FWHM is the spectral full width at half maximum. In Appendix~\ref{appendix:spectra} we present examples of fitted spectra for reference.
The physical properties of the ionized gas for each galaxy, derived from the emission-line fitting technique described in this section, are presented in Sect.~\ref{sec:results}.

\subsection{Spatially resolved star formation histories and stellar masses}
\label{sec:sinopsis}
We model the spatially resolved star formation history (SFH) of our galaxies using the spectrophotometric code \sinopsis \citep{Fritz2007, Fritz2011, Fritz2014, Fritz2017}, which uses the stellar redshifts derived from \ppxf as input.
In order to model the stellar emission, \sinopsis searches for the combination of SSP models that best reproduces the observed spectra, based on the measurement of the equivalent widths of emission and absorption features covering all the spectrum.
Average star formation rates, as well as the mass of stars formed, are computed within four main age bins:
SFR1 (ongoing star formation)$\equiv$ $t < 2 \times 10^7$\,yr, 
SFR2$\equiv$ $2 \times 10^7 \leqslant t\,[\mathrm{yr}]< 5.7 \times 10^8$, 
SFR3$\equiv$ $5.7 \times 10^8 \leqslant t\,[\mathrm{yr}] < 5.7 \times 10^9$, and 
SFR4$\equiv$ $t \geqslant 5.7 \times 10^9$\,yr. For a detailed discussion regarding the choice of these age intervals, we refer the reader to Sect.~5 of \citet{Fritz2007}.

For the youngest age bin, \sinopsis incorporates predicted Balmer line intensities into the models, obtained from the Spectral Energy Distribution (SED) of $t \leqslant 20\,$Myr stellar ionizing sources using \textsc{cloudy} \citep{Ferland2013}. Dust extinction is estimated from the observed H$\alpha$/H$\beta$ emission line ratio and applied to the models following the selective extinction hypothesis \citep{Calzetti1994}. The full-spectral fitting is performed on a spaxel-by-spaxel basis, and spaxels without detectable H$\alpha$ emission (S/N$_{\mathrm{H}\alpha} \lesssim 3$) are set to SFR1\,$\equiv 0$. To avoid including low S/N data, only spaxels with S/N~$\geqslant$~2 over the featureless continuum are considered for SFR2, SFR3 and SFR4.

We use the spatially resolved estimates of the stellar masses from \sinopsis to compute total stellar masses. These values are obtained by summing the stellar mass over all spaxels within the galaxy boundaries (see Sect.~\ref{subsec:gal_boundaries}) with continuum S/N~$\geqslant 2$.
Additionally, we compute the luminosity- and mass-weighted ages of the galaxies \citep[for a formal definition of these quantities, we refer the reader to equations 7 and 8 of][]{Fritz2011}.

\subsection{Physical properties derived from the emission-line fluxes}
\subsubsection{Dust attenuation correction}
\label{sec:dustcorr}
In order to correct the measured emission-line fluxes by dust attenuation, we calculate the color excess $E_{(B-V)}$ and its uncertainty via Balmer decrement:

\begin{align}
    E_{(B - V)} &= \frac{2.5}{k(\lambda_{\mathrm{H}\beta}) - k(\lambda_{\mathrm{H}\alpha})} \ \mathrm{log}_{10} \bigg[\frac{(\mathrm{H}\alpha/\mathrm{H}\beta)_{\mathrm{obs}}}{(\mathrm{H}\alpha/\mathrm{H}\beta)_{\mathrm{int}}} \bigg]
    \label{eq:ebv}  \\
    \delta E_{(B - V)} &= \frac{1.086}{k(\lambda_{\mathrm{H}\beta}) - k(\lambda_{\mathrm{H}\alpha})} \ \sqrt{\bigg(\frac{\delta \rm{H}\alpha}   {\rm{H}\alpha}\bigg)_{\mathrm{obs}}^{2} + \bigg(\frac{\delta \rm{H}\beta}{\rm{H}\beta}\bigg)_{\mathrm{obs}}^{2}}
    \label{eq:unc_ebv}
\end{align}

\noindent where H$\alpha$ and H$\beta$ are the emission-line fluxes, $\delta$H$\alpha$ and $\delta$H$\beta$ are their corresponding uncertainties, and $k(\lambda_{\mathrm{H}\alpha})$ and $k(\lambda_{\mathrm{H}\beta})$ are the values of the adopted dust reddening law evaluated at the wavelengths of H$\alpha$ and H$\beta$, respectively. As mentioned earlier, we adopt the reddening law of \citet{Cardelli1989} with $R_V$\,=~3.1.
Assuming case B recombination (i.e., photons are reabsorbed immediately after being emitted within the nebula), an electron density of $N_e$\,=~100\,cm$^{-3}$, and an electron temperature of $T_e$\,=~$10^{4}$\,K, the intrinsic Balmer decrement is $(\mathrm{H}\alpha/\mathrm{H}\beta)_{\mathrm{int}}$\,=~2.863 \citep{Osterbrock2006, Lopez-Sanchez2015}.

\subsubsection{Ionization mechanisms}
\label{subsec:bpt}
To determine the dominant ionization mechanism across the ISM, we use the emission-line fluxes to construct spatially resolved diagnostic diagrams, commonly referred to as the BPT diagrams \citep{Baldwin1981, Veilleux1987}.
We use three diagnostic diagrams, namely BPT--[\ion{N}{ii}], BPT--[\ion{S}{ii}], BPT--[\ion{O}{i}].
In all diagrams we adopt the theoretical dividing line from \citet{Kewley2001} to identify spaxels ionized predominantly by OB stars. For BPT--[\ion{N}{ii}], we use the empirical line from \citet{Kauffmann2003} to identify spaxels with composite ionization. To separate between AGN and LINER-like ionization, we use the dividing lines from \citet{Sharp2010} and \citet{Kewley2006} in BPT--[\ion{N}{ii}] and BPT--[\ion{S}{ii}]/BPT--[\ion{O}{i}], respectively. Only emission lines with S/N\,$\geqslant 4$ are used to classify the ionization mechanism in each galaxy.
The identification of the dominant ionizing mechanism is done following the classification of each spaxel according to the BPT--[\ion{N}{ii}] diagram.

\subsubsection{Star Formation Rates}
\label{subsec:SFRs}
Since the SFR is proportional to the production rate of ionizing photons \citep{Kennicutt1998}, Balmer recombination lines are particularly useful to trace the SFR over short ($t \lesssim 10\,$Myr) timescales. This timescale corresponds to the typical lifetime of OB stars, the main responsible for the ionizing flux budget in SF regions. We adopt the \citet{Kennicutt1998} calibration, considering a \citet{Chabrier2003} IMF:

\begin{equation}
    \log\bigg[\frac{\mathrm{SFR}}{M_{\odot}\,\mathrm{yr}^{-1}} \bigg] = \log\bigg[\frac{L_{\mathrm{H}\alpha, 0}}{\mathrm{erg}\,\mathrm{s}^{-1}} \bigg] - 41.34
    \label{eq:SFR}
\end{equation}

\noindent where $L_{\mathrm{H}\alpha, 0}$ is the luminosity of de-reddened H$\alpha$ emission line. To compute $L_{\mathrm{H}\alpha, 0}$ we calculate the luminosity distance $d_L$ considering the redshift of each galaxy. We apply Eq.~\ref{eq:SFR} only to those spaxels identified as SF by the BPT--[\ion{N}{ii}] diagnostic diagram. We compute total SFRs by summing the contribution of all star forming spaxels with S/N~$\geqslant 4$ in H$\alpha$.

\begin{table*}[!t]
    \centering
    \caption{Global properties of UG101, its companion UG101b, and UG103.}
    \label{tab:galaxy_props}
    \begin{tabular}{llll}
        \toprule
        \toprule
        \textbf{Property} & \textbf{UG101} & \textbf{UG101b} & \textbf{UG103} \\
        \midrule
        $z$ &
        $0.04548 \pm 4.6\times10^{-5}$ &
        $0.04871 \pm 4.6\times10^{-5}$ &
        $0.05387 \pm 4.7\times10^{-5}$ \\
        
        RA (ICRS)\tablefootmark{a} &
        13:27:16.67 &
        13:27:14.61 &
        13:27:03.41 \\
        
        DEC (ICRS) &
        -31:49:14.67 &
        -31:49:24.24 &
        -31:13:31.49 \\
        
        $z_{\rm cl}$\tablefootmark{b} &
        0.04829 & 
        0.04829 & 
        0.04829 \\

        $\sigma_{\rm cl}$\,[\si{\kilo\metre\per\second}] &
        $910^{+44}_{-46}$ &
        $910^{+44}_{-46}$ &
        $910^{+44}_{-46}$ \\
        
        $\log (M_{\bigstar}/M_\odot)$ &
        $9.94^{+0.05}_{-0.08}$ &
        $9.90^{+0.03}_{-0.06}$ &
        $10.80^{+0.04}_{-0.04}$ \\
        
        $\mathrm{SFR}$ [$M_{\odot}/\mathrm{yr}$]\tablefootmark{c} &
        $0.77 \pm 0.08$ &
        $(9.1 \pm 0.9)\times10^{-3}$ &
        $2.4 \pm 0.2$ \\
        
        $R_e$ [kpc] &
        $5.2 \pm 0.4$ &
        $1.75 \pm 0.07$ &
        $8.9 \pm 0.6$ \\
        
        PA [$^\circ$] &
        $90.8 \pm 2.1$ &
        $39.8 \pm 0.6$ &
        $160.0 \pm 1.4$ \\
        
        $i$ [$^\circ$] &
        $42.5 \pm 1.1$ &
        $41.7 \pm 0.3$ &
        $34.4 \pm 1.2$ \\
        \bottomrule
    \end{tabular}
    \tablefoot{
    For a description of how each property has been derived, see Sect.~\ref{sec:results}.
    \tablefoottext{a}Coordinates of galaxy center, determined in this work (see Sect.~\ref{subsec:gal_boundaries}).
    \tablefoottext{b} Properties of Abell 3558, the cluster to which all galaxies belong, are  from \citet{Biviano2017}. \tablefoottext{c}For UG101b, the reported SFR is not corrected for dust attenuation, given that H$\beta$ emission is not detected in this galaxy.}
\end{table*}

\subsection{Structural parameters}
\subsubsection{Elliptical isophote fitting}
\label{subsec:isophotes}

To derive structural parameters of each galaxy, we use Cousins/$I$ images generated from the MUSE datacube to model surface brightness profile of each galaxy using elliptical isophote fitting, employing the iterative method extensively described by \citet{Jedrzejewski1987} and as implemented by \textsc{photutils}\footnote{\label{foot:photutils}\url{https://photutils.readthedocs.io/en/stable/}}.
In this approach, the intensity profile is expressed in terms of a Fourier series expansion in azimuthal angle, with lower-order ($\leqslant2$) coefficients corresponding to physical structural parameters---i.e. ellipse center, position angle (PA) and ellipticity ($\epsilon$)---and higher-order coefficients accounting for deviations from purely elliptical shapes \citep{Monteiro-Oliveira2025}. 
Before performing the fit, \textsc{sep} is applied to the $I$-band images. The generated segmentation maps are used to mask foreground and background sources, and the resulting global background level defines the intensity threshold at which the elliptical isophote fitting stops. To estimate the characteristic extent of the stellar disk, we derive the half-light radius ($R_e$) from the luminosity profile $L(R)$, which can be calculated as follows:

\begin{equation}
    L(R) = 2\pi \int\limits_{0}^{r_{\mathrm{max}}} I(r)\,
    [1 - \epsilon (r)]\,r\mathrm{d}r
    \label{eq:calc_re}
\end{equation}

\noindent where $I(r)$ is the azimuthally averaged surface-brightness profile, $\epsilon (r)$ is the ellipticity profile, and $r$ denotes the semi-major axis. The integration is carried out up to $r_{\mathrm{max}}$, corresponding to the semi-major axis of the outermost fitted isophote, within which $I (r=r_{\mathrm{max}})$ approaches the background level.
$R_e$ is then computed from the expression $L(r=R_e) / L(r=r_{\mathrm{max}})$~=~0.5. To calculate the disk inclination $i$, we adopt the expression:

\begin{equation}
    \cos^2 i = \frac{(1 - \epsilon)^2 - q_0^2}{1 - q_0^2}
    \label{eq:inclination}
\end{equation}

\noindent where we assume an intrinsic ellipticity for galaxies of $q_0 = 0.13$ \citep{Giovanelli1994}. To derive a single inclination value we use the mean ellipticity beyond $R_e$.
Inner isophotes are not considered for the mean because they are derived from fewer data points and are also more sensitive to central structural features such as bars. The derived properties for each galaxy and their estimated uncertainties are listed in Table~\ref{tab:galaxy_props}.

\subsubsection{Galaxy boundaries}
\label{subsec:gal_boundaries}
To estimate the approximate extent of each galaxy's stellar body, we adapt the method described by \citet{Gullieuszik2020} as follows. First, we determine the center of each galaxy by computing the flux-weighted position of all spaxels within an aperture large enough to cover almost the entire galaxy, after masking stars. This mask, determined by visual inspection, ensures no overlapping sources or excessive background contribution can shift the white-light flux-weighted position. The sky coordinates of the center of each galaxy are listed in Table~\ref{tab:galaxy_props}.

Galaxy contours are then computed down to the background + 3$\sigma$ level using the same Cousins/$I$ image as in Sect.~\ref{subsec:isophotes}. This  band is chosen for this analysis because it lies entirely within the MUSE spectral range and, being in the redder part of the spectrum, its continuum emission is dominated by older stellar population. The resulting isophotes therefore trace the stellar light distribution independently of the one derived via the \ppxf full spectral fitting described in Sect.~\ref{subsec:ppxf}. Background subtraction is performed with \textsc{sep} \citep{Bertin1996, Barbary2016}.

In \citet{Gullieuszik2020}, this approach was applied to jellyfish galaxies with extended optical tails. To prevent the tails from dominating the isophotes, an ellipse was fitted to the truncated disk edge and this elliptical shape was then preserved on the opposite side of the disk
In this work, in contrast, we model galaxies without prominent optical tails. Therefore, we adapt the method and fit the ellipses on disk regions free of spiral arms to prevent the unwound arms from driving the shape of the isophotes.  
The derived contours provide a visual reference for the extent of the main stellar body, particularly in Sects.~\ref{sec:tidal_results} and \ref{subsec:diagnosis_RPS}, where we assess the galactocentric radius at which tidal forces become significant.

\section{Selection of representative candidates}
\label{sec:obs_justification} 
Within the sample of 13 galaxies observed with MUSE by the ESO program 109.23DA (PI: B. Vulcani), all characterized by strong perturbations and a nearly face-on orientation, we now describe the selection of two unwinding galaxies, chosen as candidates for RPS- and tidal-driven perturbations, whose stellar and ionized-gas properties will be used to distinguish between these two mechanisms.

\subsection{UG101: A tidal interactions candidate}
\label{subsec:selection_tidal}
To select a tidal candidate, we require the presence of at least one nearby companion in projection, preferentially fully covered by the VLT/MUSE observations to enable a more complete characterization of the potentially interacting system. UG101 (top panel of Fig.~\ref{fig:rgb_figs}, also known as WINGS J132716.69-314914.4), satisfies this criterion: it has a companion, UG101b (WINGS J132714.64-314924.4), located at a projected distance of  $\sim$28\arcsec\ to the south-west (right bottom corner in the top panel of Fig.~\ref{fig:rgb_figs}), corresponding to a physical distance of $\sim$26.5\,kpc at the redshift of the host cluster \citep[Abell 3558 in the Shapley supercluster, $z_{\mathrm{cl}}$~=~0.04829, $\sigma_{\mathrm{cl}} = 910 \pm 44$\,km/s, $R_{200} = 1.95_{-0.11}^{+0.16}\,$Mpc;][]{Biviano2017}. The system lies at projected cluster-centric distances of  1.23\,Mpc (i.e., 0.63\,$R_{200}$).

Using the VLT/MUSE observations, we estimate $z=0.04548\pm0.0005$ for UG101 and $0.04871\pm0.00005$ for UG101b, in good agreement with previous literature values \citep[$0.04512\pm0.0002$ and $0.04874 \pm 0.0001$, respectively;][]{Moretti2017, Haines2018}. The main physical properties of the two galaxies, including stellar mass, redshift, and projected distance, are listed in Table~\ref{tab:galaxy_props}. The projected velocity of UG101 implies $\left | v/\sigma_{\rm cl} \right |$\,$\approx$\,0.9, while the relative speed between the two galaxies ($\left | v_{\rm diff} \right |$\,$\sim$\,$923 \pm 19$\,km/s) does not exclude a short-lived flyby. The uncertainties and assumptions behind these estimates are discussed in Sect.~\ref{sec:tidal_results}.
Morphologically, the disturbed arms of UG101 are preferentially extended eastward, opposite to UG101b. Although no tidal bridge is detected, one-sided tidal arms can arise in galaxy pairs \citep[e.g.,][]{Struck2012, Wen2016}. A quantitative assessment of the tidal force potentially exerted by UG101b on UG101 is presented in Sect.~\ref{sec:tidal_results}.
The location of UG101 in the projected phase-space diagram \citep{Rhee2017} places it in a relatively uncertain region: it is most likely ($\sim$40\%) a recent cluster infaller, but there is a non-negligible probability ($\sim$15\%) that it is either an intermediate or ancient infaller. In these regions of the phase space, infall times ($t_{\rm inf}$) can vary between 2 and 8 Gyr depending on the galaxy’s accretion history \citep[see Fig. 6 in][]{Rhee2017}.

\subsection{UG103: A RPS candidate}
\label{subsec:selection_rps}
To select the RPS-induced unwinding candidate, we first search for galaxies without a clear nearby companion.
We select UG103 (bottom panel of Fig.~\ref{fig:rgb_figs}), also known as WINGS J132704.25-311338.5, which exhibits a clear unwinding spiral arm southeast. Despite lying in a relatively overdense region \citep[1.4262 galaxies per Mpc$^{2}$;][]{Vulcani2023}, we find that, within a projected radius of 200\,kpc and a velocity difference of $\leqslant$1500\,km/s, its closest neighbor (WINGS J132707.35-311138.1) lies at a projected distance of $\sim$124\arcsec\ ($\sim$117\,kpc at the cluster redshift). To first order, the apparent lack of close companions with the strongly disturbed spiral structure make UG103 a good candidate for RPS-driven unwinding galaxy. 

UG103 is also member of Abell 3558, located at a projected cluster-centric distance of $\sim$1.08\,Mpc (i.e.,  0.55\,$R_{200}$). From its projected velocity, we derive $ \left | v/\sigma_{\rm cl} \right | \approx$ 1.7 for UG103. The location of this on the projected phase-space diagram suggests that it is likely ($\sim$70\%) a recent cluster infaller, with an estimated time since infall of $t_{\rm inf} \sim 1.6$\,Gyr. More details on the galaxy properties are given in Table~\ref{tab:galaxy_props}.

\subsection{Pitch angles}
\label{subsec:pitch_angles}
Until now, unwinding galaxies have been identified primarily through visual inspection, based on the apparent loosening of their spiral arms. UG101 was classified as UClass~=~3 and UG103 as UClass~=~4 according to the visual scheme of \citet{Vulcani2022}.
To move beyond this qualitative selection, we now adopt a quantitative approach by measuring the pitch angle, which characterizes the tightness of the spiral arms. Pitch angles are defined as the angle between each spiral arm and the tangent to a circle in the plane of the disk. A fully wound spiral exhibits a global pitch angle of 0\textdegree, corresponding to a circular morphology, while unwinding arms open up under external perturbations, showing radially increasing pitch angles.
B21 found that, in their sample, pitch angles show a radial dependence, with the inner arms---usually defined as those confined within a galactocentric radius of $2\,R_e$---exhibiting systematically lower pitch angles than arms extending to larger radii. Rather than a progressive opening of the arms, they reported an abrupt increase in the pitch angles beyond $\sim$$2\,R_e$. For RPS-driven unwinding, the numerical simulations presented in B21 show that the resulting unwinding patterns also depend on the ICM wind direction.

To measure the pitch angles, we use both the H$\alpha$ emission-line maps (see Sects.~\ref{subsec:linefitting}, \ref{subsec:SFRs_UG101}, \ref{subsec:SFRs_UG103}) and the color-composite images to trace the spiral arms of each galaxy. 
The maps are first deprojected with \textsc{astwarp} \citep{Akhlaghi2015} by rotating them around the galaxy center according to each galaxy's PA, followed by a 1/$\cos i$ stretching along the minor axis to correct for inclination. 
We then mark points along the spiral arms by visual inspection of the deprojected images; these positions are shown as crosses in Fig.~\ref{fig:pitch_angles}. The spiral arms are traced outward from the galaxy center, and in UG103, we additionally mark clear bifurcations at larger radii. 
The pitch angles of each arm are measured starting from their outermost point. For each marked position, we identify the three nearest points (in polar coordinates) belonging to that same spiral arm and perform a linear fit, iteratively moving toward the galaxy center. The pitch angles are derived from the slope of this fit in polar space. The resulting curves are shown in red in the left-hand panels of Fig.~\ref{fig:pitch_angles}, smoothed using a spline function for improved visualization. 

To confirm the unwinding nature of UG101 and UG103, we measure and analyze their global and radially resolved pitch angles. In Table~\ref{tab:pitch_angles} we report the global mean pitch angles, as well as the mean inner and outer pitch angles, considering all the identified spiral arms of UG101 and UG103 shown in Fig.~\ref{fig:pitch_angles}. For UG101, we measure pitch angles of 28.4$^{\circ}$ and 34.8$^{\circ}$ for the inner and outer arms, respectively, while for UG103 the corresponding values are 28.8$^{\circ}$ and 44.4$^{\circ}$. We compare these measurements with the results reported in B21.

\begin{figure}[!t]
    \centering
    \begin{tabular}{lc}
    \includegraphics[width=\columnwidth]{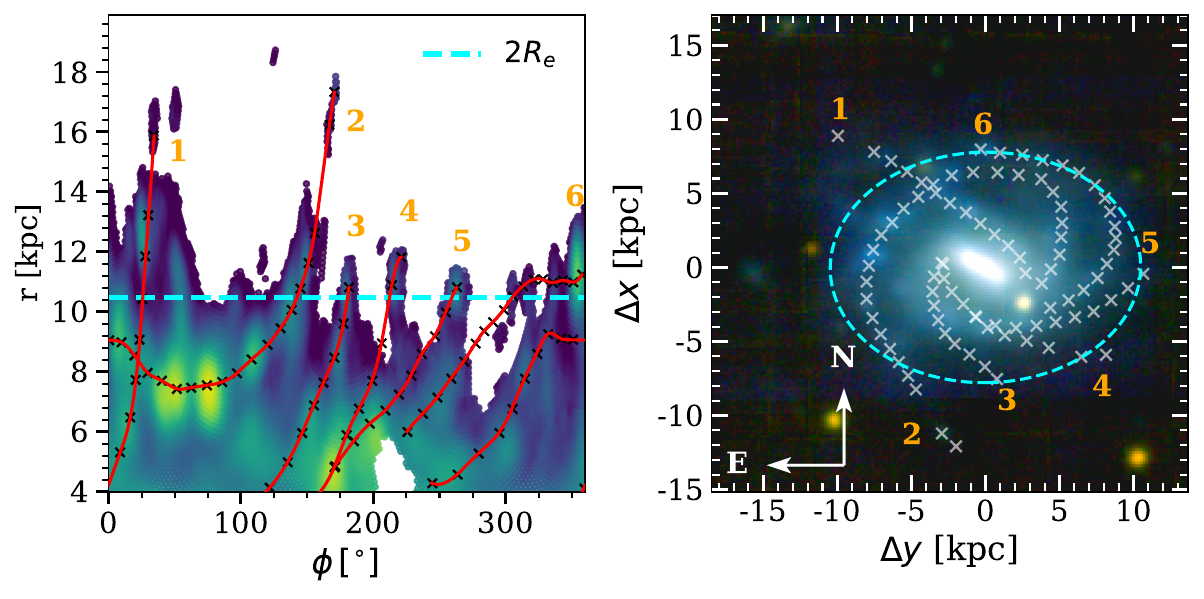} \\
    \includegraphics[width=\columnwidth]{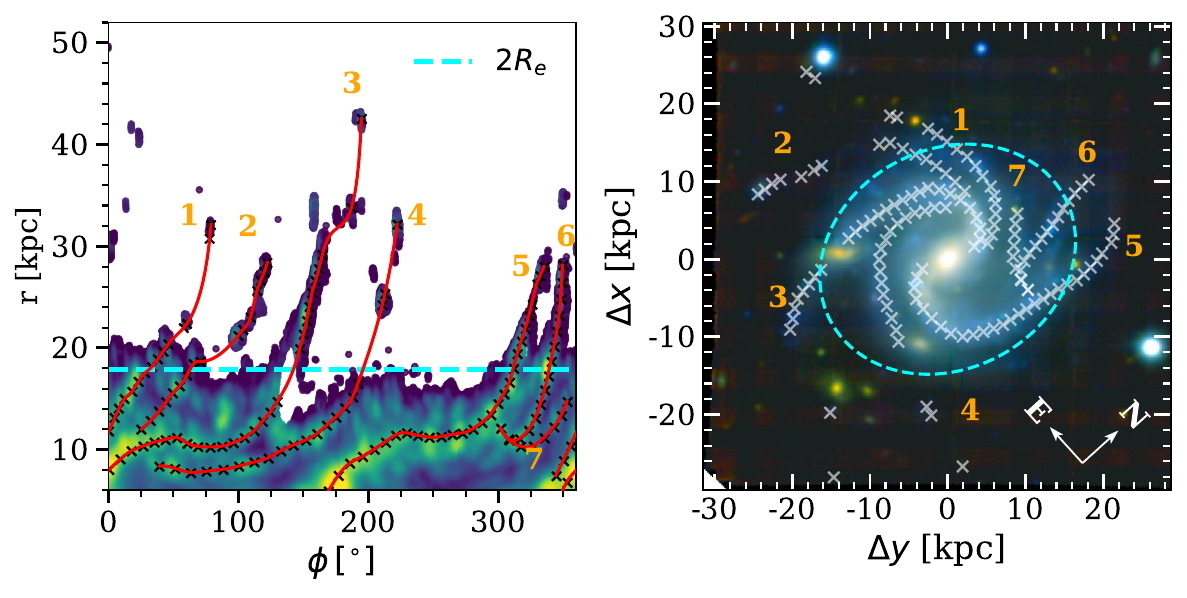} 
    \end{tabular}
    \caption{\textit{Left panels}: Deprojected H$\alpha$ emission-line maps of UG101 (top) and UG103 (bottom), in polar coordinates. The angular coordinate ($\phi$) starts from north direction and increases in the counter-clockwise direction. Only spaxels with S/N~$\geqslant 4$ are displayed, and stars are masked. Positions are marked along the spiral arms of each galaxy (``x'' marks). The inclination of the curve connecting these points is computed to measure the pitch angle of each spiral arm, and the red curves show the smoothed fit. Cyan horizontal line marks $2\,R_e$.
    \textit{Right panels}: Color-composite images of each galaxy with reprojected positions overlaid to highlight the unwound spiral arms. Cyan dashed line shows the elliptical isophote with a semi-major axis equal to $2\,R_e$. In all panels, the orange text marks the labels assigned to each identified spiral arm.}
    \label{fig:pitch_angles}
\end{figure}

For their control sample of undisturbed spiral galaxies, B21 found similarly low pitch angles for inner and outer arms (21.1$^{\circ}$ and 21.3$^{\circ}$, respectively). In contrast, RPS-induced unwinding galaxies show a clear radial increase in pitch angle, from a mean value of 18.2$^{\circ}$ in the inner arms to 37.3$^{\circ}$ in the spiral arms beyond $2\,R_e$. The pitch angles measured for both UG101 and UG103 exceed those of the control sample and show a radial increase consistent with the unwinding scenario, quantitatively supporting their classification as unwinding galaxies, as initially suggested by the visual inspection of \citet{Vulcani2022}.

\section{Results}
\label{sec:results}
We are now in the position of presenting the spatially resolved properties of UG101 and UG103 to identify the dominant external mechanism responsible for their disturbed spiral arms.

\subsection{Confirming the tidal nature of UG101}
\label{sec:tidal_results}
As mentioned in Sect.~\ref{sec:obs_justification}, the projected proximity and apparent fly-by encounter with UG101b make UG101 a suitable system to investigate whether and how gravitational disturbances can drive the spiral-arm unwinding observed in this galaxy. First, we quantify the strength of the tidal forces exerted by UG101b. To this end, we adopt the impulse approximation to express the ratio between tidal and centripetal acceleration as \citep{Henriksen1996, Vollmer2005, Watson2025}:

\begin{equation}
    \left| \frac{a_{\mathrm{tid}}}{a_{\mathrm{gal}}} \right| = \frac{M_{\mathrm{pert}}\,R^2}{M_{\mathrm{gal}}} \times \left| \frac{1}{r^2} - \frac{1}{(r - R)^2} \right|
    \label{eq:aratio}
\end{equation}

\noindent where $r$ is the projected separation between the interacting galaxies, $R$ is the projected galactocentric radius at which tidal features appear visible, and $M_{\mathrm{pert}}$ and $M_{\mathrm{gal}}$ are the stellar masses of the perturbing and main galaxies, respectively\footnote{In all cases considered in this work, we also tested the use of $M_{\rm dyn}$ instead of $M_{\bigstar}$, adopting the stellar-to-halo mass relation from \citet{Girelli2020}. For UG101 and UG101b the results are very similar, primarily because the two galaxies have nearly identical stellar masses. For UG103, the inferred $R_{\rm tid}$ decreases to $\lesssim 1\,$kpc, leaving both the results and their interpretation unchanged.}. 
Following previous works \citep[e.g.,][]{Merluzzi2016, Vulcani2021}, we consider tidal interactions to significantly perturb the morphology of a galaxy when $\left| a_{\mathrm{tid}} / a_{\mathrm{gal}} \right| \geqslant 0.15$. By inverting Eq.~\ref{eq:aratio}, we derive the tidal radius $R_{\rm tid}$ beyond which galaxy regions are expected to be affected by the gravitational field of companion. The region of tidal influence of UG101b on UG101 is indicated by the blue dashed ellipse in Fig.~\ref{fig:td_acc_tidal}, whose semi-major axis corresponds to $R_{\rm tid}$. We find $R_{\rm tid} \approx 1.5\,R_e$, with the ellipse entirely contained within the main body of UG101. Notably, the unwinding features are predominantly observed outside this region. In particular, a few prominent clumpy features are visible on the eastern side of the disk, opposite to the location of the companion.

\begin{figure}[t]
    \centering
    \includegraphics[scale=0.7]{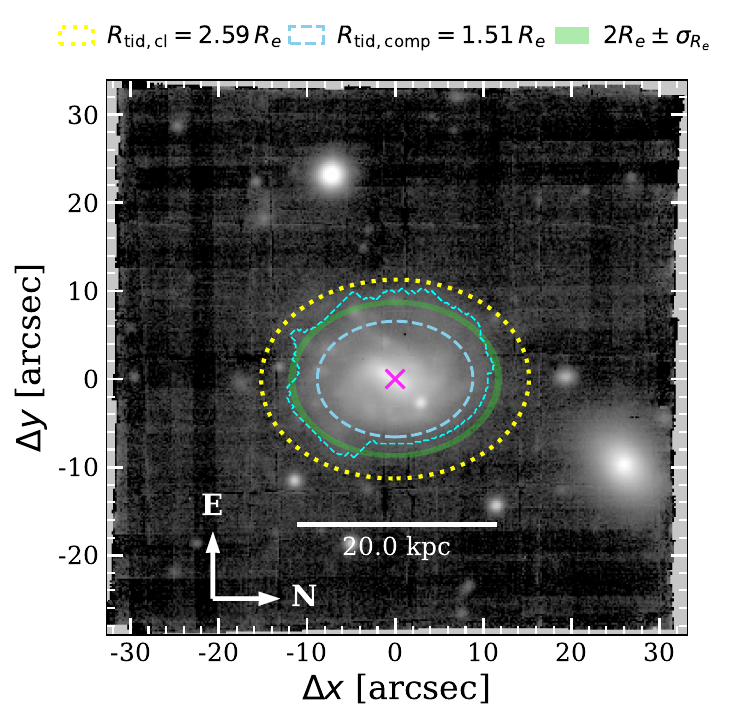}
    \caption{Grayscale Cousins/$I$ image of UG101 and UG101b, generated from the MUSE datacube. The center of UG101 is marked with a magenta cross. The thin cyan line represents UG101 main body, derived as described in Sect.~\ref{subsec:gal_boundaries}. The dashed blue ellipse indicates the location where tidal forces from UG101b reach $\sim15$\% of the centripetal force from the main galaxy ($R_{\rm tid}$). The dotted yellow ellipse shows this value now considering the net tidal force from the host cluster, A3558. For reference, the shaded green region marks the galactocentric distance of $2R_e \pm \sigma_{R_e}$.}
\label{fig:td_acc_tidal}
\end{figure}

We caution against over-interpreting the derived $R_{\mathrm{tid}}$, as it is based on the present-day projected separation between the two systems, that is a lower limit of their 3D separation. Furthermore, if UG101 and UG101b were closer in the past, the region enclosed by the blue dashed ellipse shown in Fig.~\ref{fig:td_acc_tidal} would have been smaller, implying that the tidal influence of UG101b on UG101 would be underestimated. Conversely, if the two galaxies are currently approaching each other, the tidal interaction would have been weaker at earlier times.
Nevertheless, the true three-dimensional separation is uncertain, and the derived value of $v_{\rm diff}$\,$\sim$\,923\,km/s is only a lower limit, as it does not account for the transverse motions of the galaxies in the plane of the sky.

We also consider the presence of alternative companions for UG101, which resides in an overdense region with a local density of $\sim$1.43 galaxies per Mpc$^{2}$\,\citep{Vulcani2023}. Using the OmegaWINGS spectroscopic catalog \citep{Moretti2017}, we search for confirmed cluster members within a projected distance of $\sim$200\,kpc from UG101, applying a velocity separation threshold of 1500\,km/s. Besides UG101b, we find only one cluster member satisfying these criteria, which is WINGS J132705.52-315035.1. Although its LOS velocity difference relative to UG101 is small ($\sim$12\,km/s), its large projected distance ($\gtrsim$150\,kpc) makes it unlikely to currently exert a gravitational influence comparable to UG101b. Finally, we consider the net tidal force exerted by the host cluster, following a similar approach as in \citet{Watson2025}. We apply Eq.~\ref{eq:aratio} using $r$ as the projected clustercentric distance of UG101 and $M_{\rm pert}$ as the cluster mass enclosed within this radius (i.e., $M_{\rm cl} (< r)$). 
We estimate this quantity using the mass profile of \citet{Burkert1995}, with the A3558 properties derived by \citet{Biviano2017}. For UG101, we obtain $M_{\rm cl}(<r) = (4.0_{-2.3}^{+2.4}) \times 10^{14}\,M_{\odot}$ ($\sim$0.63\,$M_{200}$), which yields $R_{\rm tid, cl} \approx 13.5\,$kpc ($\simeq$2.59\,$R_e$). For reference, this region is shown in Fig.~\ref{fig:td_acc_tidal} as a dotted yellow ellipse. Although the cluster contribution is non-negligible, the corresponding tidal radius is nearly twice that derived for UG101b. 
Taken together, these results indicate UG101b is the main candidate for the spiral-arm perturbations observed in UG101. In the following sections, we inspect whether UG101's kinematics and spatially resolved properties support this scenario.

\subsubsection{Stellar and ionized gas kinematics}
\label{subsec:tidal_gas_st_kin}
As mentioned before, gravitational perturbations, such as tidal interactions, are expected to affect both the stellar and gaseous components of galaxies, although with different signatures \citep{Barton1999, Fuentes-Carrera2004, Pedrosa2008, Smith2025, Lassen2026}. While stars behave collectively as a collisionless component \citep{Binney2008}, the ionized gas is collisional and dissipative, and therefore more sensitive to compression and turbulence. 
A comparison between stellar and ionized-gas kinematics thus provides a powerful diagnostic of the nature and impact of environmental processes. Given that both components are expected to become increasingly irregular under the action of tidal perturbations, in Fig.~\ref{fig:ug101_velocities} we present the ionized gas ($\Delta v_{\mathrm{H}\alpha}$ and $\sigma_{\mathrm{H}\alpha}$) and stellar kinematics ($\Delta v_{\bigstar}$ and $\sigma_{\bigstar}$) of UG101 and UG101b. 

\begin{figure}
    \centering
    \includegraphics[width=\columnwidth]{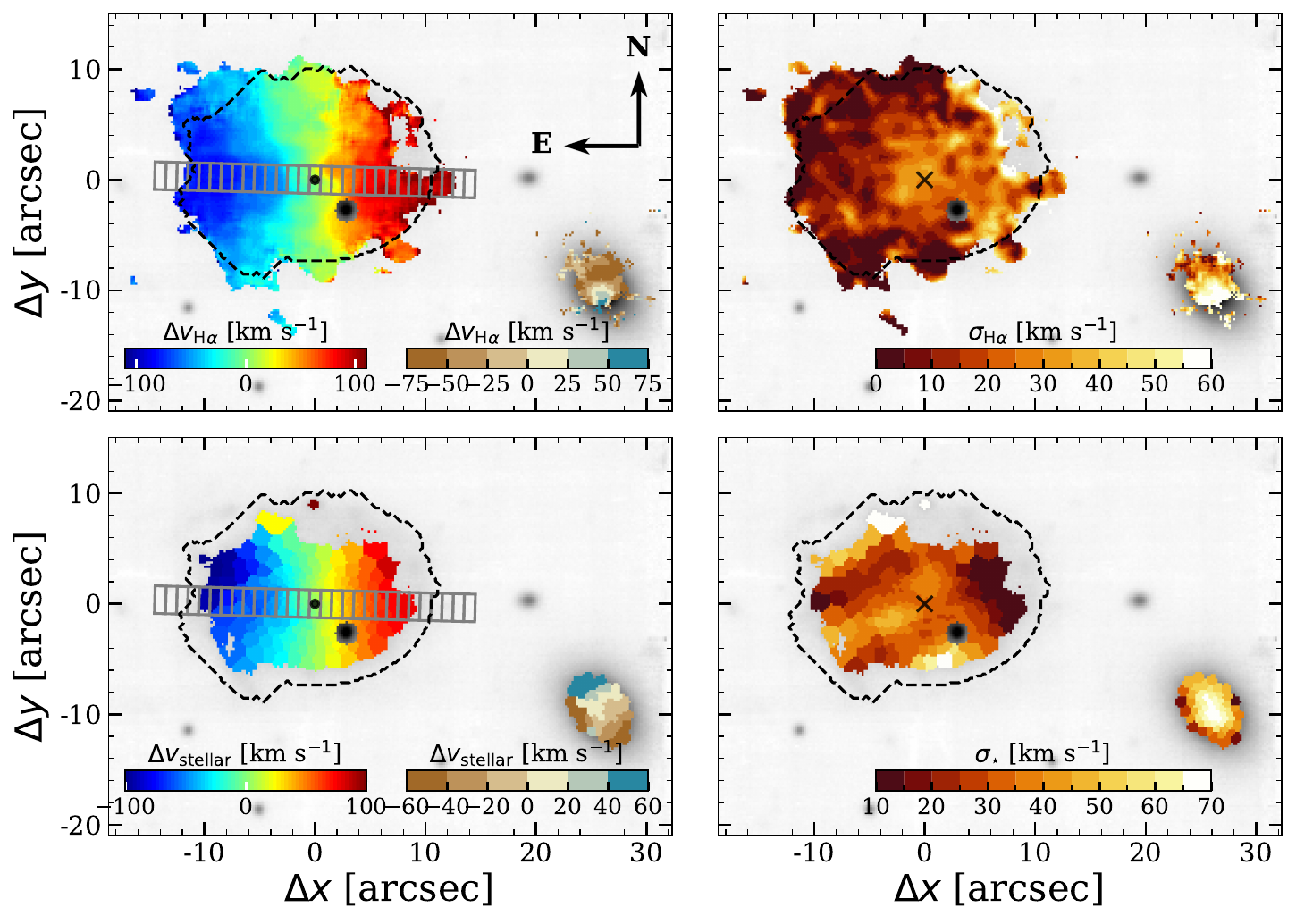}
    \caption{\textit{Top left panel:} Ionized gas kinematics of UG101 and UG101b. Only spaxels with (S/N)$_{\mathrm{H}\alpha} \geqslant 4$ are shown, and the systemic velocity of each galaxy is subtracted. In gray, we represent the slit element used to derive rotation curves (see text), which is aligned to the kinematic semi-major axis.
    \textit{Top right panel:} Gas-phase velocity dispersion of both galaxies, traced by their rest-frame H$\alpha$ emission. Displayed values are corrected for instrumental broadening.
    \textit{Bottom left panel:} Stellar velocity field of both galaxies, derived with \ppxf (see text). 
    \textit{Bottom right panel:} Stellar velocity dispersion. For better visualization, the stellar velocity and velocity dispersion maps are smoothed using  a local weighted regression, as implemented by \textsc{loess2D} Python package \citep{Cappellari2013}.}
    \label{fig:ug101_velocities}
\end{figure}

The velocity ranges spanned by $\Delta v_{\bigstar}$ and $\Delta v_{\mathrm{H}\alpha}$ are similar, generally confined within $\pm100$\,km/s across most locations. For UG101b, the velocity range is lower than in UG101 for both stars and ionized gas. Unlike UG101, its ionized gas is detected at individual spaxel resolution only at the innermost regions of the galaxy.

The dispersion maps of UG101 ($\sigma_{\mathrm{H}\alpha}$ and $\sigma_{\bigstar}$) display irregular, non-axisymmetric patterns. Values are generally low ($\lesssim$\,50\,km\,/s), except on the western edge of UG101, where $\sigma_{\mathrm{H}\alpha}$ can reach values above 100\,km/s. Although the increase of dispersion values at the disk edges may partially result from a noisier continuum --- leading to less accurate pseudo-continuum modeling--- their preferential location on the side facing the companion could also reflect enhanced turbulence due to gas compression. Despite having nearly identical stellar masses, UG101 and UG101b display clearly distinct morphologies. Their kinematics further highlight these differences: UG101 is rotationally supported, whereas UG101b is pressure supported.

To quantitatively assess the degree of asymmetry in the ionized-gas morphology and in the velocity fields shown in Fig.~\ref{fig:ug101_velocities}, we adopt the asymmetry parameter introduced by \citet{Lelli2014, Vulcani2021}:

\begin{equation}
    A = \frac{1}{N}\,\sum\limits_{i,j}^{N} \sqrt{\frac{\big[ \left | I(i,j) \right | - \left | I_{180^{\circ}}(i,j)\right | \big]^2}{\big[ \left | I(i,j) \right | + \left | I_{180^{\circ}}(i,j)\right | \big]^2}}
    \label{eq:asym}
\end{equation}

\noindent where $I_0$ are the original images, $I_{180}$ are the image rotated by 180\textdegree\ around the galaxy center and $N$ is the total number of valid pixels. The summation is performed over $i,j$ pixels, and is normalized such that a perfectly symmetric system---i.e., an image unchanged under a 180\textdegree\ rotation--- yields $A=0$, while a fully asymmetric system reaches $A=1$. Pixels in the rotated image without a corresponding counterpart in the original image are assigned $A_{i,j} \equiv 1$. 
Uncertainties are estimated via a Monte Carlo approach with 2000 realizations, in which the input images are perturbed by spatially correlated noise modeled with a Gaussian kernel. For H$\alpha$ and $\Delta v_{\mathrm{H}\alpha}$, only spaxels with (S/N)$_{\mathrm{H}\alpha} \geqslant 4$ are considered.
For UG101 we measure:
$A(\mathrm{H}\alpha) = 0.64\pm0.09$,
$A (\Delta v_{\mathrm{H}\alpha}) = 0.41\pm0.04$,
$A(\Delta v_{\bigstar}) = 0.40 \pm 0.1$.
These values indicate a markedly asymmetric ionized-gas distribution and kinematics, with asymmetries that are less pronounced but still significant in the stellar component.

Within this context, the rotation curves of the stellar and ionized-gas components provide a complementary and independent probe of the perturbations inferred from the two-dimensional kinematic maps. We analyze the rotation curves by extracting average gas and stellar velocity values along an artificial slit passing through the galaxy center (the slit is shown in Fig.~\ref{fig:ha_mosaic_ug101}). Since velocity fields are not necessarily aligned with the $I$–band photometric major axis, we define the slit orientation using the kinematic PAs (PA$_{\mathrm{kin}}$) measured from the stellar and ionized-gas velocity fields. Using the Python package \textsc{PAfit} \citep{Krajnovic2006, Cappellari2007}, we obtain $88.5 \pm 1^\circ$ and $93.5 \pm 2.8^\circ$, both in good agreement with the photometric value of $\mathrm{PA} = 90.8 \pm 2.1^\circ$.
The slit is sampled in elements of 2\farcs\,5 perpendicular to the slit and  1\arcsec along it, over which the stellar and gas velocity values are averaged. To derive both rotation curves across the semi-major axis, as well as their corresponding uncertainties, we adopt the equations below \citep{Begeman1989}:

\begin{align}
    & v_{\mathrm{rot}} (R) = \frac{\Delta v_{\mathrm{LOS}}}{\sin i} \label{eq:vrot}\\
    & \sigma_{v_{\mathrm{rot}}}  \approx \sqrt{\frac{\sigma_{v_{\mathrm{LOS}}}^2}{\sin^2 i} \ + \ \bigg(\frac{\Delta v_{\mathrm{LOS}} \,\,\cos i \,\,\sigma_i}{\sin^2 i}\bigg)^2} \label{eq:vrot_unc}
\end{align}

\begin{figure}[!t]
    \centering
    \includegraphics[width=0.9\columnwidth]{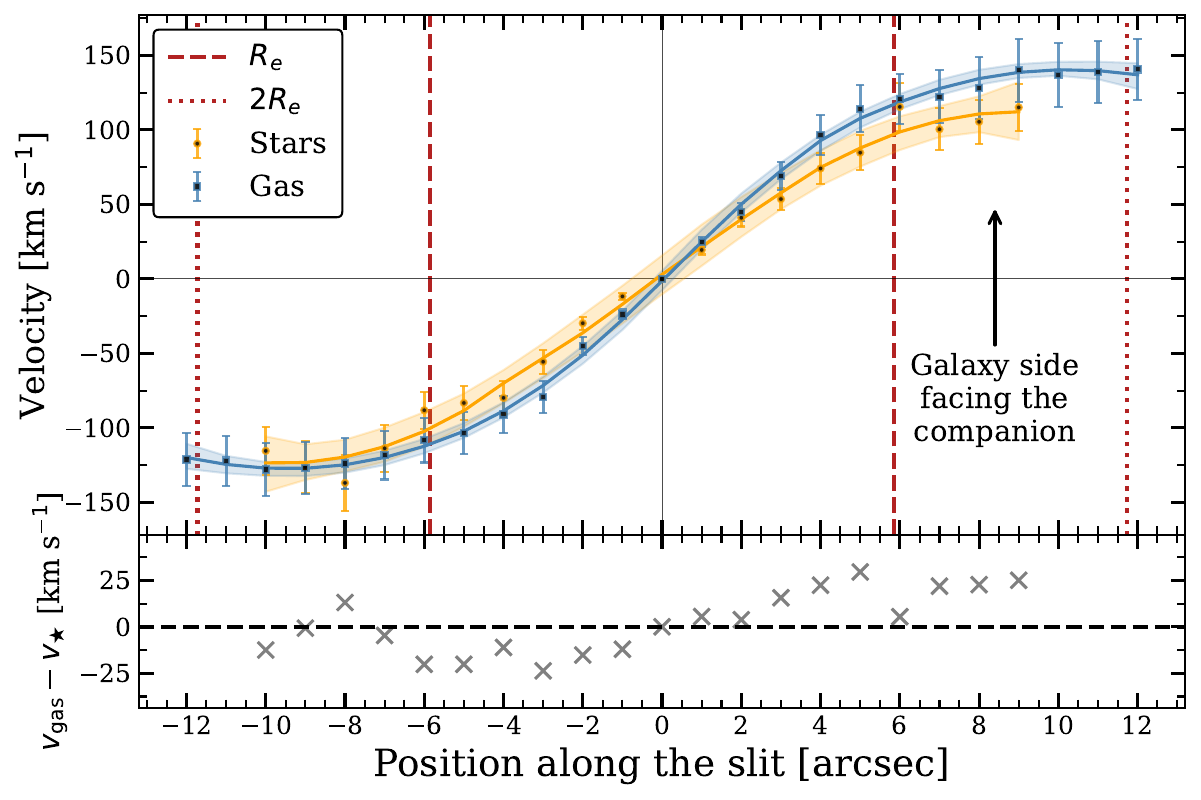}
    \caption{Stellar (orange) and ionized gas (blue) rotation curves in UG101, derived from a slit oriented according to the stellar kinematic PA (see text and also Fig.~\ref{fig:ha_mosaic_ug101}). Points show the average velocities within each slit element, while solid curves correspond to a 1D \textsc{loess} local regression \citep{Cappellari2013} with shaded areas indicating the 3$\sigma$ uncertainty. 
    The x-axis corresponds to the distance along the slit from the galaxy center, with positive values corresponding to the side facing the companion (indicated in the figure).
    The dashed vertical lines mark a distance of $R_e$, while the dotted line indicate a distance of $2 R_e$. Bottom frame shows the difference between gas and stellar velocities.}
\label{fig:rot_curve_UG101}
\end{figure}

\noindent where $\Delta v_{\mathrm{LOS}} \equiv v_{\mathrm{LOS}} (x,y) - v_{\mathrm{sys}}$ and $i$, $\sigma_i$ are the disk inclination and its associated uncertainty. We adopt the average velocity at the galaxy center as the systemic velocity $v_{\mathrm{sys}}$. The resulting rotation curves are shown in Fig.~\ref{fig:rot_curve_UG101}, exhibiting a clear  asymmetry between the approaching and receding sides of the disk. On the side opposite to the companion, the gas and stars exhibit nearly identical rotation velocities out to the last measured point, beyond $2R_e$. In contrast, on the companion-facing side, the gas rotates systematically faster than the stars by $\sim$25\,km/s. At a reference distance of 8\arcsec\ ($\sim$\,1.4$R_e$) along the slit, the gas velocity is similar on both sides of the disk ($\sim \pm120\,$km/s), whereas the stellar velocity ranges from $-130 \pm 7\,$km/s on the far side to $+100 \pm 6\,$km/s on the companion-facing  side. The moderately lower stellar rotation on this side can be attributed to the gas being more rotationally supported than the stars, an effect commonly referred to as asymmetric drift \citep{Nordstrom2004, Binney2008}.

Overall, the stellar and gas kinematics of UG101 appear disturbed, and the galaxy presents a highly irregular $\sigma_{\bigstar}$ distribution. Collectively, the evidence presented here points toward gravitational perturbations as the dominant external mechanism, rather than RPS. Nonetheless, additional analysis is needed to confirm this interpretation and to evaluate whether both processes might act simultaneously in UG101.

\subsubsection{Spatially resolved emission-line diagnostic diagrams}
\label{subsec:BPT_UG101}
In Fig.~\ref{fig:bpt_tidal} we present the spatially resolved BPT diagrams of UG101.\footnote{This analysis is not extended for UG101b as not all the necessary emission lines are detected.}
The ionized gas is predominantly excited by OB stars: the fractions of spaxels classified as star-forming (red points) are 99.7\%, 93.4\% and 86.5\% in the BPT--[\ion{N}{ii}], BPT--[\ion{S}{ii}], BPT--[\ion{O}{i}], respectively. In the BPT--[\ion{S}{ii}], 3.0\% of spaxels fall in the AGN-like region and 3.6\% in the LINER-like region, while in the BPT--[\ion{O}{i}] these fractions are 3.6\% and 9.9\%.

\begin{figure}[!ht]
    \centering
    \includegraphics[width=\columnwidth]{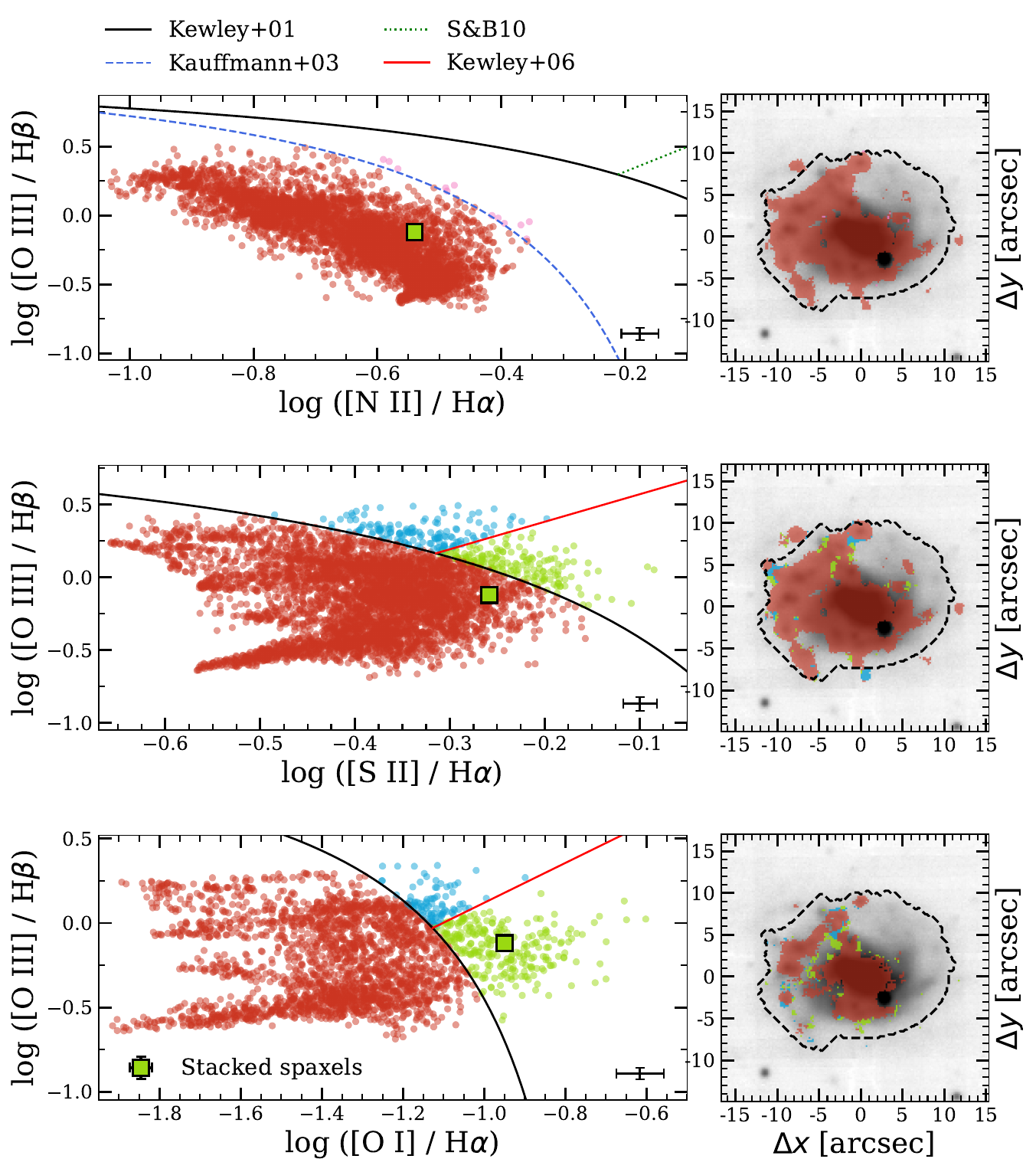}
    \caption{Emission-line diagnostic diagrams of UG101. \textit{Left panels}: From top to bottom, BPT--[\ion{N}{ii}], BPT--[\ion{S}{ii}], BPT--[\ion{O}{i}]. 
    The solid black and red curves correspond to the classification lines introduced by \citet{Kewley2001} and \citet{Kewley2006}, respectively. In the BPT--[\ion{N}{ii}] diagram, the dotted green and blue dashed curves indicate the divisions proposed by \citet{Sharp2010} and \citet{Kauffmann2003}, respectively.
    Each point corresponds to an individual spaxel, and only those with S/N~$\geqslant 4$ in all required lines are shown. Mean uncertainties in the line ratios are indicated in the bottom right corner of each panel. The green square shows the position of the stacked spectrum considering all spaxels classified as LINERs in the BPT--[\ion{O}{i}] diagram. \textit{Right}: Spatial distribution of spaxels color-coded according to the left-hand diagrams, overlaid on the MUSE white-light image of UG101. Dashed contours shown in the right-hand panels indicate the extent of the galaxy main body, for reference.}
    \label{fig:bpt_tidal}
\end{figure}

In contrast to BPT--[\ion{N}{ii}], the other two diagnostic diagrams are more sensitive to the presence of shocks, especially the BPT--[\ion{O}{i}] \citep{Monreal-Ibero2010}. This can be particularly informative in galaxy interactions/collisions, where mechanical shocks can be produced by the tidal forces during the interaction, sometimes resulting in off-nuclear regions classified as ionized by LINER-like mechanism \citep{Monreal-Ibero2006, Monreal-Ibero2010, Rich2011, Law2021}.
In Fig.~\ref{fig:bpt_tidal} we note a substantial number of spaxels classified as LINER-like in the BPT--[\ion{O}{i}] diagram. These are mostly located near the galaxy edge, and---because [\ion{O}{i}]$\lambda$6300 is intrinsically fainter than the other emission lines used in this diagram---are likely to have S/N\,([\ion{O}{i}]$\lambda$6300) $\approx4$. 
To assess whether these LINER-like classifications arise from uncertain [\ion{O}{i}] measurements, we stack the pure-gas spectra of these spaxels and refit the emission lines in the stacked spectrum following the procedure described in Sect.~\ref{subsec:linefitting}. The position of the stacked spectrum in all three diagrams is shown by the green square in Fig.~\ref{fig:bpt_tidal}: while the stacked spectrum is classified as SF in both BPT--[\ion{N}{ii}] and BPT--[\ion{S}{ii}] diagrams, it lies consistently in the LINER locus of the BPT--[\ion{O}{i}] diagram. 
This finding suggests that the classification of these spaxels is not driven by low S/N measurements, but rather likely reflects the genuine physical condition of the ionized gas at those locations \citep{Poggianti2025}.

\subsubsection{Spatially resolved star formation rates}
\label{subsec:SFRs_UG101}
When tidal forces become significant, they can leave imprints on the ionized gas, such as the formation of tidal tails formed from the material pulled away from the galaxy, and bridges connecting the galaxy pair. As discussed throughout this work, tidal interactions can also affect the spiral structure, causing the arms to lose coherence and produce the ``unwound'' morphologies \citep{Dobbs2010} like those observed in our sample of unwinding galaxies. Moreover, tidal interactions can funnel gas inwards, leading to temporary SFR enhancements \citep{Barnes1996}. While these enhancements can be global, tidal compression and shocks, combined with the inward gas flow, generally trigger central SFR bursts \citep{Ellison2008, Moreno2015}.

In Fig.~\ref{fig:ha_mosaic_ug101}, we show the ionized gas morphology of UG101, traced by H$\alpha$ emission. The figure includes two colorbars: one indicating the H$\alpha$ flux and the other corresponding to the star formation rate (SFR) surface density, $\Sigma_{\mathrm{SFR}}$, derived from Eq.~\ref{eq:SFR} and normalized by the physical area of each spaxel.

The recent star formation peaks in the center, along a stellar bar that is clearly visible in the color-composite image of UG101.
Along spiral arm \#2 (see Fig.~\ref{fig:pitch_angles}), a prominent arm-like structure with several star-forming clumps is visible at $\sim$2$R_e$. This feature, located on the galaxy side opposite to the companion, clearly traces the unwinding spiral structure previously identified in the optical images. On the side of the disk facing the companion, at least two SF regions extending beyond the stellar body (dashed line) are observed. A qualitative comparison between the stellar distribution and the ionized gas morphology in Fig.~\ref{fig:ha_mosaic_ug101} shows that the gas reaches larger radii toward the northeast and southeast, while it is less extended toward the northwest. This asymmetric extension suggests that material has been displaced toward the eastern direction. Notably, the region where the ionized gas is less extended than the stellar component spatially coincides with the region exhibiting systematically elevated $\sigma_{\mathrm{H}\alpha}$ values seen in Fig.~\ref{fig:ug101_velocities}.

UG101b is also shown in Fig.~\ref{fig:ha_mosaic_ug101}. Its H$\alpha$ map is patchy in the galaxy outskirts, and no significant trends in $\Sigma_{\mathrm{SFR}}$ are observed, likely due to its small size and the non-detection of H$\beta$ even in the integrated spectrum. Therefore, the SFR values shown for the companion should be regarded as lower limits.

\begin{figure}[!t]
    \centering
    \includegraphics[scale=0.5]{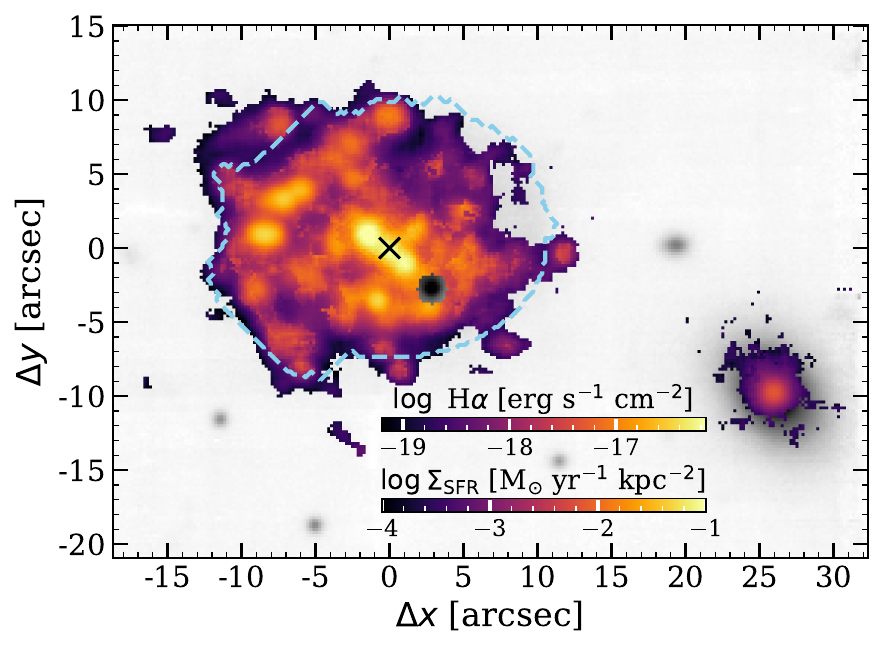}
    \caption{Ionized gas morphology of UG101 and its companion UG101b, traced by rest-frame H$\alpha$ emission. Flux values have been corrected for dust attenuation (see text), except for UG101b, where H$\beta$ is not detected. Spaxels with (S/N)$_{\mathrm{H}\alpha} < 4$ have been masked. A second colorbar shows the corresponding levels of surface density SFR ($\Sigma_{\mathrm{SFR}}$), derived from the H$\alpha$ emission line shown in the top panel (see text). The black cross marks the galaxy center, while light blue contour shows the galaxy $I$-band boundaries (see text).}
    \label{fig:ha_mosaic_ug101}
\end{figure}

\subsubsection{Spatially resolved star formation history}
\label{subsec:SFH_UG101}
Maps of SFH are particularly useful to compare the distribution of stellar populations with different ages. In the case of tidal interactions, significantly different distributions are not expected, as the gravitational perturbation exerted by the companion acts on the gas and stars simultaneously, whereas RPS is expected to affect the gas alone \citep{Gnedin2003, Bellhouse2021, Smith2025, Lassen2026}.

In the top panels of Fig.~\ref{fig:SFH_maps_UG101}, we show maps of $\Sigma_{\mathrm{SFR}}$ for UG101 and its companion across the four  age bins of \sinopsis. The galaxy exhibits  an inside-out development: it is characterized by enhanced star formation in its central regions in the second oldest bin (SFR3), while in the subsequent bin (SFR2) the SFR shifts outward, with elevated values toward the eastern side of the galaxy, coinciding with several spiral arms (e.g., \#1, \#2, \#4, \#5; see Figs.~\ref{fig:pitch_angles} and \ref{fig:ha_mosaic_ug101}). The companion shows a compact SFR distribution, with no clear trends across age bins. The bottom panel of Fig.~\ref{fig:SFH_maps_UG101} reports  the integrated SFR value for each map.

\begin{figure}
    \centering
    \begin{tabular}{c}
        \includegraphics[width=0.9\columnwidth]{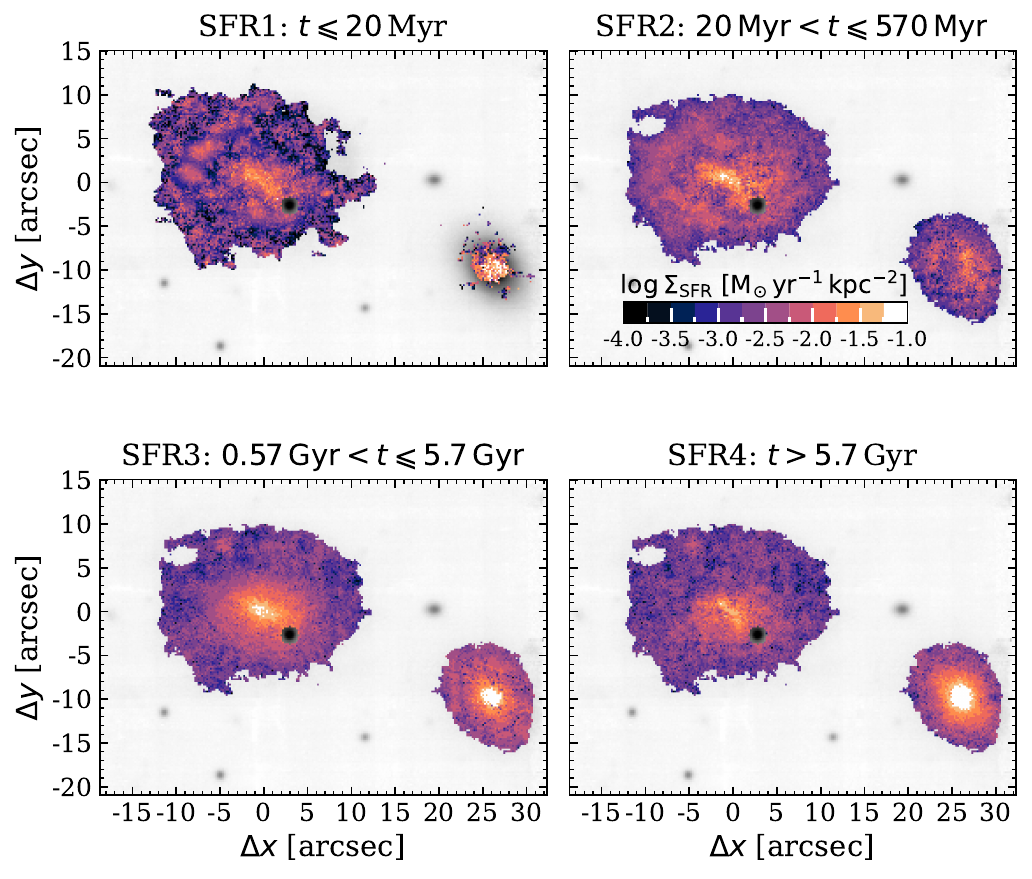} \\
        \includegraphics[width=0.95\columnwidth]{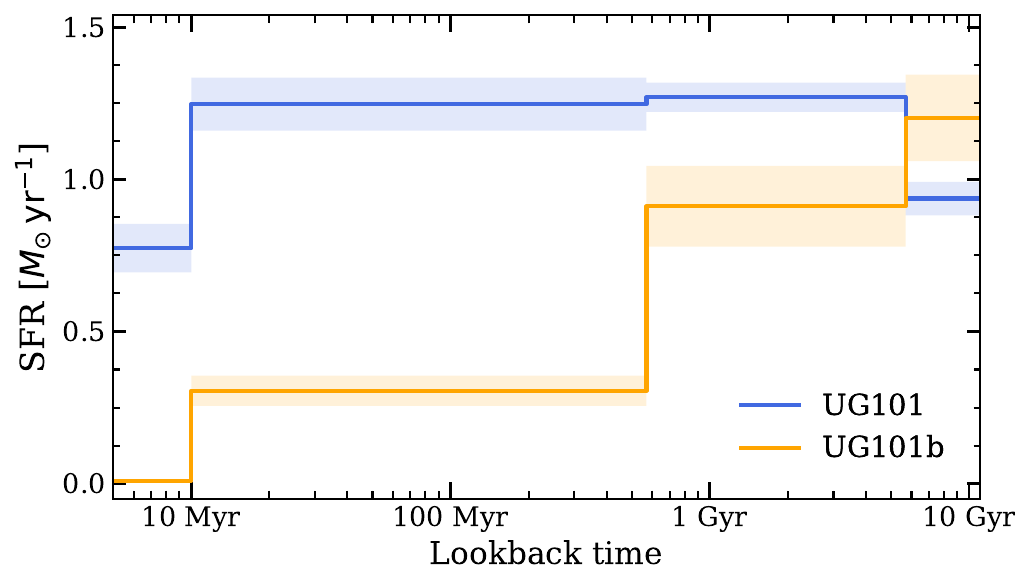}
    \end{tabular}
    \caption{\textit{Top panels}: SFH maps of UG101 and UG101b derived with \sinopsis in three age bins and from H$\alpha$ emission, as denoted in the figures. All four panels share the same color scale for $\Sigma_{\rm{SFR}}$. 
    \textit{Bottom panel}: Integrated SFH of UG101 (blue) and UG101b (orange). Shaded areas indicate the uncertainties on the SFR at each age bin.}
\label{fig:SFH_maps_UG101}
\end{figure}

UG101 shows a $\sim$35\% increase in SFR from the oldest bin (SFR4) to  SFR3, followed by a nearly constant rate of $\sim$1.2\,$M_{\odot}\,\mathrm{yr}^{-1}$ over the subsequent $\sim$5\,Gyr. The recent SFR shows a significant ($\gtrsim$\,30\%) decline. Compared to SFR2 ($1.24 \pm0.09\,M_{\odot}\,\mathrm{yr}^{-1}$), SFR1 from \sinopsis\ over the past $\sim$20\,Myr gives $0.85 \pm0.06\,M_{\odot}\,\mathrm{yr}^{-1}$, while the H$\alpha$-derived SFR, tracing star formation over the past $\sim$10\,Myr and including only spaxels with reliable SF emission, yields $0.77 \pm0.08\,M_{\odot}\,\mathrm{yr}^{-1}$. Both estimates yield consistent and comparable declines, with the H$\alpha$ measurement adopted as the fiducial value for the most recent SFR.
In contrast, UG101b follows a clearly different trend: its SFH steadily declines over its lifetime, consistent with the galaxy being currently passive and exhibiting a spheroidal morphology.

The luminosity-weighted age map shown in Fig.~\ref{fig:lw_age_UG101} provides additional evidence that the clumps seen along the unwinding feature \#2 are very young. UG101 is overall young, with $\overline{t_L} = 439 \pm 2\,$Myr, implying that most of its stellar mass was assembled during the time spanned by the SFR3 and SFR2 age bins. In contrast, the clumps along feature \#2 are only $\overline{t_L} \lesssim 20\,$Myr, approximately an order of magnitude younger than the galaxy average. For UG101b, we measure $\overline{t_L} = 1.40 \pm0.02\,$Gyr, consistent with the SFH presented in Fig.~\ref{fig:SFH_maps_UG101}, which shows higher integrated SFR at $t \gtrsim 600\,$Myr, and a significant decline in SFR2. At earlier epochs, star formation in UG101b was concentrated in the central regions, becoming uniformly low in the second age bin. Considering mass-weighted ages (plots not shown), which are less sensitive to the younger stellar populations and better traces the formation of  the bulk of the stellar mass \citep{Hopkins2018}, we obtain $\overline{t_M} = 4.44 \pm0.01\,$Gyr and $\overline{t_M} = 6.38 \pm0.02\,$Gyr for UG101 and UG101b, respectively.

\begin{figure}
    \centering
    \includegraphics[scale=0.6]{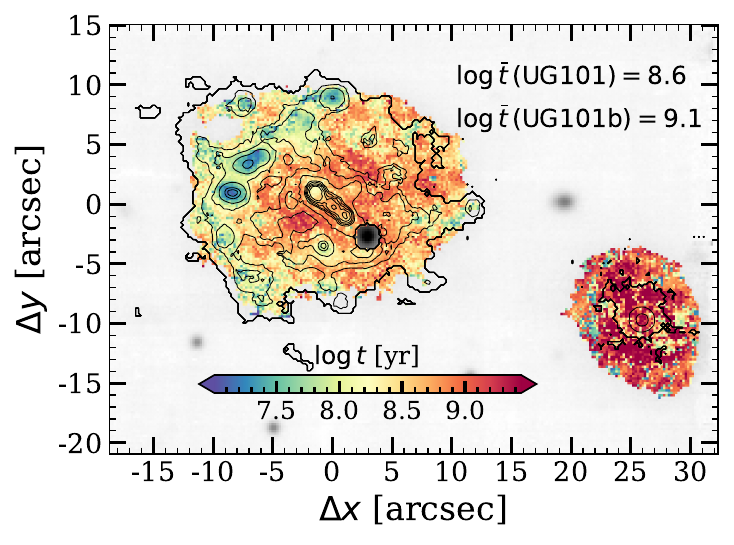}
    \caption{Maps of luminosity-weighted ages of UG101 and UG101b derived from \sinopsis. Mean values obtained for each galaxy are outlined in the figure, for reference.}
    \label{fig:lw_age_UG101}
\end{figure}

To further illustrate the spatial distribution of stellar populations of different ages in UG101 and constrain the timescales involved in the unwinding effect, in Fig.~\ref{fig:sfh_contours} we show contours derived from the SFR maps of the four main \sinopsis age bins, following the approach of B21.
The red contours indicate that the old stars are predominantly concentrated toward the galaxy center, with the exception of a compact structure located to the north, approximately where spiral arms \#1 and \#2 overlap in projection (see Fig.~\ref{fig:pitch_angles}). This feature is also clearly visible in the color-composite image of the galaxy, where it contrasts with the nearby blue SF regions along spiral arm \#2. This feature is unlikely a background/foreground source, given that it exhibits a velocity consistent with that of the surrounding regions (see, e.g., Fig.~\ref{fig:ug101_velocities}).
The youngest stellar populations (SFR1) extend to the largest galactocentric distances but appear fragmented, tracing the star-forming regions along the unwound arms of UG101. Intermediate-age stars (SFR2) reach farther out than the old and oldest ones (SFR3 and SFR4) in most directions, indicating a gradual inside-out growth. Notably, SFR2 extends beyond SFR3 and SFR4, suggesting that these stars have also been displaced by the external perturbation that affected UG101.

\begin{figure}
    \centering
    \includegraphics[scale=0.6]{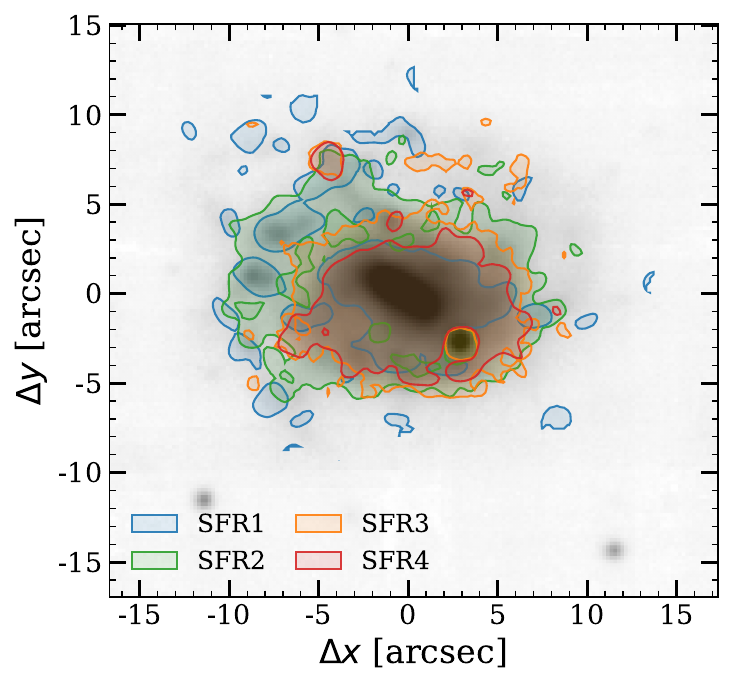}
    \caption{White-light image of UG101 (grayscale), overlaid with contours showing the distribution of the oldest (red -- SFR4; $t \geqslant 5.7\,$Gyr), old (orange -- SFR3; $5.7\,\mathrm{Gyr} < t \leqslant 0.57\,$Gyr), intermediate-age (green -- SFR2; $20\,\mathrm{Myr} < t \leqslant 570\,$Myr) and youngest (blue -- SFR1; $t \leqslant 20\,$Myr) stellar populations derived by \sinopsis.}
    \label{fig:sfh_contours}
\end{figure}

In Sect.~\ref{subsubsec:summary_tidal}, we summarize the results obtained for UG101 and discuss their implications regarding the external mechanism responsible for the unwinding of its spiral arms. We now proceed to the analysis of the next galaxy.

\subsection{Confirming the RPS nature of UG103}
\label{subsec:diagnosis_RPS}
As discussed in Sect.~\ref{sec:obs_justification}, the spiral arms of UG103 are strongly unwound, despite the absence of a close companion. We emphasize that, given UG103 is $\sim$0.9\,dex more massive than UG101, it would be unlikely to overlook a galaxy capable of exerting sufficiently strong tidal forces to induce the perturbations seen in UG103.

Considering WINGS J132707.35-311138.1, which is the closest identified possible companion and  the most likely disturbing galaxy, we calculate $R_{\rm tid}$ for this pair using Eq.~\ref{eq:aratio}, where $\log\,(M_{\rm pert}/M_{\odot})$\,$\approx$\,9.3 \citep{Vulcani2022}. The resulting $R_{\rm tid}$ is indicated by the blue dashed ellipse in Fig.~\ref{fig:td_acc_rps}.
Unlike the case of UG101, $R_{\rm tid}$ is much larger than the galaxy, with $R_{\rm tid}$\,$\approx$\,3.7\,$R_e$ (which corresponds to $\sim$30\,kpc at the cluster redshift). 
Therefore, it is unlikely that this companion can be the responsible for unwinding the spiral arms of UG103. As in Sect.~\ref{sec:tidal_results}, we also investigate the tidal contribution from the host cluster. We obtain $M_{\rm cl}(<r) = (3.4_{-1.9}^{+2.1}) \times 10^{14}\,M_{\odot}$ ($\sim$0.54\,$M_{200}$), which yields $R_{\rm tid,cl}$$\sim$26\,kpc. This region is shown in Fig.~\ref{fig:td_acc_rps} as a dotted yellow ellipse. Unlike in the case of UG101, here we find $R_{\rm tid,cl} < R_{\rm tid,comp}$, suggesting that the main candidate source to exert tidal forces in UG103 is the net tidal force from the host cluster. Still, the estimated $R_{\rm tid, cl}$ exceeds substantially the extent of the observed unwinding features, confirming that cluster tides are not their primary driver. Moreover, if UG103 is currently infalling into the cluster, presumably this effect would have been weaker in the recent past.

\begin{figure}
    \centering
    \includegraphics[scale=0.7]{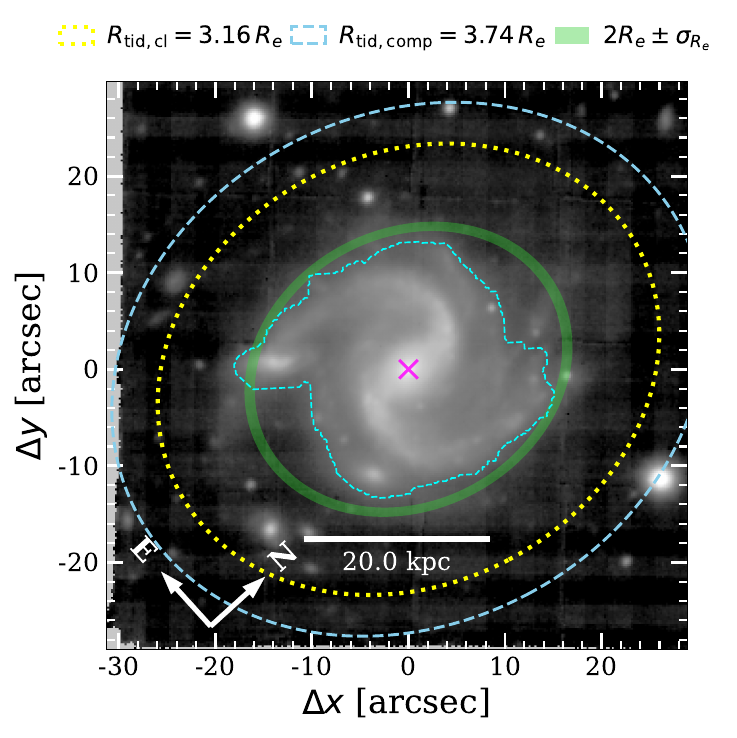}
    \caption{Grayscale Cousins/$I$ image of UG103. Symbols and colors are as in  Fig.~\ref{fig:td_acc_tidal}.}
\label{fig:td_acc_rps}
\end{figure}

\subsubsection{Stellar and ionized gas kinematics}
\label{subsec:rps_gas_st_kin}
In Fig.~\ref{fig:ug103_velocities}, we present the LOS stellar and gas-phase velocity fields of UG103. The stellar velocity field displays a regular, ellipsoidal pattern with symmetric rotation. Its stellar velocity dispersion map shows a compact central peak, smoothly decreasing outwards. In contrast, the ionized gas extends well beyond the stellar disk toward southeast and northeast, while it appears truncated on the south and southwest sides, with the gas less extended than the stellar light distribution.

\begin{figure}
    \centering
    \includegraphics[width=\columnwidth]{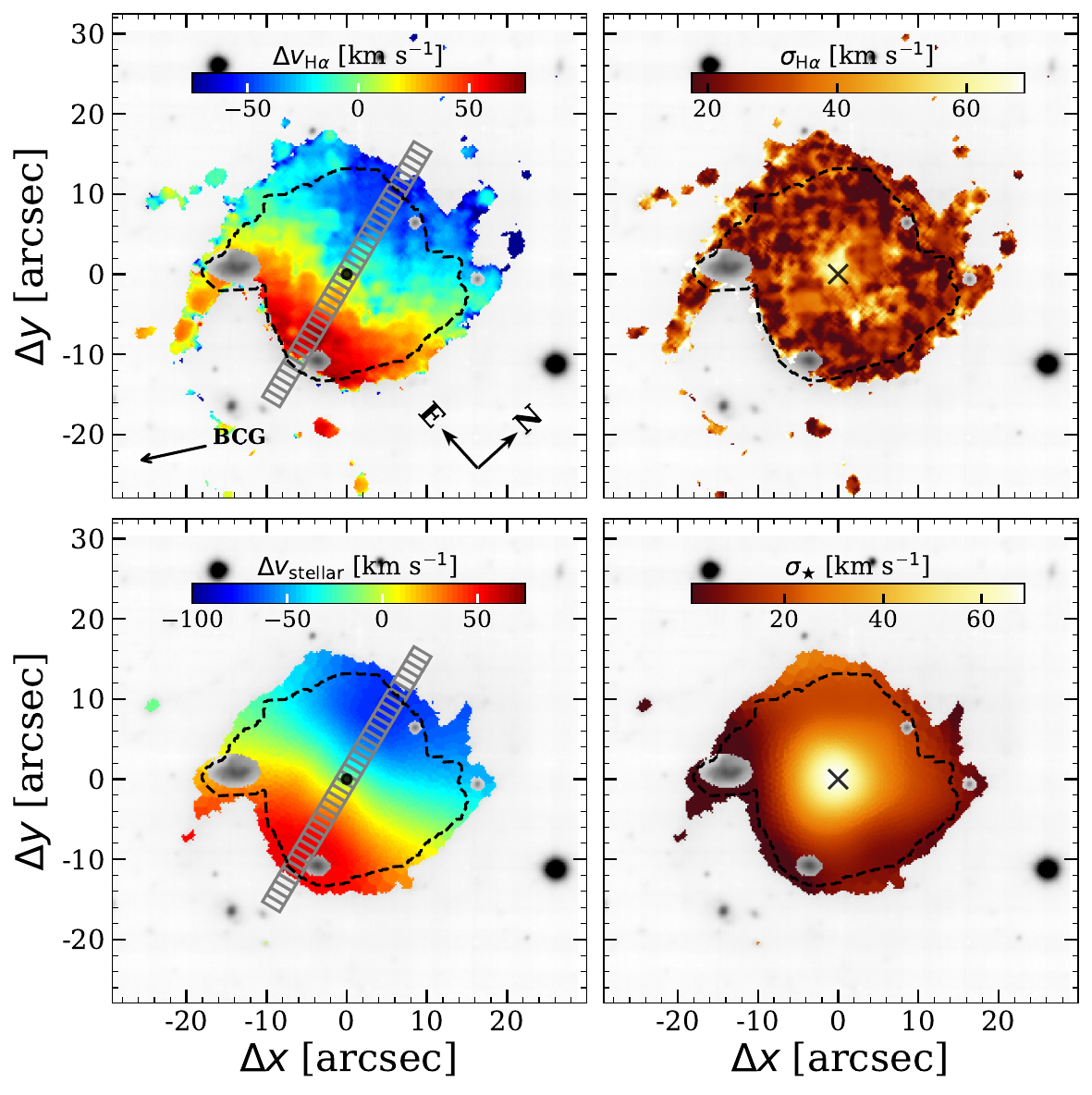}
    \caption{Velocity fields for UG103. Figure details are as in Fig.~\ref{fig:ug101_velocities}. An arrow pointing towards the brightest cluster galaxy (BCG) has been added on the bottom left corner of the left-hand top panel for reference.}
    \label{fig:ug103_velocities}
\end{figure}

We quantify the kinematic asymmetries, obtaining 
$A (\Delta v_{\bigstar}) = 0.16\pm0.02$ for the stellar component, confirming its largely undisturbed and symmetric rotation. The ionized gas, however, shows higher asymmetries:
$A(\mathrm{H}\alpha) = 0.674 \pm 0.05$ and 
$A(\Delta v_{\mathrm{H}\alpha}) = 0.56 \pm 0.02$, suggesting that the gas responds more strongly to external forces.

Rotation curves extracted along a slit aligned with the kinematic axes (Fig.~\ref{fig:rot_curve_UG103}) show that on the southern side of the disk---presumably facing the ICM wind toward the brightest cluster galaxy---gas and stellar rotation curves behave similarly up to 12\arcsec\ along the slit ($\sim$1.4\,$R_e$), beyond which H$\alpha$ emission drops below the detection threshold. The stellar component remains detectable farther out, maintaining an approximately constant rotation velocity of $\sim$75\,km/s. On the opposite side, the stellar rotation remains above $\gtrsim$ 50\,km/s beyond $R_e$, while the gas reaches higher velocities than on the southern side, particularly at radii larger than $R_e$. 

The contrast between the largely undisturbed stellar kinematics and the asymmetric gas behavior is consistent with an early or mild stage of ram-pressure stripping (RPS). Similar signatures have been observed in other RPS candidates, such as SOS90630 in the Shapley supercluster \citep{Merluzzi2016}. While the kinematics alone cannot definitively confirm RPS, the combination of truncated gas, asymmetric gas rotation, and regular stellar motion strongly supports RPS as the main environmental mechanism affecting UG103.

\begin{figure}
\centering
    \includegraphics[width=0.85\columnwidth]{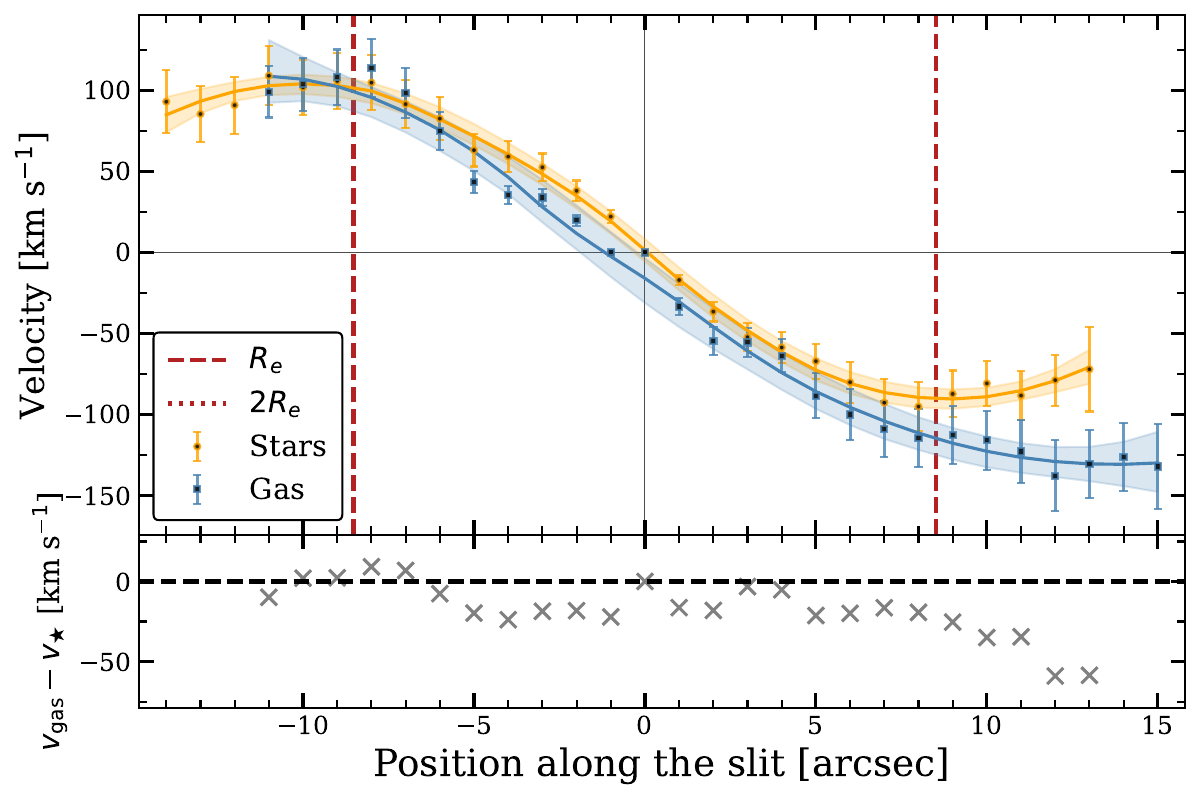} \\ 
    \caption{Inclination-corrected rotation curve of the gas and stellar components of UG103. Curve colors and point symbols are as in Fig.~\ref{fig:rot_curve_UG101}.}
\label{fig:rot_curve_UG103}
\end{figure}

\subsubsection{Spatially resolved emission-line diagnostic diagrams}
\label{subsec:BPT_UG103}
The vast majority of spaxels are classified as SF in the BPT--[\ion{N}{ii}] diagram, although the fraction of spaxels falling in the composite region is not negligible. They are concentrated around the galaxy center, which is ionized by SF. The fraction of spaxels classified as SF in BPT--[\ion{N}{ii}], BPT--[\ion{S}{ii}] and BPT--[\ion{O}{i}] are 83.9\%, 97.6\% and 83.9\%, respectively. For LINER-like ionization, the corresponding fractions are 0.2\%, 2.0\% and 14.7\%, while AGN-like ionization accounts for 0.0\%, 0.5\% and 1.4\% of the spaxels.

Stacking all the spectra classified as LINER-like ionization according to the BPT--[\ion{O}{i}] diagram, we find that this stacked spectra is classified as LINER only in the BPT--[\ion{O}{i}] diagram. This suggests an [\ion{O}{i}] excess, as reported in a few other RPS galaxies \citep{Pedrini2022, Poggianti2025}. The fact that this is consistently found in both tidal and RPS galaxies suggests that this excess is not originated due to a particular perturbing mechanisms, but may rather reflect a physical process related to the cluster environment.

\begin{figure}
    \centering
    \includegraphics[width=\columnwidth]{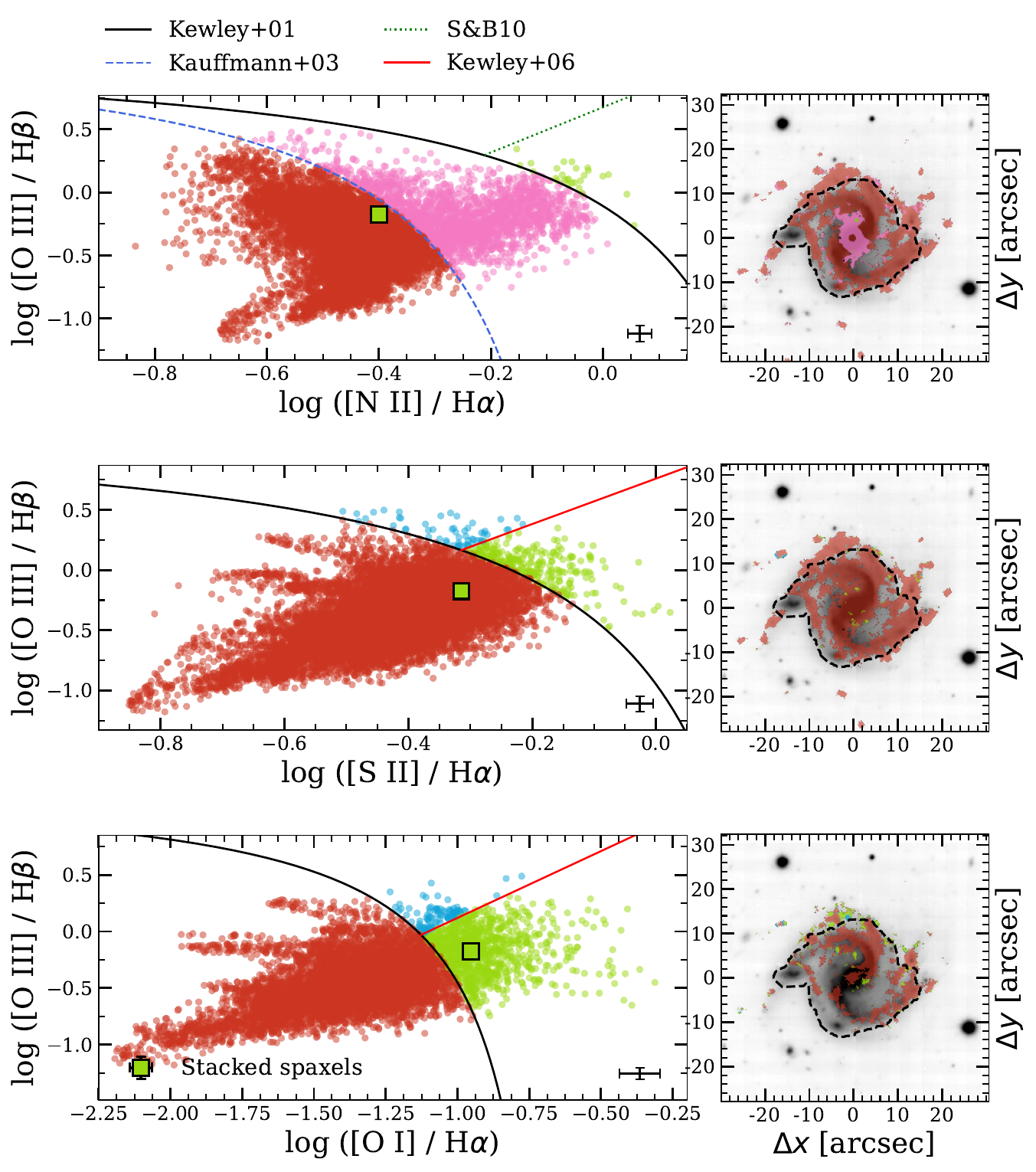}
    \caption{Emission-line diagnostic diagrams of UG103. Panels, symbols and colors are as in Fig.~\ref{fig:bpt_tidal}.}
    \label{fig:bpt_rps}
\end{figure}

\subsubsection{Spatially resolved star formation rates and histories}
\label{subsec:SFRs_UG103}

In Fig.~\ref{fig:ha_mosaic_ug103}, we show the ionized gas morphology of UG103, traced by H$\alpha$ emission, which reveals the spiral structure of the galaxy. Using the galaxy boundaries as visual reference, the major spiral arms are clearly seen opening toward the north/northwest and southeast sides of the disk. Fragmented gas structures extend well beyond the stellar body on the southern side of the disk, in some region extending beyond the VLT/MUSE FoV.

\begin{figure}[!ht]
    \centering
    \includegraphics[scale=0.45]{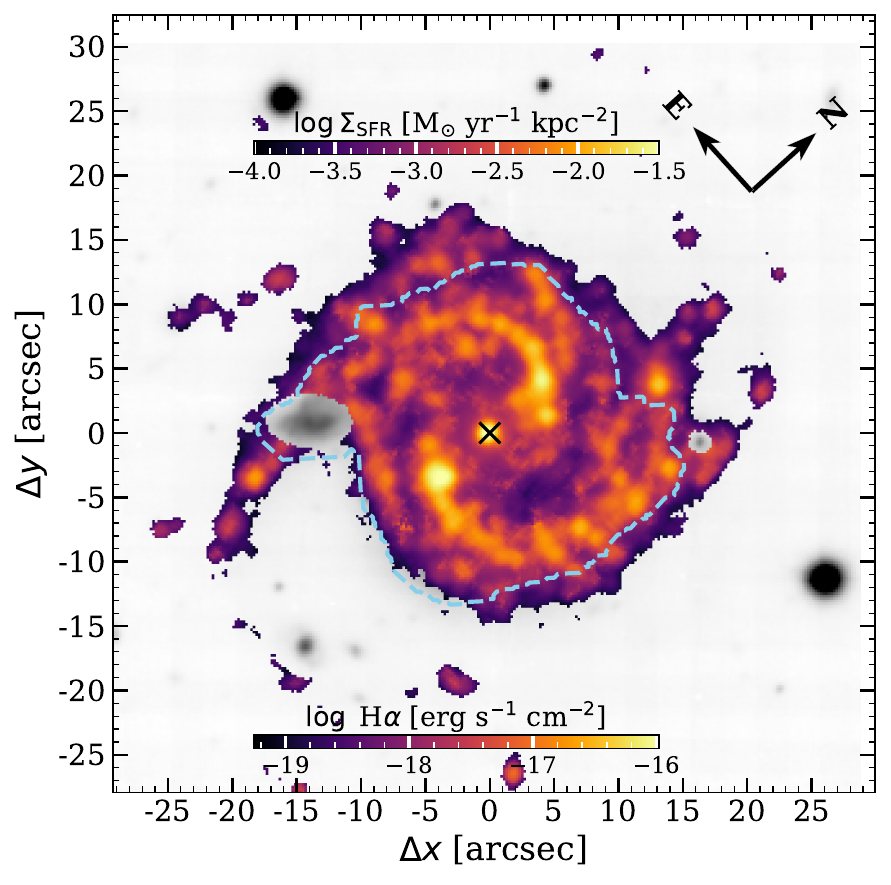}
    \caption{Ionized gas morphology of UG103. Panels, symbols and colors are as in Fig.~\ref{fig:ha_mosaic_ug101}.}
    \label{fig:ha_mosaic_ug103}
\end{figure}

\begin{figure}
    \centering
    \begin{tabular}{c}
    \includegraphics[width=0.9\columnwidth]{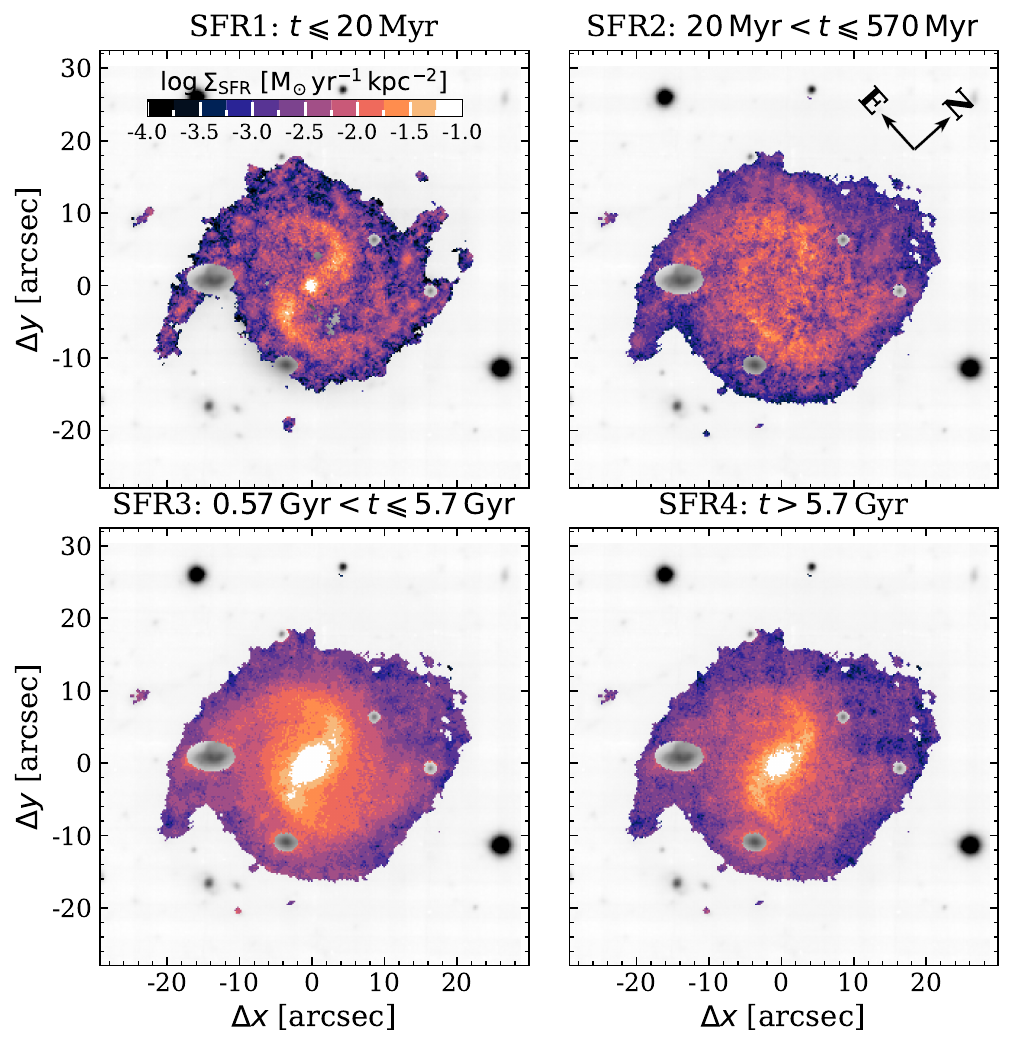} \\
    \includegraphics[width=0.9\columnwidth]{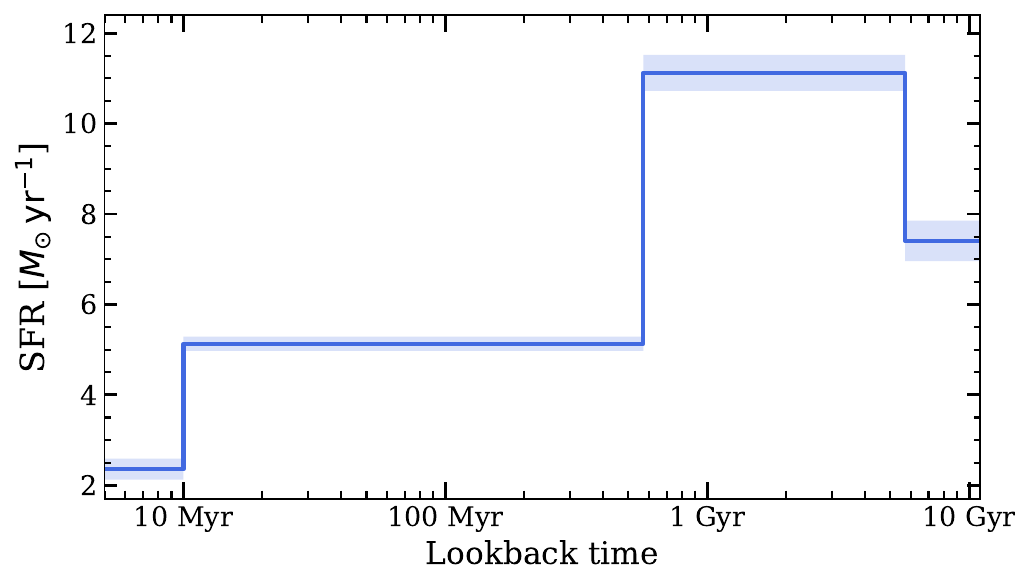}
    \end{tabular}
    \caption{SFH maps (top) and integrated SFH (bottom) of UG103. Panels, symbols and colors are as in  Fig.~\ref{fig:SFH_maps_UG101}.}
\label{fig:SFH_maps_UG103}
\end{figure}

We present the spatially resolved SFH in Fig.~\ref{fig:SFH_maps_UG103}, along with the integrated SFR. UG103 reached its SFR peak at 570\,Myr\,$\leqslant t \leqslant$\,5.7\,Gyr ($11.1 \pm 0.4\,M_{\odot}\,\mathrm{yr}^{-1}$), with star formation concentrated preferentially in the galaxy center, where a stellar bar is clearly visible in the color-composite image. The central region of the galaxy is prominent in the SFR4 map, while it is no longer visible in the SFR2 age bin. A comparison across the SFR maps suggests an outside-in truncation: although the spatial extent of star formation remains similar from SFR4 to SFR2, a clear decrease in SFR is visible in the outer regions at later times. In the SFR3 map, there is a notable central enhancement of star formation, consistent with the increase observed in the integrated SFR values. By SFR2, the star formation remains spatially extended, but the central regions exhibit a decrease in SFR, possibly indicating the onset of quenching. Interestingly, in the most recent age bin (SFR1), star formation shows a renewed central enhancement, while the outer disk appears truncated.
This pattern is consistent with expectations from ram-pressure stripping, as previously reported by \citet{Vulcani2018b, Vulcani2020b}, where the central gas compression triggers a temporary SFR enhancement before ultimately leading to quenching \citep{Vulcani2020a}.
Notably, Fig.~\ref{fig:SFH_maps_UG103} shows a relatively steep decline of the SFR in UG103: from SFR3 to SFR2, it decreases to $5.1 \pm 0.2\,M_{\odot}\,\mathrm{yr}^{-1}$ ($\sim$54\%), with SFR(H$\alpha$) = $2.4 \pm 0.2\,M_{\odot}\,\mathrm{yr}^{-1}$, which corresponds to a relative decrease of 54\% and 79\% with respect to SFR3 and SFR2, respectively.

The luminosity-weighted mean age map (Fig.~\ref{fig:lw_age_UG103}), shows the difference in shape between the youngest and older stellar populations. Clumpy structures along the spiral arms that are located closer to the center have mean ages on the order of $\log\,(t /\rm{yr}) \sim 8.5$, while SF regions closer to the ionized gas edge or along the unwound spiral arms are generally younger (blue spots in Fig.~\ref{fig:lw_age_UG103}), with $\log\,(t /\rm{yr}) \lesssim 7.5$.

\begin{figure}
    \centering
    \includegraphics[scale=0.55]{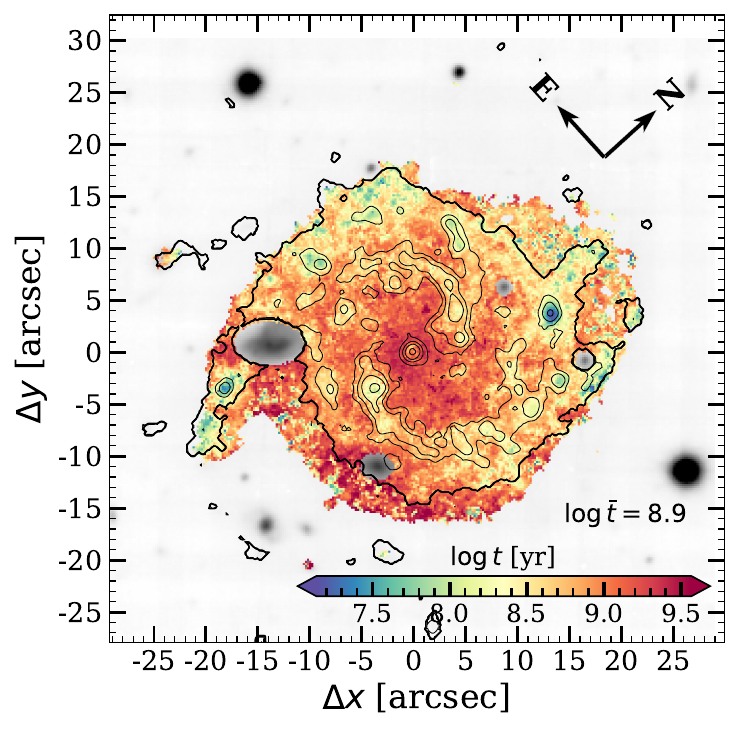}
    \caption{Maps of luminosity-weighted ages for UG103. Colors and symbols are as in  Fig.~\ref{fig:lw_age_UG101}.}
    \label{fig:lw_age_UG103}
\end{figure}

In Fig.~\ref{fig:sfh_contours_rps} we present contours tracing the spatial distribution of different stellar populations in UG103, which allows us to reconstruct how its morphology has progressively evolved. The oldest stellar populations (SFR4) are distributed more symmetrically, while in the old stars (SFR3) the spiral structure begins to emerge, becoming more extended toward the south-east and north-west directions. In contrast, the intermediate-age (SFR2) and young (blue) populations appear less extended along the southern edge of disk. Assuming this side is likely facing the ICM wind, this may reflect the disk truncation process. On the opposite side of the disk, SFR1 and SFR2 seem to extend farther out with respect to SFR3 and SFR4. A gradual transformation of the main spiral arms of UG103 (arms \#3, \#5 and \#6; see Fig.~\ref{fig:pitch_angles}) is also apparent. Over time, these arms seem to become narrower while progressively extending outward. This apparent gradual opening of the spiral arms in UG103 is consistent with what has been reported for other RPS-driven unwinding galaxies in B21. Since signs of morphological disturbance are already visible in SFR3, UG103 likely began experiencing RPS during that epoch (0.57\,Gyr $ \leqslant t \leqslant $ 5.7\,Gyr), which is consistent with the $t_{\rm inf}$\,$\sim$\,1.6\,Gyr inferred from its position on the projected phase-space diagram \citep{Rhee2017}. 

\begin{figure}
    \centering
    \includegraphics[scale=0.57]{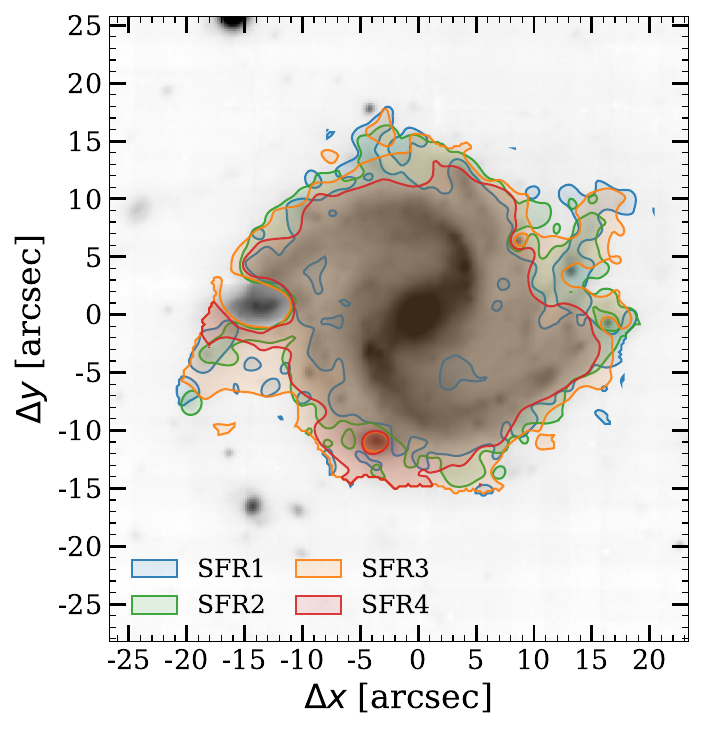}
    \caption{Spatial distribution of different stellar populations in UG103. Contours are as in Fig.~\ref{fig:sfh_contours}.}
    \label{fig:sfh_contours_rps}
\end{figure}

\section{Summary and Discussion}
\label{sec:discussion}
The main goal of this work is to provide a methodological framework to identify the external physical mechanisms responsible for unwinding the spiral arms of cluster galaxies. The most commonly invoked mechanisms are RPS and gravitational interactions. These processes leave distinct imprints on the spatially resolved properties of galaxies, making IFS observations a powerful tool to disentangle them. Motivated by this, ESO program ID 109.23DA (P.I.: Vulcani, B.) obtained VLT/MUSE IFS data for 13 unwinding cluster galaxies from the visually selected sample of \citet{Vulcani2022}.

As a proof of concept, we focused on two galaxies that are candidates of tidal- and RPS-driven unwinding spiral arms. We illustrated a consistent approach to exploit their spatially resolved properties in order to identify the external processes shaping their morphologies. To quantify the relative importance of tidal forces, we estimated the galactocentric distance where the tidal acceleration reaches 15\% of the centripetal acceleration, defining it as the tidal influence radius ($R_{\rm tid}$). Beyond this distance, galaxy regions are expected to be influenced by the gravitational field of a companion.

We investigated the kinematics of both gas and stars to distinguish between gravitational and hydrodynamical perturbations. Gravitational interactions affect both components, producing irregular velocity fields and enhanced velocity dispersions, while RPS primarily affects the gas, leaving the stellar kinematics largely undisturbed. We also analyzed the integrated star formation rates and histories (SFHs) derived with \sinopsis. Central bursts of star formation could occur in both scenarios, but the spatial distribution and temporal evolution of the stellar populations provided additional diagnostic power. In the following sections, we summarize the main results and place them in context.

\subsection{UG101}
\label{subsubsec:summary_tidal}
UG101 was chosen to illustrate the effects of tidal interactions. It exhibits visibly perturbed spiral arms and has a close companion, UG101b.
We find $R_{\rm tid}$\,$\approx$\,1.5\,$R_e$, with the unwinding features extending well beyond this radius.
This supports a tidal origin for the unwinding arms of UG101, possibly driven by an unbound encounter with UG101b.
The contribution from the cluster net tidal force was also investigated, using the galaxy clustercentric distance and $M_{\rm cl} (<r)$, finding $R_{\rm tid, cl}$$\sim$2.6\,$R_e$. This analysis has important limitations, as it relies on projected quantities and the current configuration of the pair, while $R_{\rm tid}$ is inherently time-dependent and closely linked to the dynamical history of the interaction.

Since the proper motions of both galaxies are unknown, the inferred interaction strength and duration remain uncertain. If the 3D separation of the pair is significantly larger than their projected distance, the derived $R_{\rm tid}$ would be underestimated.
To provide a tentative assessment of this uncertainty, we consider the outermost detected point within the unwound structures of UG101 and compute the maximum separation at which tidal forces would still contribute significantly at that location. The farthest identified point in UG101 lies along spiral arm \#2 (see Fig.~\ref{fig:pitch_angles}) at $R_{\rm max}$$\sim$ 12.3\,kpc ($\sim$2.35\,$R_e$). Using this value in Eq.~\ref{eq:aratio}, we obtain a corresponding separation of $r_{\rm max}$$\sim$36.1\,kpc.
We emphasize that this estimate is still based solely on the present-day configuration of the system. A proper time-dependent treatment of the  interaction, including orbital evolution and cumulative tidal effects, would require dedicated numerical simulations, which are beyond the scope of this work.
Furthermore, UG101 resides in an overdense cluster region \citep{Vulcani2023}, and the cumulative effect of unbound encounters with other members of the cluster was not considered. Finally, the tidal contribution from the cluster potential is also variable, intrinsically linked to the galaxy orbit.
The position of UG101 in the projected phase-space diagram suggests that it has been accreted at least 2\,Gyr ago, therefore UG101 likely had smaller clustercentric distances than it currently has, when the tidal contribution from the cluster was more relevant.

The stellar and ionized gas kinematics support the tidal-interaction scenario, as both components display disturbed velocity fields (see Fig.~\ref{fig:ug101_velocities}) and irregular velocity dispersions. We stress that this observation suggests the action of gravitational perturbations regardless of the source responsible for producing them. The rotation curve along the major kinematic axis shows similar trends for gas and stars, except for a $\sim$25\,km/s slower stellar rotation on the companion-facing side, likely due to asymmetric drift \citep{Nordstrom2004, Binney2008}.

Even though the signs of tidal interactions are quite compelling, an additional effect from RPS in UG101 cannot be ruled out. For example, the gas appears displaced on the disk side facing the companion, while it extends beyond the stellar body on the opposite side, which could be interpreted as an aftermath of RPS. However, we note that the intermediate-age stars in UG101 seem to have been displaced toward the same direction, corresponding to the BCG direction (see Fig.~\ref{fig:sfh_contours}). Although in many cluster galaxies the stripped material is oriented on the opposite direction of the BCG, this observation alone cannot confirm or rule out RPS effects. For instance, if UG101 has passed the pericenter but still experiences an ICM wind opposite its travel direction, the removed material could then flow in the direction of the cluster center. In fact, there is at least one confirmed RPS galaxy where a similar orientation of the stripped tail was observed \citep[JO201; e.g., see][]{Bellhouse2017, Poggianti2017b}. Nonetheless, given the nearly constant SFR sustained by UG101 during the period of 10\,Myr\,$< t <$\,5.7\,Gyr (see Fig.~\ref{fig:SFH_maps_UG101}), either an additional mechanism or very tangential orbits would have been required to prevent RPS from stripping the cold gas from the galaxy ISM, which would have resulted in a declining SFH from this period on.
Although the combined effect of RPS and tidal forces cannot be ruled out, stronger evidence for the RPS is still necessary to strengthen this hypothesis.

The nearly constant SFR UG101 sustained through the SFR3 and SFR2 epochs is followed by a $\sim$30\% decline in the last 10\,Myr. While the integrated SFR values in the SFR2 and SFR3 age bins are similar, their spatial distributions differ: intermediate-age populations extend farther out along the eastern disk (opposite the companion), while the youngest populations are concentrated in the center and along the unwound arms, particularly spiral arms \#1 and \#2. Luminosity-weighted ages confirm the recent formation ($\log(t/\mathrm{yr}) \sim 7.5$) of the star-forming knots along these features.

Overall, the spatially resolved properties indicate that gravitational interactions are the primary driver of UG101’s morphology, although additional RPS contributions cannot be excluded.

\subsection{UG103}
\label{subsubsec:summary_RPS}
UG103 exhibits clear evidence of opening spiral arms, but lacks both a close companion and a detectable extended optical tail, making it a suitable candidate for RPS-driven unwinding. The nearest plausible companion lies at a projected distance of $\sim$124\arcsec\ ($\sim$117\,kpc) and has a low stellar mass ($\log(M/M_\odot) \sim 9.3$). The corresponding tidal influence radius, $R_{\rm tid}$$\sim$3.7\,$R_e$ (see Sect.~\ref{subsec:diagnosis_RPS}), indicates that this companion is unlikely to be responsible for the pronounced unwinding features. Although the cluster potential is more relevant, with $R_{\rm tid, cl}$$\sim$3.2\,$R_e$, it is still likely not the main driver of the unwinding arms of UG103.

The kinematics strongly support RPS as the dominant external mechanism. The gas velocity field is perturbed, exhibiting higher rotational velocities ($\sim$50\,km/s) on one side of the disk, while the stellar velocity field and velocity dispersion maps remain largely regular. Assuming the ICM wind is oriented opposite to the BCG, the gas appears accelerated along the disk side aligned with the inferred wind direction.

The integrated SFH shows that UG103 reached its peak SFR during the SFR3 age bin, followed by a steep decline. From its location in the projected phase-space diagram, we infer an infall time $t_{\rm inf} \sim 1.6\,$Gyr. This is consistent with RPS beginning to remove cold gas around that epoch, causing the rapid decrease in SFR observed in Fig.~\ref{fig:SFH_maps_UG103}.
The spatial distribution of stellar populations in different age bins further illustrates the RPS effects: the spiral arms progressively emerge from SFR4 to SFR3, becoming narrower and more unwound over time. The gas disk shows truncation along the southern edge, while the northern side extends outward, consistent with stripping along the inferred wind direction.

Taken together, these spatially resolved properties indicate that RPS has played a significant role in shaping UG103’s morphology and is the most plausible mechanism driving the unwinding of its spiral arms.

\subsection{Comparison between UG101 and UG103}
\label{subsubsec:comparison}

Comparing the two galaxies highlights the distinctive signatures of tidal interactions and RPS. While the disturbed spiral arms of UG101 extend well beyond its $R_{\rm tid}$, supporting a tidal origin likely driven by the flyby of UG101b, UG103 lies entirely within its tidal radius, ruling out significant gravitational perturbations from nearby companions.
The stellar and gas kinematics reinforce this picture. In UG101, both components exhibit irregular velocity fields and enhanced velocity dispersions, consistent with tidal disturbances. In UG103, the stellar kinematics remain largely regular ($A(\Delta v_\bigstar)=0.16\pm0.02$), whereas the gas is strongly perturbed ($A(\mathrm{H}\alpha)=0.674\pm0.05$, $A(\Delta v_{\mathrm{H}\alpha})=0.56\pm0.02$), with asymmetric velocities along the disk, particularly in the wake side, indicative of ongoing RPS. The rotation curves further support this distinction: UG101 shows similar stellar and gas rotation aside from minor asymmetric drift, while in UG103 the gas reaches higher velocities on the disk side opposite the wind, contrasting with the more symmetric stellar rotation.

The integrated SFHs provide complementary insights. UG101 formed stars at an approximately constant rate over the past $\sim$5\,Gyr, followed by a $\sim$30\% decline in the last 10\,Myr. This does not seem to have been caused by RPS, given that the position of UG101 in the projected phase-space diagram suggests $t_{\rm inf} \gtrsim 2\,$Gyr. In contrast, in UG103 most of its stellar mass formed earlier ($0.57\,\mathrm{Gyr} < t \leqslant 5.7\,\mathrm{Gyr}$), followed by a more abrupt ($\gtrsim 50\%$) decline in recent SFR, consistent with rapid gas removal during cluster infall. 
Finally, the spatial distribution of stellar populations reveals different evolutionary pathways. In UG101, intermediate- and young-age stars are displaced along the unwound arms, with the youngest populations ($t \lesssim 10$\,Myr) appearing fragmented and clumpy, suggesting recent dynamical perturbations by the companion. In UG103, the spiral arms progressively unwind from earlier to later epochs, with truncation of the southern gas disk and extension toward the northern side, reflecting ongoing stripping along the inferred ICM wind direction.

Altogether, these quantitative and spatially resolved comparisons provide a clear contrast: UG101 exemplifies a system affected by tidal forces, with both stars and gas responding to the external gravitational perturbation, while UG103 demonstrates the signature of RPS, affecting primarily the gas component and leaving the stellar disk largely undisturbed. Both galaxies also host stellar bars, with UG103 exhibiting particularly prominent spiral arms emerging from the bar. While assessing the role of bars is beyond the scope of this work, the analysis of the full sample will allow us to quantify the fraction of barred galaxies and explore a possible connection between bar-driven dynamics and the presence of unwinding features.

Nonetheless, it is important to keep in mind that any inference based on just two galaxies must be taken with caution. Differences in total stellar mass, in the evolutionary stage at which each galaxy started experiencing the external process, and in local environmental conditions within the cluster, can all modulate the observed properties. These factors should be kept in mind when interpreting the results and in the broader context of applying this framework to larger samples.

\subsection{Future prospects}
The methodological framework developed here to disentangle tidal and RPS-driven unwinding from their spatially resolved properties can now be applied to the full sample of 13 galaxies. Combining optical IFS observations with multi-wavelength data---probing different gas phases and star formation timescales---and confronting them with numerical simulations will enable a systematic classification of unwinding galaxies into tidal- or RPS-dominated systems.
This approach will provide a statistically robust assessment of the incidence of RPS among unwinding spirals in clusters and enable a deeper understanding of the relative importance of gravitational versus hydrodynamical processes across diverse cluster environments. By extending the analysis to larger samples, it will also be possible to investigate how galaxy properties (e.g., stellar mass, evolutionary stage, presence of a stellar bar) and local environmental conditions modulate the observed responses, offering insight into the interplay between internal and external drivers of morphological transformation.

\begin{acknowledgements}
The authors thank Dr. Curtis Struck for his constructive comments, which improved the manuscript. A.E.L. and B.V. acknowledge support from the INAF GO grant 2023 ``Identifying ram pressure induced unwinding arms in cluster spirals'' (P.I. Vulcani). A.E.L. and B.V. acknowledge ISCRA for awarding this project access to the LEONARDO supercomputer, owned by the EuroHPC Joint Undertaking and hosted by CINECA (Italy). This project has received funding from the European Research Council (ERC) under the European Union’s Horizon 2020 research and innovation program (grant agreement No.\@~833824).
N.T. and L.M. acknowledge support from the Croatian Science Foundation under the project number HRZZ--MOBDOK--2023-8006.
RS acknowledges financial support from FONDECYT Regular 2023 project No.\@~1230441 and also gratefully acknowledges financial support from ANID--MILENIO NCN2024\_112.
\end{acknowledgements}
%
%
\bibliographystyle{aa}
\bibliography{ref}

@ARTICLE{Bellhouse2021,
       author = {{Bellhouse}, Callum and {McGee}, Sean L. and {Smith}, Rory and {Poggianti}, Bianca M. and {Jaff{\'e}}, Yara L. and {Kraljic}, Katarina and {Franchetto}, Andrea and {Fritz}, Jacopo and {Vulcani}, Benedetta and {Tonnesen}, Stephanie and {Roediger}, Elke and {Moretti}, Alessia and {Gullieuszik}, Marco and {Shin}, Jihye},
        title = "{GASP XXIX - unwinding the arms of spiral galaxies via ram-pressure stripping}",
      journal = {\mnras},
         year = 2021,
        month = jan,
       volume = {500},
       number = {1},
        pages = {1285-1312},
          doi = {10.1093/mnras/staa3298},
archivePrefix = {arXiv},
       eprint = {2010.09733},
 primaryClass = {astro-ph.GA},
       adsurl = {https://ui.adsabs.harvard.edu/abs/2021MNRAS.500.1285B}
}

@ARTICLE{Matijevic2025,
       author = {{Matijevi{\'c}}, Luka and {Tomi{\v{c}}i{\'c}}, Neven and {Marasco}, Antonino and {Ignesti}, Alessandro and {Lassen}, Augusto E. and {Smith}, Rory and {Sell}, Paul and {Roberts}, Ian D. and {Zezas}, Andreas and {Anastasopoulou}, Konstantina and {Kotoulas}, Panagiotis and {Ba{\v{s}}i{\'c}}, Roko},
        title = "{The Competing Influence of Ram Pressure and Tidal Interaction in NGC 2276}",
      journal = {arXiv e-prints},
         year = 2025,
        month = dec,
          eid = {arXiv:2512.17486},
        pages = {arXiv:2512.17486},
          doi = {10.48550/arXiv.2512.17486},
archivePrefix = {arXiv},
       eprint = {2512.17486},
 primaryClass = {astro-ph.GA},
       adsurl = {https://ui.adsabs.harvard.edu/abs/2025arXiv251217486M}
}

@ARTICLE{Ellison2008,
       author = {{Ellison}, Sara L. and {Patton}, David R. and {Simard}, Luc and {McConnachie}, Alan W.},
        title = "{Galaxy Pairs in the Sloan Digital Sky Survey. I. Star Formation, Active Galactic Nucleus Fraction, and the Mass-Metallicity Relation}",
      journal = {\aj},
         year = 2008,
        month = may,
       volume = {135},
       number = {5},
        pages = {1877-1899},
          doi = {10.1088/0004-6256/135/5/1877},
archivePrefix = {arXiv},
       eprint = {0803.0161},
 primaryClass = {astro-ph},
       adsurl = {https://ui.adsabs.harvard.edu/abs/2008AJ....135.1877E}
}

@ARTICLE{Smith2025,
       author = {{Smith}, R. and {Tonnesen}, S. and {Kraljic}, K. and {Calder{\'o}n-Castillo}, P. and {Marasco}, A. and {Jaffe}, Y. and {Vulcani}, B. and {Poggianti}, B.~M.},
        title = "{Distinguishing ram pressure from tidal interactions: Size-shape difference measure}",
      journal = {\aap},
         year = 2025,
        month = sep,
       volume = {701},
          eid = {A6},
        pages = {A6},
          doi = {10.1051/0004-6361/202554190},
archivePrefix = {arXiv},
       eprint = {2506.13884},
 primaryClass = {astro-ph.GA},
       adsurl = {https://ui.adsabs.harvard.edu/abs/2025A&A...701A...6S}
}

@ARTICLE{Gnedin2003,
       author = {{Gnedin}, Oleg Y.},
        title = "{Dynamical Evolution of Galaxies in Clusters}",
      journal = {\apj},
         year = 2003,
        month = jun,
       volume = {589},
       number = {2},
        pages = {752-769},
          doi = {10.1086/374774},
archivePrefix = {arXiv},
       eprint = {astro-ph/0302498},
 primaryClass = {astro-ph},
       adsurl = {https://ui.adsabs.harvard.edu/abs/2003ApJ...589..752G}
}

@ARTICLE{Moreno2015,
       author = {{Moreno}, Jorge and {Torrey}, Paul and {Ellison}, Sara L. and {Patton}, David R. and {Bluck}, Asa F.~L. and {Bansal}, Gunjan and {Hernquist}, Lars},
        title = "{Mapping galaxy encounters in numerical simulations: the spatial extent of induced star formation}",
      journal = {\mnras},
         year = 2015,
        month = apr,
       volume = {448},
       number = {2},
        pages = {1107-1117},
          doi = {10.1093/mnras/stv094},
archivePrefix = {arXiv},
       eprint = {1501.03573},
 primaryClass = {astro-ph.GA},
       adsurl = {https://ui.adsabs.harvard.edu/abs/2015MNRAS.448.1107M}
}

@INPROCEEDINGS{Bacon2010,
       author = {{Bacon}, R. and {Accardo}, M. and {Adjali}, L. and {Anwand}, H. and {Bauer}, S. and {Biswas}, I. and {Blaizot}, J. and {Boudon}, D. and {Brau-Nogue}, S. and {Brinchmann}, J. and {Caillier}, P. and {Capoani}, L. and {Carollo}, C.~M. and {Contini}, T. and {Couderc}, P. and {Daguis{\'e}}, E. and {Deiries}, S. and {Delabre}, B. and {Dreizler}, S. and {Dubois}, J. and {Dupieux}, M. and {Dupuy}, C. and {Emsellem}, E. and {Fechner}, T. and {Fleischmann}, A. and {Fran{\c{c}}ois}, M. and {Gallou}, G. and {Gharsa}, T. and {Glindemann}, A. and {Gojak}, D. and {Guiderdoni}, B. and {Hansali}, G. and {Hahn}, T. and {Jarno}, A. and {Kelz}, A. and {Koehler}, C. and {Kosmalski}, J. and {Laurent}, F. and {Le Floch}, M. and {Lilly}, S.~J. and {Lizon}, J.-L. and {Loupias}, M. and {Manescau}, A. and {Monstein}, C. and {Nicklas}, H. and {Olaya}, J.-C. and {Pares}, L. and {Pasquini}, L. and {P{\'e}contal-Rousset}, A. and {Pell{\'o}}, R. and {Petit}, C. and {Popow}, E. and {Reiss}, R. and {Remillieux}, A. and {Renault}, E. and {Roth}, M. and {Rupprecht}, G. and {Serre}, D. and {Schaye}, J. and {Soucail}, G. and {Steinmetz}, M. and {Streicher}, O. and {Stuik}, R. and {Valentin}, H. and {Vernet}, J. and {Weilbacher}, P. and {Wisotzki}, L. and {Yerle}, N.},
        title = "{The MUSE second-generation VLT instrument}",
    booktitle = {Ground-based and Airborne Instrumentation for Astronomy III},
         year = 2010,
       editor = {{McLean}, Ian S. and {Ramsay}, Suzanne K. and {Takami}, Hideki},
       series = {Society of Photo-Optical Instrumentation Engineers (SPIE) Conference Series},
       volume = {7735},
        month = jul,
          eid = {773508},
        pages = {773508},
          doi = {10.1117/12.856027},
archivePrefix = {arXiv},
       eprint = {2211.16795},
 primaryClass = {astro-ph.IM},
       adsurl = {https://ui.adsabs.harvard.edu/abs/2010SPIE.7735E..08B}
}

@ARTICLE{Dobbs2010,
       author = {{Dobbs}, C.~L. and {Theis}, C. and {Pringle}, J.~E. and {Bate}, M.~R.},
        title = "{Simulations of the grand design galaxy M51: a case study for analysing tidally induced spiral structure}",
      journal = {\mnras},
         year = 2010,
        month = apr,
       volume = {403},
       number = {2},
        pages = {625-645},
          doi = {10.1111/j.1365-2966.2009.16161.x},
archivePrefix = {arXiv},
       eprint = {0912.1201},
 primaryClass = {astro-ph.GA},
       adsurl = {https://ui.adsabs.harvard.edu/abs/2010MNRAS.403..625D}
}

@ARTICLE{Poggianti2017,
       author = {{Poggianti}, Bianca M. and {Moretti}, Alessia and {Gullieuszik}, Marco and {Fritz}, Jacopo and {Jaff{\'e}}, Yara and {Bettoni}, Daniela and {Fasano}, Giovanni and {Bellhouse}, Callum and {Hau}, George and {Vulcani}, Benedetta and {Biviano}, Andrea and {Omizzolo}, Alessandro and {Paccagnella}, Angela and {D'Onofrio}, Mauro and {Cava}, Antonio and {Sheen}, Y. -K. and {Couch}, Warrick and {Owers}, Matt},
        title = "{GASP. I. Gas Stripping Phenomena in Galaxies with MUSE}",
      journal = {\apj},
         year = 2017,
        month = jul,
       volume = {844},
       number = {1},
          eid = {48},
        pages = {48},
          doi = {10.3847/1538-4357/aa78ed},
archivePrefix = {arXiv},
       eprint = {1704.05086},
 primaryClass = {astro-ph.GA},
       adsurl = {https://ui.adsabs.harvard.edu/abs/2017ApJ...844...48P}
}

@ARTICLE{Fritz2007,
       author = {{Fritz}, J. and {Poggianti}, B.~M. and {Bettoni}, D. and {Cava}, A. and {Couch}, W.~J. and {D'Onofrio}, M. and {Dressler}, A. and {Fasano}, G. and {Kj{\ae}rgaard}, P. and {Moles}, M. and {Varela}, J.},
        title = "{A spectrophotometric model applied to cluster galaxies: the WINGS dataset}",
      journal = {\aap},
         year = 2007,
        month = jul,
       volume = {470},
       number = {1},
        pages = {137-152},
          doi = {10.1051/0004-6361:20077097},
archivePrefix = {arXiv},
       eprint = {0705.1513},
 primaryClass = {astro-ph},
       adsurl = {https://ui.adsabs.harvard.edu/abs/2007A&A...470..137F}
}

@ARTICLE{Calzetti1994,
       author = {{Calzetti}, Daniela and {Kinney}, Anne L. and {Storchi-Bergmann}, Thaisa},
        title = "{Dust Extinction of the Stellar Continua in Starburst Galaxies: The Ultraviolet and Optical Extinction Law}",
      journal = {\apj},
         year = 1994,
        month = jul,
       volume = {429},
        pages = {582},
          doi = {10.1086/174346},
       adsurl = {https://ui.adsabs.harvard.edu/abs/1994ApJ...429..582C}
}

@ARTICLE{Watson2025,
       author = {{Watson}, Peter J. and {Vulcani}, Benedetta and {Werle}, Ariel and {Poggianti}, Bianca and {Gullieuszik}, Marco and {Trenti}, Michele and {Wang}, Xin and {Roy}, Namrata},
        title = "{Unveiling multiple physical processes on a cluster galaxy at z = 0.3 using JWST}",
      journal = {\aap},
         year = 2025,
        month = jul,
       volume = {699},
          eid = {A365},
        pages = {A365},
          doi = {10.1051/0004-6361/202452348},
archivePrefix = {arXiv},
       eprint = {2409.15215},
 primaryClass = {astro-ph.GA},
       adsurl = {https://ui.adsabs.harvard.edu/abs/2025A&A...699A.365W}
}

@ARTICLE{Pedrini2022,
       author = {{Pedrini}, Alex and {Fossati}, Matteo and {Gavazzi}, Giuseppe and {Fumagalli}, Michele and {Boselli}, Alessandro and {Consolandi}, Guido and {Sun}, Ming and {Yagi}, Masafumi and {Yoshida}, Michitoshi},
        title = "{MUSE sneaks a peek at extreme ram-pressure stripping events - V. Towards a complete view of the galaxy cluster A1367}",
      journal = {\mnras},
         year = 2022,
        month = apr,
       volume = {511},
       number = {4},
        pages = {5180-5197},
          doi = {10.1093/mnras/stac345},
archivePrefix = {arXiv},
       eprint = {2202.03443},
 primaryClass = {astro-ph.GA},
       adsurl = {https://ui.adsabs.harvard.edu/abs/2022MNRAS.511.5180P}
}

@ARTICLE{Begeman1989,
       author = {{Begeman}, K.~G.},
        title = "{HI rotation curves of spiral galaxies. I. NGC 3198.}",
      journal = {\aap},
         year = 1989,
        month = oct,
       volume = {223},
        pages = {47-60},
       adsurl = {https://ui.adsabs.harvard.edu/abs/1989A&A...223...47B}
}

@ARTICLE{Giovanelli1994,
       author = {{Giovanelli}, Riccardo and {Haynes}, Martha P. and {Salzer}, John J. and {Wegner}, Gary and {da Costa}, Luiz N. and {Freudling}, Wolfram},
        title = "{Extinction in SC Galaxies}",
      journal = {\aj},
         year = 1994,
        month = jun,
       volume = {107},
        pages = {2036},
          doi = {10.1086/117014},
       adsurl = {https://ui.adsabs.harvard.edu/abs/1994AJ....107.2036G}
}

@ARTICLE{Krajnovic2006,
       author = {{Krajnovi{\'c}}, Davor and {Cappellari}, Michele and {de Zeeuw}, P. Tim and {Copin}, Yannick},
        title = "{Kinemetry: a generalization of photometry to the higher moments of the line-of-sight velocity distribution}",
      journal = {\mnras},
         year = 2006,
        month = mar,
       volume = {366},
       number = {3},
        pages = {787-802},
          doi = {10.1111/j.1365-2966.2005.09902.x},
archivePrefix = {arXiv},
       eprint = {astro-ph/0512200},
 primaryClass = {astro-ph},
       adsurl = {https://ui.adsabs.harvard.edu/abs/2006MNRAS.366..787K}
}

@ARTICLE{Cappellari2007,
       author = {{Cappellari}, Michele and {Emsellem}, Eric and {Bacon}, R. and {Bureau}, M. and {Davies}, Roger L. and {de Zeeuw}, P.~T. and {Falc{\'o}n-Barroso}, Jes{\'u}s and {Krajnovi{\'c}}, Davor and {Kuntschner}, Harald and {McDermid}, Richard M. and {Peletier}, Reynier F. and {Sarzi}, Marc and {van den Bosch}, Remco C.~E. and {van de Ven}, Glenn},
        title = "{The SAURON project - X. The orbital anisotropy of elliptical and lenticular galaxies: revisiting the (V/{\ensuremath{\sigma}}, {\ensuremath{\varepsilon}}) diagram with integral-field stellar kinematics}",
      journal = {MNRAS},
         year = 2007,
        month = aug,
       volume = {379},
       number = {2},
        pages = {418-444},
          doi = {10.1111/j.1365-2966.2007.11963.x},
archivePrefix = {arXiv},
       eprint = {astro-ph/0703533},
 primaryClass = {astro-ph},
       adsurl = {https://ui.adsabs.harvard.edu/abs/2007MNRAS.379..418C}
}

@ARTICLE{Marasco2026,
       author = {{Marasco}, A. and {Poggianti}, B.~M. and {Vulcani}, B. and {Moretti}, A. and {Gullieuszik}, M. and {Fritz}, J.},
        title = "{Modelling the photometric and morphological evolution of disc galaxies in the cluster environment}",
      journal = {arXiv e-prints},
         year = 2026,
        month = feb,
          eid = {arXiv:2602.06119},
        pages = {arXiv:2602.06119},
archivePrefix = {arXiv},
       eprint = {2602.06119},
 primaryClass = {astro-ph.GA},
       adsurl = {https://ui.adsabs.harvard.edu/abs/2026arXiv260206119M}
}

@ARTICLE{Marasco2023,
       author = {{Marasco}, A. and {Poggianti}, B.~M. and {Fritz}, J. and {Werle}, A. and {Vulcani}, B. and {Moretti}, A. and {Gullieuszik}, M. and {Kulier}, A.},
        title = "{The morphological transformation of ram pressure stripped galaxies: a pathway from late to early galaxy types}",
      journal = {\mnras},
         year = 2023,
        month = nov,
       volume = {525},
       number = {4},
        pages = {5359-5377},
          doi = {10.1093/mnras/stad2604},
archivePrefix = {arXiv},
       eprint = {2308.14791},
 primaryClass = {astro-ph.GA},
       adsurl = {https://ui.adsabs.harvard.edu/abs/2023MNRAS.525.5359M}
}

@ARTICLE{Poggianti2025,
       author = {{Poggianti}, Bianca M. and {Vulcani}, Benedetta and {Tomicic}, Neven and {Moretti}, Alessia and {Gullieuszik}, Marco and {Bacchini}, Cecilia and {Fritz}, Jacopo and {George}, Koshy and {Gitti}, Myriam and {Ignesti}, Alessandro and {Jaff{\'e}}, Yara and {Lassen}, Augusto and {Marasco}, Antonino and {Radovich}, Mario and {Serra}, Paolo and {Smith}, Rory and {Tonnesen}, Stephanie and {Wolter}, Anna},
        title = "{The MUSE view of ram pressure stripped galaxies in clusters: The GASP sample}",
      journal = {\aap},
         year = 2025,
        month = jul,
       volume = {699},
          eid = {A357},
        pages = {A357},
          doi = {10.1051/0004-6361/202554200},
archivePrefix = {arXiv},
       eprint = {2505.21107},
 primaryClass = {astro-ph.GA},
       adsurl = {https://ui.adsabs.harvard.edu/abs/2025A&A...699A.357P}
}

@ARTICLE{Kewley2006,
       author = {{Kewley}, Lisa J. and {Groves}, Brent and {Kauffmann}, Guinevere and {Heckman}, Tim},
        title = "{The host galaxies and classification of active galactic nuclei}",
      journal = {\mnras},
         year = 2006,
        month = nov,
       volume = {372},
       number = {3},
        pages = {961-976},
          doi = {10.1111/j.1365-2966.2006.10859.x},
archivePrefix = {arXiv},
       eprint = {astro-ph/0605681},
 primaryClass = {astro-ph},
       adsurl = {https://ui.adsabs.harvard.edu/abs/2006MNRAS.372..961K}
}

@ARTICLE{Kewley2001,
       author = {{Kewley}, L.~J. and {Dopita}, M.~A. and {Sutherland}, R.~S. and {Heisler}, C.~A. and {Trevena}, J.},
        title = "{Theoretical Modeling of Starburst Galaxies}",
      journal = {\apj},
         year = 2001,
        month = jul,
       volume = {556},
       number = {1},
        pages = {121-140},
          doi = {10.1086/321545},
archivePrefix = {arXiv},
       eprint = {astro-ph/0106324},
 primaryClass = {astro-ph},
       adsurl = {https://ui.adsabs.harvard.edu/abs/2001ApJ...556..121K}
}

@ARTICLE{Kauffmann2003,
       author = {{Kauffmann}, Guinevere and {Heckman}, Timothy M. and {Tremonti}, Christy and {Brinchmann}, Jarle and {Charlot}, St{\'e}phane and {White}, Simon D.~M. and {Ridgway}, Susan E. and {Brinkmann}, Jon and {Fukugita}, Masataka and {Hall}, Patrick B. and {Ivezi{\'c}}, {\v{Z}}eljko and {Richards}, Gordon T. and {Schneider}, Donald P.},
        title = "{The host galaxies of active galactic nuclei}",
      journal = {\mnras},
         year = 2003,
        month = dec,
       volume = {346},
       number = {4},
        pages = {1055-1077},
          doi = {10.1111/j.1365-2966.2003.07154.x},
archivePrefix = {arXiv},
       eprint = {astro-ph/0304239},
 primaryClass = {astro-ph},
       adsurl = {https://ui.adsabs.harvard.edu/abs/2003MNRAS.346.1055K}
}

@ARTICLE{Sharp2010,
       author = {{Sharp}, R.~G. and {Bland-Hawthorn}, J.},
        title = "{Three-Dimensional Integral Field Observations of 10 Galactic Winds. I. Extended Phase (gsim10 Myr) of Mass/Energy Injection Before the Wind Blows}",
      journal = {\apj},
         year = 2010,
        month = mar,
       volume = {711},
       number = {2},
        pages = {818-852},
          doi = {10.1088/0004-637X/711/2/818},
archivePrefix = {arXiv},
       eprint = {1001.4315},
 primaryClass = {astro-ph.CO},
       adsurl = {https://ui.adsabs.harvard.edu/abs/2010ApJ...711..818S}
}

@ARTICLE{Veilleux1987,
       author = {{Veilleux}, Sylvain and {Osterbrock}, Donald E.},
        title = "{Spectral Classification of Emission-Line Galaxies}",
      journal = {\apjs},
         year = 1987,
        month = feb,
       volume = {63},
        pages = {295},
          doi = {10.1086/191166},
       adsurl = {https://ui.adsabs.harvard.edu/abs/1987ApJS...63..295V}
}

@ARTICLE{Finlator2008,
       author = {{Finlator}, Kristian and {Dav{\'e}}, Romeel},
        title = "{The origin of the galaxy mass-metallicity relation and implications for galactic outflows}",
      journal = {\mnras},
         year = 2008,
        month = apr,
       volume = {385},
       number = {4},
        pages = {2181-2204},
          doi = {10.1111/j.1365-2966.2008.12991.x},
archivePrefix = {arXiv},
       eprint = {0704.3100},
 primaryClass = {astro-ph},
       adsurl = {https://ui.adsabs.harvard.edu/abs/2008MNRAS.385.2181F}
}

@ARTICLE{Tremonti2004,
       author = {{Tremonti}, Christy A. and {Heckman}, Timothy M. and {Kauffmann}, Guinevere and {Brinchmann}, Jarle and {Charlot}, St{\'e}phane and {White}, Simon D.~M. and {Seibert}, Mark and {Peng}, Eric W. and {Schlegel}, David J. and {Uomoto}, Alan and {Fukugita}, Masataka and {Brinkmann}, Jon},
        title = "{The Origin of the Mass-Metallicity Relation: Insights from 53,000 Star-forming Galaxies in the Sloan Digital Sky Survey}",
      journal = {\apj},
         year = 2004,
        month = oct,
       volume = {613},
       number = {2},
        pages = {898-913},
          doi = {10.1086/423264},
archivePrefix = {arXiv},
       eprint = {astro-ph/0405537},
 primaryClass = {astro-ph},
       adsurl = {https://ui.adsabs.harvard.edu/abs/2004ApJ...613..898T}
}

@ARTICLE{Smith2010,
       author = {{Smith}, Russell J. and {Lucey}, John R. and {Hammer}, Derek and {Hornschemeier}, Ann E. and {Carter}, David and {Hudson}, Michael J. and {Marzke}, Ronald O. and {Mouhcine}, Mustapha and {Eftekharzadeh}, Sareh and {James}, Phil and {Khosroshahi}, Habib and {Kourkchi}, Ehsan and {Karick}, Arna},
        title = "{Ultraviolet tails and trails in cluster galaxies: a sample of candidate gaseous stripping events in Coma}",
      journal = {\mnras},
         year = 2010,
        month = nov,
       volume = {408},
       number = {3},
        pages = {1417-1432},
          doi = {10.1111/j.1365-2966.2010.17253.x},
archivePrefix = {arXiv},
       eprint = {1006.4867},
 primaryClass = {astro-ph.CO},
       adsurl = {https://ui.adsabs.harvard.edu/abs/2010MNRAS.408.1417S}
}

@ARTICLE{Fumagalli2014,
       author = {{Fumagalli}, Michele and {Fossati}, Matteo and {Hau}, George K.~T. and {Gavazzi}, Giuseppe and {Bower}, Richard and {Sun}, Ming and {Boselli}, Alessandro},
        title = "{MUSE sneaks a peek at extreme ram-pressure stripping events - I. A kinematic study of the archetypal galaxy ESO137-001}",
      journal = {\mnras},
         year = 2014,
        month = dec,
       volume = {445},
       number = {4},
        pages = {4335-4344},
          doi = {10.1093/mnras/stu2092},
archivePrefix = {arXiv},
       eprint = {1407.7527},
 primaryClass = {astro-ph.GA},
       adsurl = {https://ui.adsabs.harvard.edu/abs/2014MNRAS.445.4335F}
}

@ARTICLE{Poggianti2019a,
       author = {{Poggianti}, Bianca M. and {Gullieuszik}, Marco and {Tonnesen}, Stephanie and {Moretti}, Alessia and {Vulcani}, Benedetta and {Radovich}, Mario and {Jaff{\'e}}, Yara and {Fritz}, Jacopo and {Bettoni}, Daniela and {Franchetto}, Andrea and {Fasano}, Giovanni and {Bellhouse}, Callum and {Omizzolo}, Alessandro},
        title = "{GASP XIII. Star formation in gas outside galaxies}",
      journal = {\mnras},
         year = 2019,
        month = feb,
       volume = {482},
       number = {4},
        pages = {4466-4502},
          doi = {10.1093/mnras/sty2999},
archivePrefix = {arXiv},
       eprint = {1811.00823},
 primaryClass = {astro-ph.GA},
       adsurl = {https://ui.adsabs.harvard.edu/abs/2019MNRAS.482.4466P}
}

@ARTICLE{Rasmussen2006,
       author = {{Rasmussen}, Jesper and {Ponman}, Trevor J. and {Mulchaey}, John S.},
        title = "{Gas stripping in galaxy groups - the case of the starburst spiral NGC 2276}",
      journal = {\mnras},
         year = 2006,
        month = jul,
       volume = {370},
       number = {1},
        pages = {453-467},
          doi = {10.1111/j.1365-2966.2006.10492.x},
archivePrefix = {arXiv},
       eprint = {astro-ph/0604549},
 primaryClass = {astro-ph},
       adsurl = {https://ui.adsabs.harvard.edu/abs/2006MNRAS.370..453R}
}

@ARTICLE{Poggianti2019b,
       author = {{Poggianti}, Bianca M. and {Ignesti}, Alessandro and {Gitti}, Myriam and {Wolter}, Anna and {Brighenti}, Fabrizio and {Biviano}, Andrea and {George}, Koshy and {Vulcani}, Benedetta and {Gullieuszik}, Marco and {Moretti}, Alessia and {Paladino}, Rosita and {Bettoni}, Daniela and {Franchetto}, Andrea and {Jaff{\'e}}, Yara L. and {Radovich}, Mario and {Roediger}, Elke and {Tomi{\v{c}}i{\'c}}, Neven and {Tonnesen}, Stephanie and {Bellhouse}, Callum and {Fritz}, Jacopo and {Omizzolo}, Alessandro},
        title = "{GASP XXIII: A Jellyfish Galaxy as an Astrophysical Laboratory of the Baryonic Cycle}",
      journal = {\apj},
         year = 2019,
        month = dec,
       volume = {887},
       number = {2},
          eid = {155},
        pages = {155},
          doi = {10.3847/1538-4357/ab5224},
archivePrefix = {arXiv},
       eprint = {1910.11622},
 primaryClass = {astro-ph.GA},
       adsurl = {https://ui.adsabs.harvard.edu/abs/2019ApJ...887..155P}
}

@ARTICLE{Kapferer2009,
       author = {{Kapferer}, W. and {Sluka}, C. and {Schindler}, S. and {Ferrari}, C. and {Ziegler}, B.},
        title = "{The effect of ram pressure on the star formation, mass distribution and morphology of galaxies}",
      journal = {\aap},
         year = 2009,
        month = may,
       volume = {499},
       number = {1},
        pages = {87-102},
          doi = {10.1051/0004-6361/200811551},
archivePrefix = {arXiv},
       eprint = {0903.3818},
 primaryClass = {astro-ph.CO},
       adsurl = {https://ui.adsabs.harvard.edu/abs/2009A&A...499...87K}
}

@ARTICLE{Rhee2017,
       author = {{Rhee}, Jinsu and {Smith}, Rory and {Choi}, Hoseung and {Yi}, Sukyoung K. and {Jaff{\'e}}, Yara and {Candlish}, Graeme and {S{\'a}nchez-J{\'a}nssen}, Ruben},
        title = "{Phase-space Analysis in the Group and Cluster Environment: Time Since Infall and Tidal Mass Loss}",
      journal = {\apj},
         year = 2017,
        month = jul,
       volume = {843},
       number = {2},
          eid = {128},
        pages = {128},
          doi = {10.3847/1538-4357/aa6d6c},
archivePrefix = {arXiv},
       eprint = {1704.04243},
 primaryClass = {astro-ph.GA},
       adsurl = {https://ui.adsabs.harvard.edu/abs/2017ApJ...843..128R}
}

@ARTICLE{Vulcani2018b,
       author = {{Vulcani}, Benedetta and {Poggianti}, Bianca M. and {Gullieuszik}, Marco and {Moretti}, Alessia and {Tonnesen}, Stephanie and {Jaff{\'e}}, Yara L. and {Fritz}, Jacopo and {Fasano}, Giovanni and {Bettoni}, Daniela},
        title = "{Enhanced Star Formation in Both Disks and Ram-pressure-stripped Tails of GASP Jellyfish Galaxies}",
      journal = {\apjl},
         year = 2018,
        month = oct,
       volume = {866},
       number = {2},
          eid = {L25},
        pages = {L25},
          doi = {10.3847/2041-8213/aae68b},
archivePrefix = {arXiv},
       eprint = {1810.05164},
 primaryClass = {astro-ph.GA},
       adsurl = {https://ui.adsabs.harvard.edu/abs/2018ApJ...866L..25V}
}

@ARTICLE{Vulcani2020a,
       author = {{Vulcani}, Benedetta and {Fritz}, Jacopo and {Poggianti}, Bianca M. and {Bettoni}, Daniela and {Franchetto}, Andrea and {Moretti}, Alessia and {Gullieuszik}, Marco and {Jaff{\'e}}, Yara and {Biviano}, Andrea and {Radovich}, Mario and {Mingozzi}, Matilde},
        title = "{GASP XXIV. The History of Abruptly Quenched Galaxies in Clusters}",
      journal = {\apj},
         year = 2020,
        month = apr,
       volume = {892},
       number = {2},
          eid = {146},
        pages = {146},
          doi = {10.3847/1538-4357/ab7bdd},
archivePrefix = {arXiv},
       eprint = {2003.02274},
 primaryClass = {astro-ph.GA},
       adsurl = {https://ui.adsabs.harvard.edu/abs/2020ApJ...892..146V}
}

@ARTICLE{Vulcani2020b,
       author = {{Vulcani}, Benedetta and {Poggianti}, Bianca M. and {Tonnesen}, Stephanie and {McGee}, Sean L. and {Moretti}, Alessia and {Fritz}, Jacopo and {Gullieuszik}, Marco and {Jaff{\'e}}, Yara L. and {Franchetto}, Andrea and {Tomi{\v{c}}i{\'c}}, Neven and {Mingozzi}, Matilde and {Bettoni}, Daniela and {Wolter}, Anna},
        title = "{GASP XXX. The Spatially Resolved SFR-Mass Relation in Stripping Galaxies in the Local Universe}",
      journal = {\apj},
         year = 2020,
        month = aug,
       volume = {899},
       number = {2},
          eid = {98},
        pages = {98},
          doi = {10.3847/1538-4357/aba4ae},
archivePrefix = {arXiv},
       eprint = {2007.04996},
 primaryClass = {astro-ph.GA},
       adsurl = {https://ui.adsabs.harvard.edu/abs/2020ApJ...899...98V}
}

@ARTICLE{Cortese2021,
       author = {{Cortese}, L. and {Catinella}, B. and {Smith}, R.},
        title = "{The Dawes Review 9: The role of cold gas stripping on the star formation quenching of satellite galaxies}",
      journal = {\pasa},
         year = 2021,
        month = aug,
       volume = {38},
          eid = {e035},
        pages = {e035},
          doi = {10.1017/pasa.2021.18},
archivePrefix = {arXiv},
       eprint = {2104.02193},
 primaryClass = {astro-ph.GA},
       adsurl = {https://ui.adsabs.harvard.edu/abs/2021PASA...38...35C}
}

@ARTICLE{Vulcani2021,
       author = {{Vulcani}, Benedetta and {Poggianti}, Bianca M. and {Moretti}, Alessia and {Franchetto}, Andrea and {Bacchini}, Cecilia and {McGee}, Sean and {Jaff{\'e}}, Yara L. and {Mingozzi}, Matilde and {Werle}, Ariel and {Tomi{\v{c}}i{\'c}}, Neven and {Fritz}, Jacopo and {Bettoni}, Daniela and {Wolter}, Anna and {Gullieuszik}, Marco},
        title = "{GASP. XXXIII. The Ability of Spatially Resolved Data to Distinguish among the Different Physical Mechanisms Affecting Galaxies in Low-density Environments}",
      journal = {\apj},
         year = 2021,
        month = jun,
       volume = {914},
       number = {1},
          eid = {27},
        pages = {27},
          doi = {10.3847/1538-4357/abf655},
archivePrefix = {arXiv},
       eprint = {2104.02089},
 primaryClass = {astro-ph.GA},
       adsurl = {https://ui.adsabs.harvard.edu/abs/2021ApJ...914...27V}
}

@ARTICLE{Bournaud2004,
       author = {{Bournaud}, F. and {Duc}, P. -A. and {Amram}, P. and {Combes}, F. and {Gach}, J. -L.},
        title = "{Kinematics of tidal tails in interacting galaxies: Tidal dwarf galaxies and projection effects}",
      journal = {\aap},
         year = 2004,
        month = oct,
       volume = {425},
        pages = {813-823},
          doi = {10.1051/0004-6361:20040394},
archivePrefix = {arXiv},
       eprint = {astro-ph/0406169},
 primaryClass = {astro-ph},
       adsurl = {https://ui.adsabs.harvard.edu/abs/2004A&A...425..813B}
}

@incollection{Duc2013,
  author    = {Duc, Pierre-Alain and Renaud, Florent},
  title     = {Tides in Colliding Galaxies},
  booktitle = {Lecture Notes in Physics},
  editor    = {Souchay, Jean and Mathis, Stéphane and Tokieda, Tadashi},
  publisher = {Springer},
  year      = {2013},
  volume    = {861},
  pages     = {327--372},
  doi       = {10.1007/978-3-642-32961-6_9}
}

@ARTICLE{Byrd1990,
       author = {{Byrd}, Gene and {Valtonen}, Mauri},
        title = "{Tidal Generation of Active Spirals and S0 Galaxies by Rich Clusters}",
      journal = {\apj},
         year = 1990,
        month = feb,
       volume = {350},
        pages = {89},
          doi = {10.1086/168362},
       adsurl = {https://ui.adsabs.harvard.edu/abs/1990ApJ...350...89B}
}

@ARTICLE{Valluri1993,
       author = {{Valluri}, Monica},
        title = "{Compressive Tidal Heating of a Disk Galaxy in a Rich Cluster}",
      journal = {\apj},
         year = 1993,
        month = may,
       volume = {408},
        pages = {57},
          doi = {10.1086/172569},
       adsurl = {https://ui.adsabs.harvard.edu/abs/1993ApJ...408...57V}
}

@ARTICLE{Struck1999,
       author = {{Struck}, C.},
        title = "{Galaxy collisions.}",
      journal = {\physrep},
         year = 1999,
        month = jan,
       volume = {321},
        pages = {1-137},
          doi = {10.1016/S0370-1573(99)00030-7},
archivePrefix = {arXiv},
       eprint = {astro-ph/9908269},
 primaryClass = {astro-ph},
       adsurl = {https://ui.adsabs.harvard.edu/abs/1999PhR...321....1S}
}

@ARTICLE{Barbary2016,
       author = {{Barbary}, Kyle},
        title = "{SEP: Source Extractor as a library}",
      journal = {The Journal of Open Source Software},
         year = 2016,
        month = oct,
       volume = {1},
       number = {6},
          eid = {58},
        pages = {58},
          doi = {10.21105/joss.00058},
       adsurl = {https://ui.adsabs.harvard.edu/abs/2016JOSS....1...58B}
}

@ARTICLE{Haines2018,
       author = {{Haines}, C.~P. and {Busarello}, G. and {Merluzzi}, P. and {Pimbblet}, K.~A. and {Vogt}, F.~P.~A. and {Dopita}, M.~A. and {Mercurio}, A. and {Grado}, A. and {Limatola}, L.},
        title = "{Shapley Supercluster Survey: mapping the filamentary network connecting the clusters}",
      journal = {\mnras},
         year = 2018,
        month = nov,
       volume = {481},
       number = {1},
        pages = {1055-1074},
          doi = {10.1093/mnras/sty2338},
       adsurl = {https://ui.adsabs.harvard.edu/abs/2018MNRAS.481.1055H}
}

@ARTICLE{Fasano2006,
       author = {{Fasano}, G. and {Marmo}, C. and {Varela}, J. and {D'Onofrio}, M. and {Poggianti}, B.~M. and {Moles}, M. and {Pignatelli}, E. and {Bettoni}, D. and {Kj{\ae}rgaard}, P. and {Rizzi}, L. and {Couch}, W.~J. and {Dressler}, A.},
        title = "{WINGS: a WIde-field Nearby Galaxy-cluster Survey. I. Optical imaging}",
      journal = {\aap},
         year = 2006,
        month = jan,
       volume = {445},
       number = {3},
        pages = {805-817},
          doi = {10.1051/0004-6361:20053816},
archivePrefix = {arXiv},
       eprint = {astro-ph/0507247},
 primaryClass = {astro-ph},
       adsurl = {https://ui.adsabs.harvard.edu/abs/2006A&A...445..805F}
}

@ARTICLE{Moretti2014,
       author = {{Moretti}, A. and {Poggianti}, B.~M. and {Fasano}, G. and {Bettoni}, D. and {D'Onofrio}, M. and {Fritz}, J. and {Cava}, A. and {Varela}, J. and {Vulcani}, B. and {Gullieuszik}, M. and {Couch}, W.~J. and {Omizzolo}, A. and {Valentinuzzi}, T. and {Dressler}, A. and {Moles}, M. and {Kj{\ae}rgaard}, P. and {Smareglia}, R. and {Molinaro}, M.},
        title = "{WINGS Data Release: a database of galaxies in nearby clusters}",
      journal = {\aap},
         year = 2014,
        month = apr,
       volume = {564},
          eid = {A138},
        pages = {A138},
          doi = {10.1051/0004-6361/201323098},
archivePrefix = {arXiv},
       eprint = {1403.1408},
 primaryClass = {astro-ph.GA},
       adsurl = {https://ui.adsabs.harvard.edu/abs/2014A&A...564A.138M}
}

@ARTICLE{Vulcani2023,
       author = {{Vulcani}, Benedetta and {Poggianti}, Bianca M. and {Gullieuszik}, Marco and {Moretti}, Alessia and {Fritz}, Jacopo and {Bettoni}, Daniela and {Facciolli}, Beatrice and {Fasano}, Giovanni and {Omizzolo}, Alessandro},
        title = "{Clustercentric Distance or Local Density? It Depends on Galaxy Morphology}",
      journal = {\apj},
         year = 2023,
        month = jun,
       volume = {949},
       number = {2},
          eid = {73},
        pages = {73},
          doi = {10.3847/1538-4357/acc5e2},
archivePrefix = {arXiv},
       eprint = {2302.02376},
 primaryClass = {astro-ph.GA},
       adsurl = {https://ui.adsabs.harvard.edu/abs/2023ApJ...949...73V}
}

@ARTICLE{Moretti2017,
       author = {{Moretti}, A. and {Gullieuszik}, M. and {Poggianti}, B. and {Paccagnella}, A. and {Couch}, W.~J. and {Vulcani}, B. and {Bettoni}, D. and {Fritz}, J. and {Cava}, A. and {Fasano}, G. and {D'Onofrio}, M. and {Omizzolo}, A.},
        title = "{OmegaWINGS: spectroscopy in the outskirts of local clusters of galaxies}",
      journal = {\aap},
         year = 2017,
        month = mar,
       volume = {599},
          eid = {A81},
        pages = {A81},
          doi = {10.1051/0004-6361/201630030},
archivePrefix = {arXiv},
       eprint = {1701.02590},
 primaryClass = {astro-ph.GA},
       adsurl = {https://ui.adsabs.harvard.edu/abs/2017A&A...599A..81M}
}

@ARTICLE{Poggianti2016,
       author = {{Poggianti}, B.~M. and {Fasano}, G. and {Omizzolo}, A. and {Gullieuszik}, M. and {Bettoni}, D. and {Moretti}, A. and {Paccagnella}, A. and {Jaff{\'e}}, Y.~L. and {Vulcani}, B. and {Fritz}, J. and {Couch}, W. and {D'Onofrio}, M.},
        title = "{Jellyfish Galaxy Candidates at Low Redshift}",
      journal = {\aj},
         year = 2016,
        month = mar,
       volume = {151},
       number = {3},
          eid = {78},
        pages = {78},
          doi = {10.3847/0004-6256/151/3/78},
archivePrefix = {arXiv},
       eprint = {1504.07105},
 primaryClass = {astro-ph.GA},
       adsurl = {https://ui.adsabs.harvard.edu/abs/2016AJ....151...78P}
}

@ARTICLE{Durret2021,
       author = {{Durret}, F. and {Chiche}, S. and {Lobo}, C. and {Jauzac}, M.},
        title = "{Jellyfish galaxy candidates in MACS J0717.5+3745 and 39 other clusters of the DAFT/FADA and CLASH surveys}",
      journal = {\aap},
         year = 2021,
        month = apr,
       volume = {648},
          eid = {A63},
        pages = {A63},
          doi = {10.1051/0004-6361/202039770},
archivePrefix = {arXiv},
       eprint = {2102.02595},
 primaryClass = {astro-ph.GA},
       adsurl = {https://ui.adsabs.harvard.edu/abs/2021A&A...648A..63D}
}

@ARTICLE{Vulcani2022,
       author = {{Vulcani}, Benedetta and {Poggianti}, Bianca M. and {Smith}, Rory and {Moretti}, Alessia and {Jaff{\'e}}, Yara L. and {Gullieuszik}, Marco and {Fritz}, Jacopo and {Bellhouse}, Callum},
        title = "{The Relevance of Ram Pressure Stripping for the Evolution of Blue Cluster Galaxies as Seen at Optical Wavelengths}",
      journal = {\apj},
         year = 2022,
        month = mar,
       volume = {927},
       number = {1},
          eid = {91},
        pages = {91},
          doi = {10.3847/1538-4357/ac4809},
archivePrefix = {arXiv},
       eprint = {2201.02644},
 primaryClass = {astro-ph.GA},
       adsurl = {https://ui.adsabs.harvard.edu/abs/2022ApJ...927...91V}
}

@BOOK{Binney2008,
    author = {{Binney}, James and {Tremaine}, Scott},
    title = "{Galactic Dynamics: Second Edition}",
    year = 2008,
    publisher = "{Princeton University Press}",
    adsurl = {https://ui.adsabs.harvard.edu/abs/2008gady.book.....B}
}

@ARTICLE{Fuentes-Carrera2004,
       author = {{Fuentes-Carrera}, I. and {Rosado}, M. and {Amram}, P. and {Dultzin-Hacyan}, D. and {Cruz-Gonz{\'a}lez}, I. and {Salo}, H. and {Laurikainen}, E. and {Bernal}, A. and {Ambrocio-Cruz}, P. and {Le Coarer}, E.},
        title = "{The isolated interacting galaxy pair NGC 5426/27 (Arp 271)}",
      journal = {\aap},
         year = 2004,
        month = feb,
       volume = {415},
        pages = {451-469},
          doi = {10.1051/0004-6361:20034190},
archivePrefix = {arXiv},
       eprint = {astro-ph/0312535},
 primaryClass = {astro-ph},
       adsurl = {https://ui.adsabs.harvard.edu/abs/2004A&A...415..451F}
}

@ARTICLE{Barton1999,
       author = {{Barton}, Elizabeth J. and {Bromley}, Benjamin C. and {Geller}, Margaret J.},
        title = "{Kinematic Effects of Tidal Interaction on Galaxy Rotation Curves}",
      journal = {\apjl},
         year = 1999,
        month = jan,
       volume = {511},
       number = {1},
        pages = {L25-L28},
          doi = {10.1086/311830},
archivePrefix = {arXiv},
       eprint = {astro-ph/9811340},
 primaryClass = {astro-ph},
       adsurl = {https://ui.adsabs.harvard.edu/abs/1999ApJ...511L..25B}
}

@ARTICLE{Pedrosa2008,
       author = {{Pedrosa}, S. and {Tissera}, P.~B. and {Fuentes-Carrera}, I. and {Mendes de Oliveira}, C.},
        title = "{Rotation curve bifurcations as indicators of close recent galaxy encounters}",
      journal = {\aap},
         year = 2008,
        month = jun,
       volume = {484},
       number = {2},
        pages = {299-302},
          doi = {10.1051/0004-6361:20078318},
archivePrefix = {arXiv},
       eprint = {0804.0533},
 primaryClass = {astro-ph},
       adsurl = {https://ui.adsabs.harvard.edu/abs/2008A&A...484..299P}
}

@ARTICLE{Hopkins2018,
       author = {{Hopkins}, A.~M.},
        title = "{The Dawes Review 8: Measuring the Stellar Initial Mass Function}",
      journal = {\pasa},
         year = 2018,
        month = nov,
       volume = {35},
          eid = {e039},
        pages = {e039},
          doi = {10.1017/pasa.2018.29},
archivePrefix = {arXiv},
       eprint = {1807.09949},
 primaryClass = {astro-ph.GA},
       adsurl = {https://ui.adsabs.harvard.edu/abs/2018PASA...35...39H}
}

@ARTICLE{Girelli2020,
       author = {{Girelli}, G. and {Pozzetti}, L. and {Bolzonella}, M. and {Giocoli}, C. and {Marulli}, F. and {Baldi}, M.},
        title = "{The stellar-to-halo mass relation over the past 12 Gyr. I. Standard {\ensuremath{\Lambda}}CDM model}",
      journal = {\aap},
         year = 2020,
        month = feb,
       volume = {634},
          eid = {A135},
        pages = {A135},
          doi = {10.1051/0004-6361/201936329},
archivePrefix = {arXiv},
       eprint = {2001.02230},
 primaryClass = {astro-ph.CO},
       adsurl = {https://ui.adsabs.harvard.edu/abs/2020A&A...634A.135G}
}

@ARTICLE{Nordstrom2004,
       author = {{Nordstr{\"o}m}, B. and {Mayor}, M. and {Andersen}, J. and {Holmberg}, J. and {Pont}, F. and {J{\o}rgensen}, B.~R. and {Olsen}, E.~H. and {Udry}, S. and {Mowlavi}, N.},
        title = "{The Geneva-Copenhagen survey of the Solar neighbourhood. Ages, metallicities, and kinematic properties of {\ensuremath{\sim}}14 000 F and G dwarfs}",
      journal = {\aap},
         year = 2004,
        month = may,
       volume = {418},
        pages = {989-1019},
          doi = {10.1051/0004-6361:20035959},
archivePrefix = {arXiv},
       eprint = {astro-ph/0405198},
 primaryClass = {astro-ph},
       adsurl = {https://ui.adsabs.harvard.edu/abs/2004A&A...418..989N}
}

@ARTICLE{Masters2019,
       author = {{Masters}, Karen L. and {Lintott}, Chris J. and {Hart}, Ross E. and {Kruk}, Sandor J. and {Smethurst}, Rebecca J. and {Casteels}, Kevin V. and {Keel}, William C. and {Simmons}, Brooke D. and {Stanescu}, Dennis O. and {Tate}, Jean and {Tomi}, Satoshi},
        title = "{Galaxy Zoo: unwinding the winding problem - observations of spiral bulge prominence and arm pitch angles suggest local spiral galaxies are winding}",
      journal = {\mnras},
         year = 2019,
        month = aug,
       volume = {487},
       number = {2},
        pages = {1808-1820},
          doi = {10.1093/mnras/stz1153},
archivePrefix = {arXiv},
       eprint = {1904.11436},
 primaryClass = {astro-ph.GA},
       adsurl = {https://ui.adsabs.harvard.edu/abs/2019MNRAS.487.1808M}
}

@ARTICLE{Font2019,
       author = {{Font}, Joan and {Beckman}, John E. and {James}, Phil A. and {Patsis}, Panos A.},
        title = "{Spiral structure in barred galaxies. Observational constraints to spiral arm formation mechanisms}",
      journal = {\mnras},
         year = 2019,
        month = feb,
       volume = {482},
       number = {4},
        pages = {5362-5378},
          doi = {10.1093/mnras/sty2983},
archivePrefix = {arXiv},
       eprint = {1901.04725},
 primaryClass = {astro-ph.GA},
       adsurl = {https://ui.adsabs.harvard.edu/abs/2019MNRAS.482.5362F}
}

@ARTICLE{Lingard2021,
       author = {{Lingard}, Timothy and {Masters}, Karen L. and {Krawczyk}, Coleman and {Lintott}, Chris and {Kruk}, Sandor and {Simmons}, Brooke and {Keel}, William and {Nichol}, Robert C. and {Baeten}, Elisabeth},
        title = "{Galaxy zoo builder: Morphological dependence of spiral galaxy pitch angle}",
      journal = {\mnras},
         year = 2021,
        month = jul,
       volume = {504},
       number = {3},
        pages = {3364-3374},
          doi = {10.1093/mnras/stab1072},
archivePrefix = {arXiv},
       eprint = {2105.04500},
 primaryClass = {astro-ph.GA},
       adsurl = {https://ui.adsabs.harvard.edu/abs/2021MNRAS.504.3364L}
}

@ARTICLE{Hart2017,
       author = {{Hart}, Ross E. and {Bamford}, Steven P. and {Hayes}, Wayne B. and {Cardamone}, Carolin N. and {Keel}, William C. and {Kruk}, Sandor J. and {Lintott}, Chris J. and {Masters}, Karen L. and {Simmons}, Brooke D. and {Smethurst}, Rebecca J.},
        title = "{Galaxy Zoo and SPARCFIRE: constraints on spiral arm formation mechanisms from spiral arm number and pitch angles}",
      journal = {\mnras},
         year = 2017,
        month = dec,
       volume = {472},
       number = {2},
        pages = {2263-2279},
          doi = {10.1093/mnras/stx2137},
archivePrefix = {arXiv},
       eprint = {1708.04628},
 primaryClass = {astro-ph.GA},
       adsurl = {https://ui.adsabs.harvard.edu/abs/2017MNRAS.472.2263H}
}

@ARTICLE{Romero-Gomez2007,
       author = {{Romero-G{\'o}mez}, M. and {Athanassoula}, E. and {Masdemont}, J.~J. and {Garc{\'\i}a-G{\'o}mez}, C.},
        title = "{The formation of spiral arms and rings in barred galaxies}",
      journal = {\aap},
         year = 2007,
        month = sep,
       volume = {472},
       number = {1},
        pages = {63-75},
          doi = {10.1051/0004-6361:20077504},
archivePrefix = {arXiv},
       eprint = {0705.2958},
 primaryClass = {astro-ph},
       adsurl = {https://ui.adsabs.harvard.edu/abs/2007A&A...472...63R}
}

@ARTICLE{Romero-Gomez2006,
       author = {{Romero-G{\'o}mez}, M. and {Masdemont}, J.~J. and {Athanassoula}, E. and {Garc{\'\i}a-G{\'o}mez}, C.},
        title = "{The origin of rR$_{1}$ ring structures in barred galaxies}",
      journal = {\aap},
         year = 2006,
        month = jul,
       volume = {453},
       number = {1},
        pages = {39-45},
          doi = {10.1051/0004-6361:20054653},
archivePrefix = {arXiv},
       eprint = {astro-ph/0603124},
 primaryClass = {astro-ph},
       adsurl = {https://ui.adsabs.harvard.edu/abs/2006A&A...453...39R}
}

@ARTICLE{Lelli2014,
       author = {{Lelli}, Federico and {Verheijen}, Marc and {Fraternali}, Filippo},
        title = "{The triggering of starbursts in low-mass galaxies}",
      journal = {\mnras},
         year = 2014,
        month = dec,
       volume = {445},
       number = {2},
        pages = {1694-1712},
          doi = {10.1093/mnras/stu1804},
archivePrefix = {arXiv},
       eprint = {1409.1239},
 primaryClass = {astro-ph.GA},
       adsurl = {https://ui.adsabs.harvard.edu/abs/2014MNRAS.445.1694L}
}

@ARTICLE{Lassen2026,
       author = {{Lassen}, Augusto E. and {Smith}, Rory and {Vulcani}, Benedetta and {Tonnesen}, Stephanie and {Calder{\'o}n-Castillo}, Paula and {Poggianti}, Bianca M. and {Fritz}, Jacopo and {George}, Koshy and {Ignesti}, Alessandro and {Jaff{\'e}}, Yara and {Marasco}, Antonino and {Matijevi{\'c}}, Luka and {Moretti}, Alessia and {Radovich}, Mario and {Tomi{\v{c}}i{\'c}}, Neven},
        title = "{Distinguishing ram pressure from gravitational interactions: Applying the size-shape difference method to real galaxies}",
      journal = {\aap},
         year = 2026,
        month = feb,
       volume = {706},
          eid = {A85},
        pages = {A85},
          doi = {10.1051/0004-6361/202556010},
archivePrefix = {arXiv},
       eprint = {2512.02923},
 primaryClass = {astro-ph.GA},
       adsurl = {https://ui.adsabs.harvard.edu/abs/2026A&A...706A..85L}
}

@ARTICLE{Machado2025,
       author = {{Machado}, Rubens E.~G. and {Grinberg}, Caroline F.~O. and {Mello-Terencio}, Elvis A.},
        title = "{Quantifying the Unwinding Due to Ram Pressure Stripping in Simulated Galaxies}",
      journal = {Galaxies},
         year = 2025,
        month = jul,
       volume = {13},
       number = {4},
          eid = {76},
        pages = {76},
          doi = {10.3390/galaxies13040076},
archivePrefix = {arXiv},
       eprint = {2507.02555},
 primaryClass = {astro-ph.GA},
       adsurl = {https://ui.adsabs.harvard.edu/abs/2025Galax..13...76M}
}

@ARTICLE{Biviano2017,
       author = {{Biviano}, A. and {Moretti}, A. and {Paccagnella}, A. and {Poggianti}, B.~M. and {Bettoni}, D. and {Gullieuszik}, M. and {Vulcani}, B. and {Fasano}, G. and {D'Onofrio}, M. and {Fritz}, J. and {Cava}, A.},
        title = "{The concentration-mass relation of clusters of galaxies from the OmegaWINGS survey}",
      journal = {\aap},
         year = 2017,
        month = nov,
       volume = {607},
          eid = {A81},
        pages = {A81},
          doi = {10.1051/0004-6361/201731289},
archivePrefix = {arXiv},
       eprint = {1708.07349},
 primaryClass = {astro-ph.CO},
       adsurl = {https://ui.adsabs.harvard.edu/abs/2017A&A...607A..81B}
}

@ARTICLE{Burkert1995,
       author = {{Burkert}, A.},
        title = "{The Structure of Dark Matter Halos in Dwarf Galaxies}",
      journal = {\apjl},
         year = 1995,
        month = jul,
       volume = {447},
        pages = {L25-L28},
          doi = {10.1086/309560},
archivePrefix = {arXiv},
       eprint = {astro-ph/9504041},
 primaryClass = {astro-ph},
       adsurl = {https://ui.adsabs.harvard.edu/abs/1995ApJ...447L..25B}
}

@ARTICLE{Struck2012,
       author = {{Struck}, Curtis and {Smith}, Beverly J.},
        title = "{The symmetries and scaling of tidal tails in galaxies}",
      journal = {\mnras},
         year = 2012,
        month = may,
       volume = {422},
       number = {3},
        pages = {2444-2464},
          doi = {10.1111/j.1365-2966.2012.20798.x},
archivePrefix = {arXiv},
       eprint = {1202.5280},
 primaryClass = {astro-ph.CO},
       adsurl = {https://ui.adsabs.harvard.edu/abs/2012MNRAS.422.2444S}
}

@ARTICLE{Cid2013,
       author = {{Cid Fernandes}, R. and {P{\'e}rez}, E. and {Garc{\'\i}a Benito}, R. and {Gonz{\'a}lez Delgado}, R.~M. and {de Amorim}, A.~L. and {S{\'a}nchez}, S.~F. and {Husemann}, B. and {Falc{\'o}n Barroso}, J. and {S{\'a}nchez-Bl{\'a}zquez}, P. and {Walcher}, C.~J. and {Mast}, D.},
        title = "{Resolving galaxies in time and space. I. Applying STARLIGHT to CALIFA datacubes}",
      journal = {\aap},
         year = 2013,
        month = sep,
       volume = {557},
          eid = {A86},
        pages = {A86},
          doi = {10.1051/0004-6361/201220616},
archivePrefix = {arXiv},
       eprint = {1304.5788},
 primaryClass = {astro-ph.CO},
       adsurl = {https://ui.adsabs.harvard.edu/abs/2013A&A...557A..86C}
}

@ARTICLE{Wen2016,
       author = {{Wen}, Zhang Zheng and {Zheng}, Xian Zhong},
        title = "{Merging Galaxies with Tidal Tails in COSMOS to z = 1}",
      journal = {\apj},
         year = 2016,
        month = nov,
       volume = {832},
       number = {1},
          eid = {90},
        pages = {90},
          doi = {10.3847/0004-637X/832/1/90},
archivePrefix = {arXiv},
       eprint = {1608.04298},
 primaryClass = {astro-ph.GA},
       adsurl = {https://ui.adsabs.harvard.edu/abs/2016ApJ...832...90W}
}

@ARTICLE{George2025,
       author = {{George}, K. and {Boselli}, A. and {Cuillandre}, J.-C. and {K{\"u}mmel}, M. and {Lan{\c{c}}on}, A. and {Bellhouse}, C. and {Saifollahi}, T. and {Mondelin}, M. and {Bolzonella}, M. and {Joseph}, P. and {Roberts}, I.~D. and {van Weeren}, R.~J. and {Liu}, Q. and {Sola}, E. and {Urbano}, M. and {Baes}, M. and {Peletier}, R.~F. and {Klein}, M. and {Davies}, C.~T. and {Zinchenko}, I.~A. and {Sorce}, J.~G. and {Poulain}, M. and {Aghanim}, N. and {Altieri}, B. and {Amara}, A. and {Andreon}, S. and {Auricchio}, N. and {Baccigalupi}, C. and {Baldi}, M. and {Balestra}, A. and {Bardelli}, S. and {Battaglia}, P. and {Biviano}, A. and {Bonino}, D. and {Branchini}, E. and {Brescia}, M. and {Brinchmann}, J. and {Camera}, S. and {Ca{\~n}as-Herrera}, G. and {Capobianco}, V. and {Carbone}, C. and {Carretero}, J. and {Casas}, S. and {Castellano}, M. and {Castignani}, G. and {Cavuoti}, S. and {Chambers}, K.~C. and {Cimatti}, A. and {Colodro-Conde}, C. and {Congedo}, G. and {Conselice}, C.~J. and {Conversi}, L. and {Copin}, Y. and {Courbin}, F. and {Courtois}, H.~M. and {Cropper}, M. and {Da Silva}, A. and {Degaudenzi}, H. and {De Lucia}, G. and {Di Giorgio}, A.~M. and {Dole}, H. and {Douspis}, M. and {Dubath}, F. and {Dupac}, X. and {Dusini}, S. and {Escoffier}, S. and {Farina}, M. and {Faustini}, F. and {Ferriol}, S. and {Fotopoulou}, S. and {Frailis}, M. and {Franceschi}, E. and {Galeotta}, S. and {Gillis}, B. and {Giocoli}, C. and {Gracia-Carpio}, J. and {Grazian}, A. and {Grupp}, F. and {Haugan}, S.~V.~H. and {Holmes}, W. and {Hook}, I.~M. and {Hormuth}, F. and {Hornstrup}, A. and {Hudelot}, P. and {Jahnke}, K. and {Jhabvala}, M. and {Keih{\"a}nen}, E. and {Kermiche}, S. and {Kiessling}, A. and {Kubik}, B. and {Kunz}, M. and {Kurki-Suonio}, H. and {Le Brun}, A.~M.~C. and {Le Mignant}, D. and {Ligori}, S. and {Lilje}, P.~B. and {Lindholm}, V. and {Lloro}, I. and {Mainetti}, G. and {Maino}, D. and {Maiorano}, E. and {Mansutti}, O. and {Marggraf}, O. and {Markovic}, K. and {Martinelli}, M. and {Martinet}, N. and {Marulli}, F. and {Massey}, R. and {Maurogordato}, S. and {Medinaceli}, E. and {Mei}, S. and {Mellier}, Y. and {Meneghetti}, M. and {Merlin}, E. and {Meylan}, G. and {Mohr}, J.~J. and {Mora}, A. and {Moresco}, M. and {Moscardini}, L. and {Nakajima}, R. and {Neissner}, C. and {Nichol}, R.~C. and {Niemi}, S.-M. and {Nightingale}, J.~W. and {Padilla}, C. and {Paltani}, S. and {Pasian}, F. and {Pedersen}, K. and {Percival}, W.~J. and {Pettorino}, V. and {Pires}, S. and {Polenta}, G. and {Poncet}, M. and {Popa}, L.~A. and {Pozzetti}, L. and {Raison}, F. and {Rebolo}, R. and {Renzi}, A. and {Rhodes}, J. and {Riccio}, G. and {Romelli}, E. and {Roncarelli}, M. and {Rossetti}, E. and {Saglia}, R. and {Sakr}, Z. and {Sapone}, D. and {Sartoris}, B. and {Schewtschenko}, J.~A. and {Schirmer}, M. and {Schneider}, P. and {Secroun}, A. and {Seidel}, G. and {Seiffert}, M. and {Serrano}, S. and {Sirignano}, C. and {Sirri}, G. and {Stanco}, L. and {Steinwagner}, J. and {Tallada-Cresp{\'\i}}, P. and {Taylor}, A.~N. and {Tereno}, I. and {Toft}, S. and {Toledo-Moreo}, R. and {Torradeflot}, F. and {Tutusaus}, I. and {Valentijn}, E.~A. and {Valenziano}, L. and {Valiviita}, J. and {Vassallo}, T. and {Verdoes Kleijn}, G. and {Veropalumbo}, A. and {Wang}, Y. and {Weller}, J. and {Zamorani}, G. and {Zerbi}, F.~M. and {Zucca}, E. and {Burigana}, C. and {Gabarra}, L. and {Mart{\'\i}n-Fleitas}, J. and {Scottez}, V.},
        title = "{Euclid: Early Release Observations of ram-pressure stripping in the Perseus cluster: Detection of parsec-scale star formation within the low surface brightness stripped tails of UGC 2665 and MCG +07-07-070}",
      journal = {\aap},
         year = 2025,
        month = sep,
       volume = {701},
          eid = {A40},
        pages = {A40},
          doi = {10.1051/0004-6361/202554836},
archivePrefix = {arXiv},
       eprint = {2505.23342},
 primaryClass = {astro-ph.GA},
       adsurl = {https://ui.adsabs.harvard.edu/abs/2025A&A...701A..40G}
}

@ARTICLE{Gullieuszik2020,
       author = {{Gullieuszik}, Marco and {Poggianti}, Bianca M. and {McGee}, Sean L. and {Moretti}, Alessia and {Vulcani}, Benedetta and {Tonnesen}, Stephanie and {Roediger}, Elke and {Jaff{\'e}}, Yara L. and {Fritz}, Jacopo and {Franchetto}, Andrea and {Omizzolo}, Alessandro and {Bettoni}, Daniela and {Radovich}, Mario and {Wolter}, Anna},
        title = "{GASP. XXI. Star Formation Rates in the Tails of Galaxies Undergoing Ram Pressure Stripping}",
      journal = {\apj},
         year = 2020,
        month = aug,
       volume = {899},
       number = {1},
          eid = {13},
        pages = {13},
          doi = {10.3847/1538-4357/aba3cb},
archivePrefix = {arXiv},
       eprint = {2006.16032},
 primaryClass = {astro-ph.GA},
       adsurl = {https://ui.adsabs.harvard.edu/abs/2020ApJ...899...13G}
}

@inproceedings{Bournaud2010,
  author    = {Bournaud, F.},
  title     = {Star Formation and Structure Formation in Galaxy Interactions and Mergers},
  booktitle = {Galaxy Wars: Stellar Populations and Star Formation in Interacting Galaxies},
  editor    = {Smith, B. and Higdon, J. and Higdon, S. and Bastian, N.},
  series    = {Astronomical Society of the Pacific Conference Series},
  volume    = {423},
  year      = {2010},
  month     = jun,
  pages     = {177--180},
  publisher = {Astronomical Society of the Pacific},
  doi       = {10.48550/arXiv.0909.1812}
}

@ARTICLE{Bertin1996,
       author = {{Bertin}, E. and {Arnouts}, S.},
        title = "{SExtractor: Software for source extraction.}",
      journal = {\aaps},
         year = 1996,
        month = jun,
       volume = {117},
        pages = {393-404},
          doi = {10.1051/aas:1996164},
       adsurl = {https://ui.adsabs.harvard.edu/abs/1996A&AS..117..393B}
}

@ARTICLE{Monteiro-Oliveira2025,
       author = {{Monteiro-Oliveira}, Rog{\'e}rio and {Lin}, Yen-Ting and {Chen}, Wei-Huai and {Chuang}, Chen-Yu and {Abdurro'uf} and {Wu}, Po-Feng},
        title = "{No Evidence of a Dichotomy in the Elliptical Galaxy Population}",
      journal = {\apj},
         year = 2025,
        month = jul,
       volume = {988},
       number = {1},
          eid = {138},
        pages = {138},
          doi = {10.3847/1538-4357/ade0ba},
archivePrefix = {arXiv},
       eprint = {2409.16349},
 primaryClass = {astro-ph.GA},
       adsurl = {https://ui.adsabs.harvard.edu/abs/2025ApJ...988..138M}
}

@ARTICLE{Jedrzejewski1987,
       author = {{Jedrzejewski}, Robert I.},
        title = "{CCD surface photometry of elliptical galaxies - I. Observations, reduction and results.}",
      journal = {\mnras},
         year = 1987,
        month = jun,
       volume = {226},
        pages = {747-768},
          doi = {10.1093/mnras/226.4.747},
       adsurl = {https://ui.adsabs.harvard.edu/abs/1987MNRAS.226..747J}
}

@ARTICLE{Monreal-Ibero2010,
       author = {{Monreal-Ibero}, A. and {Arribas}, S. and {Colina}, L. and {Rodr{\'\i}guez-Zaur{\'\i}n}, J. and {Alonso-Herrero}, A. and {Garc{\'\i}a-Mar{\'\i}n}, M.},
        title = "{VLT-VIMOS integral field spectroscopy of luminous and ultraluminous infrared galaxies. II. Evidence for shock ionization caused by tidal forces in the extra-nuclear regions of interacting and merging LIRGs}",
      journal = {\aap},
         year = 2010,
        month = jul,
       volume = {517},
          eid = {A28},
        pages = {A28},
          doi = {10.1051/0004-6361/200913239},
archivePrefix = {arXiv},
       eprint = {1004.3933},
 primaryClass = {astro-ph.CO},
       adsurl = {https://ui.adsabs.harvard.edu/abs/2010A&A...517A..28M}
}

@ARTICLE{Rich2011,
       author = {{Rich}, J.~A. and {Kewley}, L.~J. and {Dopita}, M.~A.},
        title = "{Galaxy-wide Shocks in Late-merger Stage Luminous Infrared Galaxies}",
      journal = {\apj},
         year = 2011,
        month = jun,
       volume = {734},
       number = {2},
          eid = {87},
        pages = {87},
          doi = {10.1088/0004-637X/734/2/87},
archivePrefix = {arXiv},
       eprint = {1104.1177},
 primaryClass = {astro-ph.CO},
       adsurl = {https://ui.adsabs.harvard.edu/abs/2011ApJ...734...87R}
}

@ARTICLE{Monreal-Ibero2006,
       author = {{Monreal-Ibero}, A. and {Arribas}, S. and {Colina}, L.},
        title = "{LINER-like Extended Nebulae in ULIRGs: Shocks Generated by Merger-Induced Flows}",
      journal = {\apj},
         year = 2006,
        month = jan,
       volume = {637},
       number = {1},
        pages = {138-146},
          doi = {10.1086/498257},
archivePrefix = {arXiv},
       eprint = {astro-ph/0509681},
 primaryClass = {astro-ph},
       adsurl = {https://ui.adsabs.harvard.edu/abs/2006ApJ...637..138M}
}

@ARTICLE{Law2021,
       author = {{Law}, David R. and {Ji}, Xihan and {Belfiore}, Francesco and {Bershady}, Matthew A. and {Cappellari}, Michele and {Westfall}, Kyle B. and {Yan}, Renbin and {Bizyaev}, Dmitry and {Brownstein}, Joel R. and {Drory}, Niv and {Andrews}, Brett H.},
        title = "{SDSS-IV MaNGA: Refining Strong Line Diagnostic Classifications Using Spatially Resolved Gas Dynamics}",
      journal = {\apj},
         year = 2021,
        month = jul,
       volume = {915},
       number = {1},
          eid = {35},
        pages = {35},
          doi = {10.3847/1538-4357/abfe0a},
archivePrefix = {arXiv},
       eprint = {2011.06012},
 primaryClass = {astro-ph.GA},
       adsurl = {https://ui.adsabs.harvard.edu/abs/2021ApJ...915...35L}
}

@ARTICLE{Barnes1996,
       author = {{Barnes}, Joshua E. and {Hernquist}, Lars},
        title = "{Transformations of Galaxies. II. Gasdynamics in Merging Disk Galaxies}",
      journal = {\apj},
         year = 1996,
        month = nov,
       volume = {471},
        pages = {115},
          doi = {10.1086/177957},
       adsurl = {https://ui.adsabs.harvard.edu/abs/1996ApJ...471..115B}
}

@ARTICLE{Cappellari2013,
       author = {{Cappellari}, Michele and {McDermid}, Richard M. and {Alatalo}, Katherine and {Blitz}, Leo and {Bois}, Maxime and {Bournaud}, Fr{\'e}d{\'e}ric and {Bureau}, M. and {Crocker}, Alison F. and {Davies}, Roger L. and {Davis}, Timothy A. and {de Zeeuw}, P.~T. and {Duc}, Pierre-Alain and {Emsellem}, Eric and {Khochfar}, Sadegh and {Krajnovi{\'c}}, Davor and {Kuntschner}, Harald and {Morganti}, Raffaella and {Naab}, Thorsten and {Oosterloo}, Tom and {Sarzi}, Marc and {Scott}, Nicholas and {Serra}, Paolo and {Weijmans}, Anne-Marie and {Young}, Lisa M.},
        title = "{The ATLAS$^{3D}$ project - XX. Mass-size and mass-{\ensuremath{\sigma}} distributions of early-type galaxies: bulge fraction drives kinematics, mass-to-light ratio, molecular gas fraction and stellar initial mass function}",
      journal = {\mnras},
         year = 2013,
        month = jul,
       volume = {432},
       number = {3},
        pages = {1862-1893},
          doi = {10.1093/mnras/stt644},
archivePrefix = {arXiv},
       eprint = {1208.3523},
 primaryClass = {astro-ph.CO},
       adsurl = {https://ui.adsabs.harvard.edu/abs/2013MNRAS.432.1862C}
}

@ARTICLE{Henriksen1996,
       author = {{Henriksen}, Mark and {Byrd}, Gene},
        title = "{Tidal Triggering of Star Formation by the Galaxy Cluster Potential}",
      journal = {\apj},
         year = 1996,
        month = mar,
       volume = {459},
        pages = {82},
          doi = {10.1086/176870},
       adsurl = {https://ui.adsabs.harvard.edu/abs/1996ApJ...459...82H}
}

@ARTICLE{Vollmer2005,
       author = {{Vollmer}, B. and {Braine}, J. and {Combes}, F. and {Sofue}, Y.},
        title = "{New CO observations and simulations of the NGC 4438/NGC 4435 system. Interaction diagnostics of the Virgo cluster galaxy NGC 4438}",
      journal = {\aap},
         year = 2005,
        month = oct,
       volume = {441},
       number = {2},
        pages = {473-489},
          doi = {10.1051/0004-6361:20041389},
       adsurl = {https://ui.adsabs.harvard.edu/abs/2005A&A...441..473V}
}

@ARTICLE{Baldwin1981,
       author = {{Baldwin}, J.~A. and {Phillips}, M.~M. and {Terlevich}, R.},
        title = "{Classification parameters for the emission-line spectra of extragalactic objects.}",
      journal = {\pasp},
         year = 1981,
        month = feb,
       volume = {93},
        pages = {5-19},
          doi = {10.1086/130766},
       adsurl = {https://ui.adsabs.harvard.edu/abs/1981PASP...93....5B}
}

@ARTICLE{Ferland2013,
       author = {{Ferland}, G.~J. and {Porter}, R.~L. and {van Hoof}, P.~A.~M. and {Williams}, R.~J.~R. and {Abel}, N.~P. and {Lykins}, M.~L. and {Shaw}, G. and {Henney}, W.~J. and {Stancil}, P.~C.},
        title = "{The 2013 Release of Cloudy}",
      journal = {\rmxaa},
         year = 2013,
        month = apr,
       volume = {49},
        pages = {137-163},
          doi = {10.48550/arXiv.1302.4485},
archivePrefix = {arXiv},
       eprint = {1302.4485},
 primaryClass = {astro-ph.GA},
       adsurl = {https://ui.adsabs.harvard.edu/abs/2013RMxAA..49..137F}
}

@ARTICLE{Bellhouse2017,
       author = {{Bellhouse}, C. and {Jaff{\'e}}, Y.~L. and {Hau}, G.~K.~T. and {McGee}, S.~L. and {Poggianti}, B.~M. and {Moretti}, A. and {Gullieuszik}, M. and {Bettoni}, D. and {Fasano}, G. and {D'Onofrio}, M. and {Fritz}, J. and {Omizzolo}, A. and {Sheen}, Y. -K. and {Vulcani}, B.},
        title = "{GASP. II. A MUSE View of Extreme Ram-Pressure Stripping along the Line of Sight: Kinematics of the Jellyfish Galaxy JO201}",
      journal = {\apj},
         year = 2017,
        month = jul,
       volume = {844},
       number = {1},
          eid = {49},
        pages = {49},
          doi = {10.3847/1538-4357/aa7875},
archivePrefix = {arXiv},
       eprint = {1704.05087},
 primaryClass = {astro-ph.GA},
       adsurl = {https://ui.adsabs.harvard.edu/abs/2017ApJ...844...49B}
}

@ARTICLE{Fritz2017,
       author = {{Fritz}, Jacopo and {Moretti}, Alessia and {Gullieuszik}, Marco and {Poggianti}, Bianca and {Bruzual}, Gustavo and {Vulcani}, Benedetta and {Nicastro}, Fabrizio and {Jaff{\'e}}, Yara and {Cervantes Sodi}, Bernardo and {Bettoni}, Daniela and {Biviano}, Andrea and {Fasano}, Giovanni and {Charlot}, St{\'e}phane and {Bellhouse}, Callum and {Hau}, George},
        title = "{GASP. III. JO36: A Case of Multiple Environmental Effects at Play?}",
      journal = {\apj},
         year = 2017,
        month = oct,
       volume = {848},
       number = {2},
          eid = {132},
        pages = {132},
          doi = {10.3847/1538-4357/aa8f51},
archivePrefix = {arXiv},
       eprint = {1704.05088},
 primaryClass = {astro-ph.GA},
       adsurl = {https://ui.adsabs.harvard.edu/abs/2017ApJ...848..132F}
}

@ARTICLE{Merluzzi2016,
       author = {{Merluzzi}, P. and {Busarello}, G. and {Dopita}, M.~A. and {Haines}, C.~P. and {Steinhauser}, D. and {Bourdin}, H. and {Mazzotta}, P.},
        title = "{Shapley Supercluster Survey: ram-pressure stripping versus tidal interactions in the Shapley supercluster}",
      journal = {\mnras},
         year = 2016,
        month = aug,
       volume = {460},
       number = {3},
        pages = {3345-3369},
          doi = {10.1093/mnras/stw1198},
archivePrefix = {arXiv},
       eprint = {1605.06329},
 primaryClass = {astro-ph.GA},
       adsurl = {https://ui.adsabs.harvard.edu/abs/2016MNRAS.460.3345M}
}

@ARTICLE{Poggianti2017b,
       author = {{Poggianti}, Bianca M. and {Jaff{\'e}}, Yara L. and {Moretti}, Alessia and {Gullieuszik}, Marco and {Radovich}, Mario and {Tonnesen}, Stephanie and {Fritz}, Jacopo and {Bettoni}, Daniela and {Vulcani}, Benedetta and {Fasano}, Giovanni and {Bellhouse}, Callum and {Hau}, George and {Omizzolo}, Alessandro},
        title = "{Ram-pressure feeding of supermassive black holes}",
      journal = {\nat},
         year = 2017,
        month = aug,
       volume = {548},
       number = {7667},
        pages = {304-309},
          doi = {10.1038/nature23462},
archivePrefix = {arXiv},
       eprint = {1708.09036},
 primaryClass = {astro-ph.GA},
       adsurl = {https://ui.adsabs.harvard.edu/abs/2017Natur.548..304P}
}

@ARTICLE{Gonzalez1999a,
       author = {{Gonz{\'a}lez Delgado}, Rosa M. and {Leitherer}, Claus},
        title = "{Synthetic Spectra of H Balmer and HE I Absorption Lines. I. Stellar Library}",
      journal = {\apjs},
         year = 1999,
        month = dec,
       volume = {125},
       number = {2},
        pages = {479-488},
          doi = {10.1086/313284},
archivePrefix = {arXiv},
       eprint = {astro-ph/9907115},
 primaryClass = {astro-ph},
       adsurl = {https://ui.adsabs.harvard.edu/abs/1999ApJS..125..479G}
}

@ARTICLE{Gonzalez1999b,
       author = {{Gonz{\'a}lez Delgado}, Rosa M. and {Leitherer}, Claus and {Heckman}, Timothy M.},
        title = "{Synthetic Spectra of H Balmer and HE I Absorption Lines. II. Evolutionary Synthesis Models for Starburst and Poststarburst Galaxies}",
      journal = {\apjs},
         year = 1999,
        month = dec,
       volume = {125},
       number = {2},
        pages = {489-509},
          doi = {10.1086/313285},
archivePrefix = {arXiv},
       eprint = {astro-ph/9907116},
 primaryClass = {astro-ph},
       adsurl = {https://ui.adsabs.harvard.edu/abs/1999ApJS..125..489G}
}

@ARTICLE{Diehl2006,
       author = {{Diehl}, Steven and {Statler}, Thomas S.},
        title = "{Adaptive binning of X-ray data with weighted Voronoi tessellations}",
      journal = {\mnras},
         year = 2006,
        month = may,
       volume = {368},
       number = {2},
        pages = {497-510},
          doi = {10.1111/j.1365-2966.2006.10125.x},
archivePrefix = {arXiv},
       eprint = {astro-ph/0512074},
 primaryClass = {astro-ph},
       adsurl = {https://ui.adsabs.harvard.edu/abs/2006MNRAS.368..497D}
}

@ARTICLE{Kennicutt1998,
       author = {{Kennicutt}, Robert C. Jr.},
        title = "{Star Formation in Galaxies Along the Hubble Sequence}",
      journal = {\araa},
         year = 1998,
        month = jan,
       volume = {36},
        pages = {189-232},
          doi = {10.1146/annurev.astro.36.1.189},
archivePrefix = {arXiv},
       eprint = {astro-ph/9807187},
 primaryClass = {astro-ph},
       adsurl = {https://ui.adsabs.harvard.edu/abs/1998ARA&A..36..189K}
}

@ARTICLE{Schulz2001,
       author = {{Schulz}, Steven and {Struck}, Curtis},
        title = "{Multi stage three-dimensional sweeping and annealing of disc galaxies in clusters}",
      journal = {\mnras},
         year = 2001,
        month = nov,
       volume = {328},
       number = {1},
        pages = {185-202},
          doi = {10.1046/j.1365-8711.2001.04847.x},
archivePrefix = {arXiv},
       eprint = {astro-ph/0107570},
 primaryClass = {astro-ph},
       adsurl = {https://ui.adsabs.harvard.edu/abs/2001MNRAS.328..185S}
}

@ARTICLE{Forbes1992,
       author = {{Forbes}, Duncan A. and {Thomson}, R.~C.},
        title = "{Shells and isophotal distortions in elliptical galaxies.}",
      journal = {\mnras},
         year = 1992,
        month = feb,
       volume = {254},
        pages = {723},
          doi = {10.1093/mnras/254.4.723},
       adsurl = {https://ui.adsabs.harvard.edu/abs/1992MNRAS.254..723F}
}

@ARTICLE{Ilina2016,
       author = {{Il'ina}, M.~A. and {Sil'chenko}, O.~K.},
        title = "{Analysis of the structure of disk galaxies in the NGC 2300 group}",
      journal = {Astronomy Reports},
         year = 2016,
        month = oct,
       volume = {60},
       number = {10},
        pages = {894-903},
          doi = {10.1134/S1063772916100036},
       adsurl = {https://ui.adsabs.harvard.edu/abs/2016ARep...60..894I}
}

@ARTICLE{Tomicic2018,
       author = {{Tomi{\v{c}}i{\'c}}, Neven and {Hughes}, Annie and {Kreckel}, Kathryn and {Renaud}, Florent and {Pety}, J{\'e}r{\^o}me and {Schinnerer}, Eva and {Saito}, Toshiki and {Querejeta}, Miguel and {Faesi}, Christopher M. and {Garcia-Burillo}, Santiago},
        title = "{Two Orders of Magnitude Variation in the Star Formation Efficiency across the Premerger Galaxy NGC 2276}",
      journal = {\apjl},
         year = 2018,
        month = dec,
       volume = {869},
       number = {2},
          eid = {L38},
        pages = {L38},
          doi = {10.3847/2041-8213/aaf810},
archivePrefix = {arXiv},
       eprint = {1812.05048},
 primaryClass = {astro-ph.GA},
       adsurl = {https://ui.adsabs.harvard.edu/abs/2018ApJ...869L..38T}
}

@ARTICLE{Wolter2015,
       author = {{Wolter}, Anna and {Esposito}, Paolo and {Mapelli}, Michela and {Pizzolato}, Fabio and {Ripamonti}, Emanuele},
        title = "{NGC 2276: a remarkable galaxy with a large number of ultraluminous X-ray sources}",
      journal = {\mnras},
         year = 2015,
        month = mar,
       volume = {448},
       number = {1},
        pages = {781-791},
          doi = {10.1093/mnras/stv054},
archivePrefix = {arXiv},
       eprint = {1501.01994},
 primaryClass = {astro-ph.HE},
       adsurl = {https://ui.adsabs.harvard.edu/abs/2015MNRAS.448..781W}
}

@ARTICLE{Chabrier2003,
       author = {{Chabrier}, Gilles},
        title = "{Galactic Stellar and Substellar Initial Mass Function}",
      journal = {\pasp},
         year = 2003,
        month = jul,
       volume = {115},
       number = {809},
        pages = {763-795},
          doi = {10.1086/376392},
archivePrefix = {arXiv},
       eprint = {astro-ph/0304382},
 primaryClass = {astro-ph},
       adsurl = {https://ui.adsabs.harvard.edu/abs/2003PASP..115..763C}
}

@ARTICLE{Vazdekis2010,
       author = {{Vazdekis}, A. and {S{\'a}nchez-Bl{\'a}zquez}, P. and {Falc{\'o}n-Barroso}, J. and {Cenarro}, A.~J. and {Beasley}, M.~A. and {Cardiel}, N. and {Gorgas}, J. and {Peletier}, R.~F.},
        title = "{Evolutionary stellar population synthesis with MILES - I. The base models and a new line index system}",
      journal = {\mnras},
         year = 2010,
        month = jun,
       volume = {404},
       number = {4},
        pages = {1639-1671},
          doi = {10.1111/j.1365-2966.2010.16407.x},
archivePrefix = {arXiv},
       eprint = {1004.4439},
 primaryClass = {astro-ph.CO},
       adsurl = {https://ui.adsabs.harvard.edu/abs/2010MNRAS.404.1639V}
}

@ARTICLE{Girardi2000,
       author = {{Girardi}, L. and {Bressan}, A. and {Bertelli}, G. and {Chiosi}, C.},
        title = "{Evolutionary tracks and isochrones for low- and intermediate-mass stars: From 0.15 to 7 M$_{sun}$, and from Z=0.0004 to 0.03}",
      journal = {\aaps},
         year = 2000,
        month = feb,
       volume = {141},
        pages = {371-383},
          doi = {10.1051/aas:2000126},
archivePrefix = {arXiv},
       eprint = {astro-ph/9910164},
 primaryClass = {astro-ph},
       adsurl = {https://ui.adsabs.harvard.edu/abs/2000A&AS..141..371G}
}

@ARTICLE{Guerou2017,
       author = {{Gu{\'e}rou}, Adrien and {Krajnovi{\'c}}, Davor and {Epinat}, Benoit and {Contini}, Thierry and {Emsellem}, Eric and {Bouch{\'e}}, Nicolas and {Bacon}, Roland and {Michel-Dansac}, Leo and {Richard}, Johan and {Weilbacher}, Peter M. and {Schaye}, Joop and {Marino}, Raffaella Anna and {den Brok}, Mark and {Erroz-Ferrer}, Santiago},
        title = "{The MUSE Hubble Ultra Deep Field Survey. V. Spatially resolved stellar kinematics of galaxies at redshift 0.2 {\ensuremath{\lesssim}} z {\ensuremath{\lesssim}} 0.8}",
      journal = {\aap},
         year = 2017,
        month = nov,
       volume = {608},
          eid = {A5},
        pages = {A5},
          doi = {10.1051/0004-6361/201730905},
archivePrefix = {arXiv},
       eprint = {1710.07694},
 primaryClass = {astro-ph.GA},
       adsurl = {https://ui.adsabs.harvard.edu/abs/2017A&A...608A...5G}
}

@ARTICLE{DellaBruna2020,
       author = {{Della Bruna}, Lorenza and {Adamo}, Angela and {Bik}, Arjan and {Fumagalli}, Michele and {Walterbos}, Rene and {{\"O}stlin}, G{\"o}ran and {Bruzual}, Gustavo and {Calzetti}, Daniela and {Charlot}, Stephane and {Grasha}, Kathryn and {Smith}, Linda J. and {Thilker}, David and {Wofford}, Aida},
        title = "{Studying the ISM at {\ensuremath{\sim}}10 pc scale in NGC 7793 with MUSE. I. Data description and properties of the ionised gas}",
      journal = {\aap},
         year = 2020,
        month = mar,
       volume = {635},
          eid = {A134},
        pages = {A134},
          doi = {10.1051/0004-6361/201937173},
archivePrefix = {arXiv},
       eprint = {2002.08966},
 primaryClass = {astro-ph.GA},
       adsurl = {https://ui.adsabs.harvard.edu/abs/2020A&A...635A.134D}
}

@ARTICLE{ifscube2021,
       author = {{Ruschel-Dutra}, D. and {Storchi-Bergmann}, T. and {Schnorr-M{\"u}ller}, A. and {Riffel}, R.~A. and {Dall'Agnol de Oliveira}, B. and {Lena}, D. and {Robinson}, A. and {Nagar}, N. and {Elvis}, M.},
        title = "{AGNIFS survey of local AGN: GMOS-IFU data and outflows in 30 sources}",
      journal = {\mnras},
         year = 2021,
        month = oct,
       volume = {507},
       number = {1},
        pages = {74-89},
          doi = {10.1093/mnras/stab2058},
archivePrefix = {arXiv},
       eprint = {2107.07635},
 primaryClass = {astro-ph.GA},
       adsurl = {https://ui.adsabs.harvard.edu/abs/2021MNRAS.507...74R}
}

@ARTICLE{Cappellari2004,
       author = {{Cappellari}, Michele and {Emsellem}, Eric},
        title = "{Parametric Recovery of Line-of-Sight Velocity Distributions from Absorption-Line Spectra of Galaxies via Penalized Likelihood}",
      journal = {\pasp},
         year = 2004,
        month = feb,
       volume = {116},
       number = {816},
        pages = {138-147},
          doi = {10.1086/381875},
archivePrefix = {arXiv},
       eprint = {astro-ph/0312201},
 primaryClass = {astro-ph},
       adsurl = {https://ui.adsabs.harvard.edu/abs/2004PASP..116..138C}
}

@ARTICLE{Cappellari2017,
       author = {{Cappellari}, Michele},
        title = "{Improving the full spectrum fitting method: accurate convolution with Gauss-Hermite functions}",
      journal = {\mnras},
         year = 2017,
        month = apr,
       volume = {466},
       number = {1},
        pages = {798-811},
          doi = {10.1093/mnras/stw3020},
archivePrefix = {arXiv},
       eprint = {1607.08538},
 primaryClass = {astro-ph.GA},
       adsurl = {https://ui.adsabs.harvard.edu/abs/2017MNRAS.466..798C}
}

@ARTICLE{Cappellari2003,
       author = {{Cappellari}, Michele and {Copin}, Yannick},
        title = "{Adaptive spatial binning of integral-field spectroscopic data using Voronoi tessellations}",
      journal = {\mnras},
         year = 2003,
        month = jun,
       volume = {342},
       number = {2},
        pages = {345-354},
          doi = {10.1046/j.1365-8711.2003.06541.x},
archivePrefix = {arXiv},
       eprint = {astro-ph/0302262},
 primaryClass = {astro-ph},
       adsurl = {https://ui.adsabs.harvard.edu/abs/2003MNRAS.342..345C}
}

@ARTICLE{Fritz2011,
       author = {{Fritz}, J. and {Poggianti}, B.~M. and {Cava}, A. and {Valentinuzzi}, T. and {Moretti}, A. and {Bettoni}, D. and {Bressan}, A. and {Couch}, W.~J. and {D'Onofrio}, M. and {Dressler}, A. and {Fasano}, G. and {Kj{\ae}rgaard}, P. and {Moles}, M. and {Omizzolo}, A. and {Varela}, J.},
        title = "{WINGS-SPE II: A catalog of stellar ages and star formation histories, stellar masses and dust extinction values for local clusters galaxies}",
      journal = {\aap},
         year = 2011,
        month = feb,
       volume = {526},
          eid = {A45},
        pages = {A45},
          doi = {10.1051/0004-6361/201015214},
archivePrefix = {arXiv},
       eprint = {1010.2214},
 primaryClass = {astro-ph.CO},
       adsurl = {https://ui.adsabs.harvard.edu/abs/2011A&A...526A..45F}
}

@ARTICLE{Fritz2014,
       author = {{Fritz}, J. and {Poggianti}, B.~M. and {Cava}, A. and {Moretti}, A. and {Varela}, J. and {Bettoni}, D. and {Couch}, W.~J. and {D'Onofrio D'Onofrio}, M. and {Dressler}, A. and {Fasano}, G. and {Kj{\ae}rgaard}, P. and {Marziani}, P. and {Moles}, M. and {Omizzolo}, A.},
        title = "{WINGS-SPE. III. Equivalent width measurements, spectral properties, and evolution of local cluster galaxies}",
      journal = {\aap},
         year = 2014,
        month = jun,
       volume = {566},
          eid = {A32},
        pages = {A32},
          doi = {10.1051/0004-6361/201323138},
archivePrefix = {arXiv},
       eprint = {1402.4131},
 primaryClass = {astro-ph.GA},
       adsurl = {https://ui.adsabs.harvard.edu/abs/2014A&A...566A..32F}
}

@ARTICLE{Gullieuszik2015,
       author = {{Gullieuszik}, M. and {Poggianti}, B. and {Fasano}, G. and {Zaggia}, S. and {Paccagnella}, A. and {Moretti}, A. and {Bettoni}, D. and {D'Onofrio}, M. and {Couch}, W.~J. and {Vulcani}, B. and {Fritz}, J. and {Omizzolo}, A. and {Baruffolo}, A. and {Schipani}, P. and {Capaccioli}, M. and {Varela}, J.},
        title = "{OmegaWINGS: OmegaCAM-VST observations of WINGS galaxy clusters}",
      journal = {\aap},
         year = 2015,
        month = sep,
       volume = {581},
          eid = {A41},
        pages = {A41},
          doi = {10.1051/0004-6361/201526061},
archivePrefix = {arXiv},
       eprint = {1503.02628},
 primaryClass = {astro-ph.GA},
       adsurl = {https://ui.adsabs.harvard.edu/abs/2015A&A...581A..41G}
}

@ARTICLE{Schlafly2011,
       author = {{Schlafly}, Edward F. and {Finkbeiner}, Douglas P.},
        title = "{Measuring Reddening with Sloan Digital Sky Survey Stellar Spectra and Recalibrating SFD}",
      journal = {\apj},
         year = 2011,
        month = aug,
       volume = {737},
       number = {2},
          eid = {103},
        pages = {103},
          doi = {10.1088/0004-637X/737/2/103},
archivePrefix = {arXiv},
       eprint = {1012.4804},
 primaryClass = {astro-ph.GA},
       adsurl = {https://ui.adsabs.harvard.edu/abs/2011ApJ...737..103S}
}

@ARTICLE{Cardelli1989,
       author = {{Cardelli}, Jason A. and {Clayton}, Geoffrey C. and {Mathis}, John S.},
        title = "{The Relationship between Infrared, Optical, and Ultraviolet Extinction}",
      journal = {\apj},
         year = 1989,
        month = oct,
       volume = {345},
        pages = {245},
          doi = {10.1086/167900},
       adsurl = {https://ui.adsabs.harvard.edu/abs/1989ApJ...345..245C}
}

@ARTICLE{Storey2000,
       author = {{Storey}, P.~J. and {Zeippen}, C.~J.},
        title = "{Theoretical values for the [OIII] 5007/4959 line-intensity ratio and homologous cases}",
      journal = {\mnras},
         year = 2000,
        month = mar,
       volume = {312},
       number = {4},
        pages = {813-816},
          doi = {10.1046/j.1365-8711.2000.03184.x},
       adsurl = {https://ui.adsabs.harvard.edu/abs/2000MNRAS.312..813S}
}

@BOOK{Osterbrock2006,
       author = {{Osterbrock}, Donald E. and {Ferland}, Gary J.},
        title = "{Astrophysics of gaseous nebulae and active galactic nuclei}",
         year = 2006,
       publisher = "{University Science Books}",
       adsurl = {https://ui.adsabs.harvard.edu/abs/2006agna.book.....O}
}

@ARTICLE{Lenz1992,
       author = {{Lenz}, Dawn D. and {Ayres}, Thomas R.},
        title = "{Errors Associated with Fitting Gaussian Profiles to Noisy Emission-Line Spectra}",
      journal = {\pasp},
         year = 1992,
        month = nov,
       volume = {104},
        pages = {1104},
          doi = {10.1086/133096},
       adsurl = {https://ui.adsabs.harvard.edu/abs/1992PASP..104.1104L}
}

@ARTICLE{Wesson2016,
       author = {{Wesson}, R.},
        title = "{ALFA: an automated line fitting algorithm}",
      journal = {\mnras},
         year = 2016,
        month = mar,
       volume = {456},
       number = {4},
        pages = {3774-3781},
          doi = {10.1093/mnras/stv2946},
archivePrefix = {arXiv},
       eprint = {1512.04539},
 primaryClass = {astro-ph.SR},
       adsurl = {https://ui.adsabs.harvard.edu/abs/2016MNRAS.456.3774W}
}

@ARTICLE{Lopez-Sanchez2015,
       author = {{L{\'o}pez-S{\'a}nchez}, {\'A}. R. and {Westmeier}, T. and {Esteban}, C. and {Koribalski}, B.~S.},
        title = "{Ionized gas in the XUV disc of the NGC 1512/1510 system}",
      journal = {\mnras},
         year = 2015,
        month = jul,
       volume = {450},
       number = {4},
        pages = {3381-3409},
          doi = {10.1093/mnras/stv703},
       adsurl = {https://ui.adsabs.harvard.edu/abs/2015MNRAS.450.3381L}
}

@ARTICLE{Rola1994,
       author = {{Rola}, C. and {Pelat}, D.},
        title = "{On the estimation of intensity for low S/N ratio narrow emission lines.}",
      journal = {\aap},
         year = 1994,
        month = jul,
       volume = {287},
        pages = {676-684},
       adsurl = {https://ui.adsabs.harvard.edu/abs/1994A&A...287..676R}
}

@ARTICLE{Gaia2023,
       author = {{Gaia Collaboration} and {Vallenari}, A. and {Brown}, A.~G.~A. and {Prusti}, T. and {de Bruijne}, J.~H.~J. and {Arenou}, F. and {Babusiaux}, C. and {Biermann}, M. and {Creevey}, O.~L. and {Ducourant}, C. and {Evans}, D.~W. and {Eyer}, L. and {Guerra}, R. and {Hutton}, A. and {Jordi}, C. and {Klioner}, S.~A. and {Lammers}, U.~L. and {Lindegren}, L. and {Luri}, X. and {Mignard}, F. and {Panem}, C. and {Pourbaix}, D. and {Randich}, S. and {Sartoretti}, P. and {Soubiran}, C. and {Tanga}, P. and {Walton}, N.~A. and {Bailer-Jones}, C.~A.~L. and {Bastian}, U. and {Drimmel}, R. and {Jansen}, F. and {Katz}, D. and {Lattanzi}, M.~G. and {van Leeuwen}, F. and {Bakker}, J. and {Cacciari}, C. and {Casta{\~n}eda}, J. and {De Angeli}, F. and {Fabricius}, C. and {Fouesneau}, M. and {Fr{\'e}mat}, Y. and {Galluccio}, L. and {Guerrier}, A. and {Heiter}, U. and {Masana}, E. and {Messineo}, R. and {Mowlavi}, N. and {Nicolas}, C. and {Nienartowicz}, K. and {Pailler}, F. and {Panuzzo}, P. and {Riclet}, F. and {Roux}, W. and {Seabroke}, G.~M. and {Sordo}, R. and {Th{\'e}venin}, F. and {Gracia-Abril}, G. and {Portell}, J. and {Teyssier}, D. and {Altmann}, M. and {Andrae}, R. and {Audard}, M. and {Bellas-Velidis}, I. and {Benson}, K. and {Berthier}, J. and {Blomme}, R. and {Burgess}, P.~W. and {Busonero}, D. and {Busso}, G. and {C{\'a}novas}, H. and {Carry}, B. and {Cellino}, A. and {Cheek}, N. and {Clementini}, G. and {Damerdji}, Y. and {Davidson}, M. and {de Teodoro}, P. and {Nu{\~n}ez Campos}, M. and {Delchambre}, L. and {Dell'Oro}, A. and {Esquej}, P. and {Fern{\'a}ndez-Hern{\'a}ndez}, J. and {Fraile}, E. and {Garabato}, D. and {Garc{\'\i}a-Lario}, P. and {Gosset}, E. and {Haigron}, R. and {Halbwachs}, J. -L. and {Hambly}, N.~C. and {Harrison}, D.~L. and {Hern{\'a}ndez}, J. and {Hestroffer}, D. and {Hodgkin}, S.~T. and {Holl}, B. and {Jan{\ss}en}, K. and {Jevardat de Fombelle}, G. and {Jordan}, S. and {Krone-Martins}, A. and {Lanzafame}, A.~C. and {L{\"o}ffler}, W. and {Marchal}, O. and {Marrese}, P.~M. and {Moitinho}, A. and {Muinonen}, K. and {Osborne}, P. and {Pancino}, E. and {Pauwels}, T. and {Recio-Blanco}, A. and {Reyl{\'e}}, C. and {Riello}, M. and {Rimoldini}, L. and {Roegiers}, T. and {Rybizki}, J. and {Sarro}, L.~M. and {Siopis}, C. and {Smith}, M. and {Sozzetti}, A. and {Utrilla}, E. and {van Leeuwen}, M. and {Abbas}, U. and {{\'A}brah{\'a}m}, P. and {Abreu Aramburu}, A. and {Aerts}, C. and {Aguado}, J.~J. and {Ajaj}, M. and {Aldea-Montero}, F. and {Altavilla}, G. and {{\'A}lvarez}, M.~A. and {Alves}, J. and {Anders}, F. and {Anderson}, R.~I. and {Anglada Varela}, E. and {Antoja}, T. and {Baines}, D. and {Baker}, S.~G. and {Balaguer-N{\'u}{\~n}ez}, L. and {Balbinot}, E. and {Balog}, Z. and {Barache}, C. and {Barbato}, D. and {Barros}, M. and {Barstow}, M.~A. and {Bartolom{\'e}}, S. and {Bassilana}, J. -L. and {Bauchet}, N. and {Becciani}, U. and {Bellazzini}, M. and {Berihuete}, A. and {Bernet}, M. and {Bertone}, S. and {Bianchi}, L. and {Binnenfeld}, A. and {Blanco-Cuaresma}, S. and {Blazere}, A. and {Boch}, T. and {Bombrun}, A. and {Bossini}, D. and {Bouquillon}, S. and {Bragaglia}, A. and {Bramante}, L. and {Breedt}, E. and {Bressan}, A. and {Brouillet}, N. and {Brugaletta}, E. and {Bucciarelli}, B. and {Burlacu}, A. and {Butkevich}, A.~G. and {Buzzi}, R. and {Caffau}, E. and {Cancelliere}, R. and {Cantat-Gaudin}, T. and {Carballo}, R. and {Carlucci}, T. and {Carnerero}, M.~I. and {Carrasco}, J.~M. and {Casamiquela}, L. and {Castellani}, M. and {Castro-Ginard}, A. and {Chaoul}, L. and {Charlot}, P. and {Chemin}, L. and {Chiaramida}, V. and {Chiavassa}, A. and {Chornay}, N. and {Comoretto}, G. and {Contursi}, G. and {Cooper}, W.~J. and {Cornez}, T. and {Cowell}, S. and {Crifo}, F. and {Cropper}, M. and {Crosta}, M. and {Crowley}, C. and {Dafonte}, C. and {Dapergolas}, A. and {David}, M. and {David}, P. and {de Laverny}, P. and {De Luise}, F. and {De March}, R.},
        title = "{Gaia Data Release 3. Summary of the content and survey properties}",
      journal = {\aap},
         year = 2023,
        month = jun,
       volume = {674},
          eid = {A1},
        pages = {A1},
          doi = {10.1051/0004-6361/202243940},
archivePrefix = {arXiv},
       eprint = {2208.00211},
 primaryClass = {astro-ph.GA},
       adsurl = {https://ui.adsabs.harvard.edu/abs/2023A&A...674A...1G}
}

@ARTICLE{Chambers2016,
       author = {{Chambers}, K.~C. and {Magnier}, E.~A. and {Metcalfe}, N. and {Flewelling}, H.~A. and {Huber}, M.~E. and {Waters}, C.~Z. and {Denneau}, L. and {Draper}, P.~W. and {Farrow}, D. and {Finkbeiner}, D.~P. and {Holmberg}, C. and {Koppenhoefer}, J. and {Price}, P.~A. and {Rest}, A. and {Saglia}, R.~P. and {Schlafly}, E.~F. and {Smartt}, S.~J. and {Sweeney}, W. and {Wainscoat}, R.~J. and {Burgett}, W.~S. and {Chastel}, S. and {Grav}, T. and {Heasley}, J.~N. and {Hodapp}, K.~W. and {Jedicke}, R. and {Kaiser}, N. and {Kudritzki}, R. -P. and {Luppino}, G.~A. and {Lupton}, R.~H. and {Monet}, D.~G. and {Morgan}, J.~S. and {Onaka}, P.~M. and {Shiao}, B. and {Stubbs}, C.~W. and {Tonry}, J.~L. and {White}, R. and {Ba{\~n}ados}, E. and {Bell}, E.~F. and {Bender}, R. and {Bernard}, E.~J. and {Boegner}, M. and {Boffi}, F. and {Botticella}, M.~T. and {Calamida}, A. and {Casertano}, S. and {Chen}, W. -P. and {Chen}, X. and {Cole}, S. and {Deacon}, N. and {Frenk}, C. and {Fitzsimmons}, A. and {Gezari}, S. and {Gibbs}, V. and {Goessl}, C. and {Goggia}, T. and {Gourgue}, R. and {Goldman}, B. and {Grant}, P. and {Grebel}, E.~K. and {Hambly}, N.~C. and {Hasinger}, G. and {Heavens}, A.~F. and {Heckman}, T.~M. and {Henderson}, R. and {Henning}, T. and {Holman}, M. and {Hopp}, U. and {Ip}, W. -H. and {Isani}, S. and {Jackson}, M. and {Keyes}, C.~D. and {Koekemoer}, A.~M. and {Kotak}, R. and {Le}, D. and {Liska}, D. and {Long}, K.~S. and {Lucey}, J.~R. and {Liu}, M. and {Martin}, N.~F. and {Masci}, G. and {McLean}, B. and {Mindel}, E. and {Misra}, P. and {Morganson}, E. and {Murphy}, D.~N.~A. and {Obaika}, A. and {Narayan}, G. and {Nieto-Santisteban}, M.~A. and {Norberg}, P. and {Peacock}, J.~A. and {Pier}, E.~A. and {Postman}, M. and {Primak}, N. and {Rae}, C. and {Rai}, A. and {Riess}, A. and {Riffeser}, A. and {Rix}, H.~W. and {R{\"o}ser}, S. and {Russel}, R. and {Rutz}, L. and {Schilbach}, E. and {Schultz}, A.~S.~B. and {Scolnic}, D. and {Strolger}, L. and {Szalay}, A. and {Seitz}, S. and {Small}, E. and {Smith}, K.~W. and {Soderblom}, D.~R. and {Taylor}, P. and {Thomson}, R. and {Taylor}, A.~N. and {Thakar}, A.~R. and {Thiel}, J. and {Thilker}, D. and {Unger}, D. and {Urata}, Y. and {Valenti}, J. and {Wagner}, J. and {Walder}, T. and {Walter}, F. and {Watters}, S.~P. and {Werner}, S. and {Wood-Vasey}, W.~M. and {Wyse}, R.},
        title = "{The Pan-STARRS1 Surveys}",
      journal = {arXiv e-prints},
         year = 2016,
        month = dec,
          eid = {\verb|arXiv|: 1612.05560},
        pages = {\verb|arXiv|: 1612.05560},
          doi = {10.48550/arXiv.1612.05560},
archivePrefix = {arXiv},
       eprint = {1612.05560},
 primaryClass = {astro-ph.IM},
       adsurl = {https://ui.adsabs.harvard.edu/abs/2016arXiv161205560C}
}

@ARTICLE{Tumlinson2017,
       author = {{Tumlinson}, Jason and {Peeples}, Molly S. and {Werk}, Jessica K.},
        title = "{The Circumgalactic Medium}",
      journal = {\araa},
         year = 2017,
        month = aug,
       volume = {55},
       number = {1},
        pages = {389-432},
          doi = {10.1146/annurev-astro-091916-055240},
archivePrefix = {arXiv},
       eprint = {1709.09180},
 primaryClass = {astro-ph.GA},
       adsurl = {https://ui.adsabs.harvard.edu/abs/2017ARA&A..55..389T}
}

@ARTICLE{Abadi1999,
       author = {{Abadi}, Mario G. and {Moore}, Ben and {Bower}, Richard G.},
        title = "{Ram pressure stripping of spiral galaxies in clusters}",
      journal = {\mnras},
         year = 1999,
        month = oct,
       volume = {308},
       number = {4},
        pages = {947-954},
          doi = {10.1046/j.1365-8711.1999.02715.x},
archivePrefix = {arXiv},
       eprint = {astro-ph/9903436},
 primaryClass = {astro-ph},
       adsurl = {https://ui.adsabs.harvard.edu/abs/1999MNRAS.308..947A}
}

@ARTICLE{Moore1996,
       author = {{Moore}, Ben and {Katz}, Neal and {Lake}, George and {Dressler}, Alan and {Oemler}, Augustus},
        title = "{Galaxy harassment and the evolution of clusters of galaxies}",
      journal = {\nat},
         year = 1996,
        month = feb,
       volume = {379},
       number = {6566},
        pages = {613-616},
          doi = {10.1038/379613a0},
archivePrefix = {arXiv},
       eprint = {astro-ph/9510034},
 primaryClass = {astro-ph},
       adsurl = {https://ui.adsabs.harvard.edu/abs/1996Natur.379..613M}
}

@ARTICLE{Moore1998,
       author = {{Moore}, Ben and {Lake}, George and {Katz}, Neal},
        title = "{Morphological Transformation from Galaxy Harassment}",
      journal = {\apj},
         year = 1998,
        month = mar,
       volume = {495},
       number = {1},
        pages = {139-151},
          doi = {10.1086/305264},
archivePrefix = {arXiv},
       eprint = {astro-ph/9701211},
 primaryClass = {astro-ph},
       adsurl = {https://ui.adsabs.harvard.edu/abs/1998ApJ...495..139M}
}

@ARTICLE{Boselli2006,
       author = {{Boselli}, Alessandro and {Gavazzi}, Giuseppe},
        title = "{Environmental Effects on Late-Type Galaxies in Nearby Clusters}",
      journal = {\pasp},
         year = 2006,
        month = apr,
       volume = {118},
       number = {842},
        pages = {517-559},
          doi = {10.1086/500691},
archivePrefix = {arXiv},
       eprint = {astro-ph/0601108},
 primaryClass = {astro-ph},
       adsurl = {https://ui.adsabs.harvard.edu/abs/2006PASP..118..517B}
}

@ARTICLE{Adams2012,
       author = {{Adams}, Scott M. and {Zaritsky}, Dennis and {Sand}, David J. and {Graham}, Melissa L. and {Bildfell}, Chris and {Hoekstra}, Henk and {Pritchet}, Chris},
        title = "{The Environmental Dependence of the Incidence of Galactic Tidal Features}",
      journal = {\aj},
         year = 2012,
        month = nov,
       volume = {144},
       number = {5},
          eid = {128},
        pages = {128},
          doi = {10.1088/0004-6256/144/5/128},
archivePrefix = {arXiv},
       eprint = {1208.4843},
 primaryClass = {astro-ph.CO},
       adsurl = {https://ui.adsabs.harvard.edu/abs/2012AJ....144..128A}
}

@ARTICLE{Boselli2022,
       author = {{Boselli}, Alessandro and {Fossati}, Matteo and {Sun}, Ming},
        title = "{Ram pressure stripping in high-density environments}",
      journal = {\aapr},
         year = 2022,
        month = dec,
       volume = {30},
       number = {1},
          eid = {3},
        pages = {3},
          doi = {10.1007/s00159-022-00140-3},
archivePrefix = {arXiv},
       eprint = {2109.13614},
 primaryClass = {astro-ph.GA},
       adsurl = {https://ui.adsabs.harvard.edu/abs/2022A&ARv..30....3B}
}

@ARTICLE{Kenney1989,
       author = {{Kenney}, Jeffrey D.~P. and {Young}, Judith S.},
        title = "{The Effects of Environment on the Molecular and Atomic Gas Properties of Large Virgo Cluster Spirals}",
      journal = {\apj},
         year = 1989,
        month = sep,
       volume = {344},
        pages = {171},
          doi = {10.1086/167787},
       adsurl = {https://ui.adsabs.harvard.edu/abs/1989ApJ...344..171K}
}

@ARTICLE{Gunn1972,
       author = {{Gunn}, James E. and {Gott}, III, J. Richard},
        title = "{On the Infall of Matter Into Clusters of Galaxies and Some Effects on Their Evolution}",
      journal = {\apj},
         year = 1972,
        month = aug,
       volume = {176},
        pages = {1},
          doi = {10.1086/151605},
       adsurl = {https://ui.adsabs.harvard.edu/abs/1972ApJ...176....1G}
}

@ARTICLE{Dressler1980,
       author = {{Dressler}, A.},
        title = "{Galaxy morphology in rich clusters: implications for the formation and evolution of galaxies.}",
      journal = {\apj},
         year = 1980,
        month = mar,
       volume = {236},
        pages = {351-365},
          doi = {10.1086/157753},
       adsurl = {https://ui.adsabs.harvard.edu/abs/1980ApJ...236..351D}
}

@ARTICLE{Akhlaghi2015,
    author = {{Akhlaghi}, Mohammad and {Ichikawa}, Takashi},
    title = "{Noise-based Detection and Segmentation of Nebulous Objects}",
    journal = {ApJS},
    archivePrefix = "arXiv",
    eprint = {1505.01664},
    primaryClass = "astro-ph.IM",
    year = 2015,
    month = sep,
    volume = 220,
    eid = {1},
    pages = {1},
    doi = {10.1088/0067-0049/220/1/1},
    adsurl = {https://ui.adsabs.harvard.edu/abs/2015ApJS..220....1A}
  }

@ARTICLE{Abbott2021,
       author = {{Abbott}, T.~M.~C. and {Adam{\'o}w}, M. and {Aguena}, M. and {Allam}, S. and {Amon}, A. and {Annis}, J. and {Avila}, S. and {Bacon}, D. and {Banerji}, M. and {Bechtol}, K. and {Becker}, M.~R. and {Bernstein}, G.~M. and {Bertin}, E. and {Bhargava}, S. and {Bridle}, S.~L. and {Brooks}, D. and {Burke}, D.~L. and {Carnero Rosell}, A. and {Carrasco Kind}, M. and {Carretero}, J. and {Castander}, F.~J. and {Cawthon}, R. and {Chang}, C. and {Choi}, A. and {Conselice}, C. and {Costanzi}, M. and {Crocce}, M. and {da Costa}, L.~N. and {Davis}, T.~M. and {De Vicente}, J. and {DeRose}, J. and {Desai}, S. and {Diehl}, H.~T. and {Dietrich}, J.~P. and {Drlica-Wagner}, A. and {Eckert}, K. and {Elvin-Poole}, J. and {Everett}, S. and {Evrard}, A.~E. and {Ferrero}, I. and {Fert{\'e}}, A. and {Flaugher}, B. and {Fosalba}, P. and {Friedel}, D. and {Frieman}, J. and {Garc{\'\i}a-Bellido}, J. and {Gaztanaga}, E. and {Gelman}, L. and {Gerdes}, D.~W. and {Giannantonio}, T. and {Gill}, M.~S.~S. and {Gruen}, D. and {Gruendl}, R.~A. and {Gschwend}, J. and {Gutierrez}, G. and {Hartley}, W.~G. and {Hinton}, S.~R. and {Hollowood}, D.~L. and {Honscheid}, K. and {Huterer}, D. and {James}, D.~J. and {Jeltema}, T. and {Johnson}, M.~D. and {Kent}, S. and {Kron}, R. and {Kuehn}, K. and {Kuropatkin}, N. and {Lahav}, O. and {Li}, T.~S. and {Lidman}, C. and {Lin}, H. and {MacCrann}, N. and {Maia}, M.~A.~G. and {Manning}, T.~A. and {Maloney}, J.~D. and {March}, M. and {Marshall}, J.~L. and {Martini}, P. and {Melchior}, P. and {Menanteau}, F. and {Miquel}, R. and {Morgan}, R. and {Myles}, J. and {Neilsen}, E. and {Ogando}, R.~L.~C. and {Palmese}, A. and {Paz-Chinch{\'o}n}, F. and {Petravick}, D. and {Pieres}, A. and {Plazas}, A.~A. and {Pond}, C. and {Rodriguez-Monroy}, M. and {Romer}, A.~K. and {Roodman}, A. and {Rykoff}, E.~S. and {Sako}, M. and {Sanchez}, E. and {Santiago}, B. and {Scarpine}, V. and {Serrano}, S. and {Sevilla-Noarbe}, I. and {Smith}, J. Allyn and {Smith}, M. and {Soares-Santos}, M. and {Suchyta}, E. and {Swanson}, M.~E.~C. and {Tarle}, G. and {Thomas}, D. and {To}, C. and {Tremblay}, P.~E. and {Troxel}, M.~A. and {Tucker}, D.~L. and {Turner}, D.~J. and {Varga}, T.~N. and {Walker}, A.~R. and {Wechsler}, R.~H. and {Weller}, J. and {Wester}, W. and {Wilkinson}, R.~D. and {Yanny}, B. and {Zhang}, Y. and {Nikutta}, R. and {Fitzpatrick}, M. and {Jacques}, A. and {Scott}, A. and {Olsen}, K. and {Huang}, L. and {Herrera}, D. and {Juneau}, S. and {Nidever}, D. and {Weaver}, B.~A. and {Adean}, C. and {Correia}, V. and {de Freitas}, M. and {Freitas}, F.~N. and {Singulani}, C. and {Vila-Verde}, G. and {Linea Science Server}},
        title = "{The Dark Energy Survey Data Release 2}",
      journal = {\apjs},
         year = 2021,
        month = aug,
       volume = {255},
       number = {2},
          eid = {20},
        pages = {20},
          doi = {10.3847/1538-4365/ac00b3},
archivePrefix = {arXiv},
       eprint = {2101.05765},
 primaryClass = {astro-ph.IM},
       adsurl = {https://ui.adsabs.harvard.edu/abs/2021ApJS..255...20A}
}

\begin{appendix}

\section{A background source superposed on UG103}
\label{appendix:back_source}
In Fig.~\ref{fig:backsource}, we assess the nature of an elongated source projected onto the spiral arms of UG103, whose morphology could otherwise be mistaken for star-forming regions associated with the galaxy. The analysis demonstrates that this emission instead originates from a background source at $z \sim 0.25$.
 
The white dashed ellipse marks the visually defined region used to extract the integrated MUSE spectrum shown in the bottom panel. Two distinct emission-line complexes are detected: one H$\alpha$ + [\ion{N}{ii}] complex is at $\lambda_{\rm obs} \sim 6910 \,$\AA\,consistent with the systemic redshift of UG103 (Sect.~\ref{sec:obs_justification}), and a second complex at $\lambda_{\rm obs} \sim 8175 \,$\AA, corresponding to $z \sim 0.25$. The latter emission line features are spatially confined to the elongated yellowish source and to a second object near the edge of the FoV, visible only in the right-hand top panel of Fig.~\ref{fig:backsource}.

\begin{figure}[!h]
    \centering
    \includegraphics[width=\linewidth]{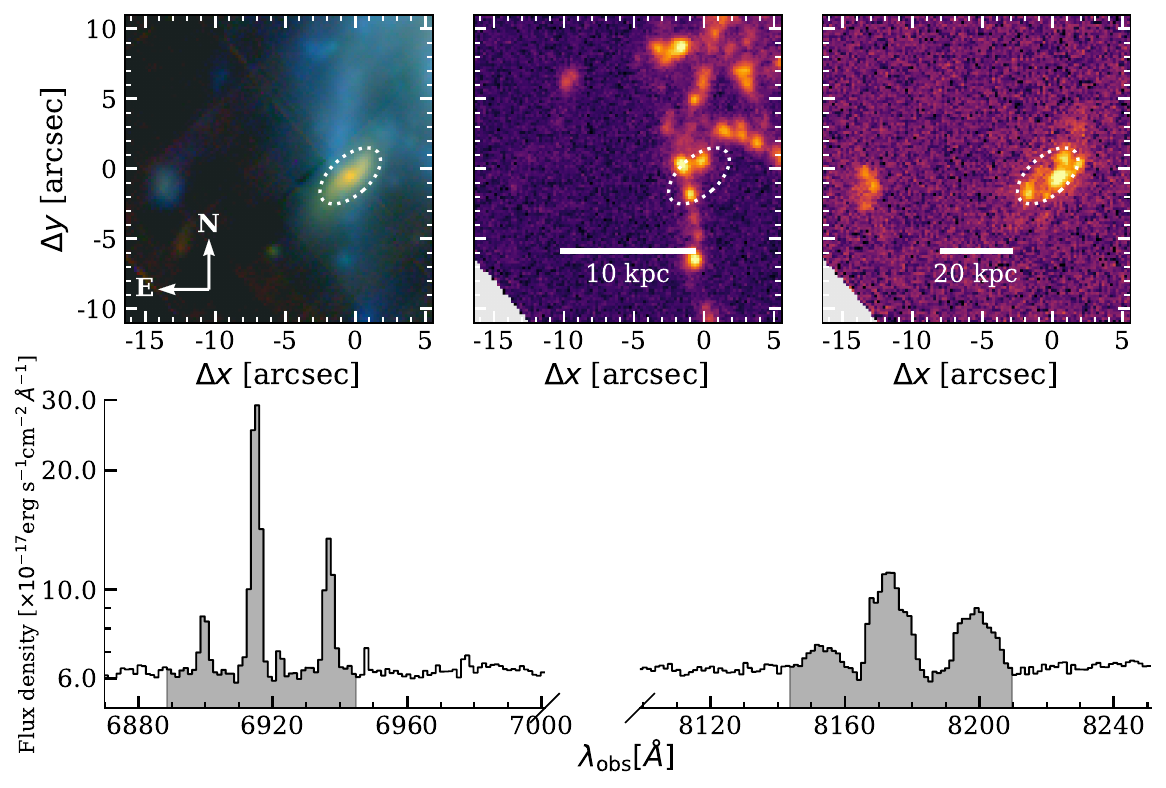}
    \caption{\textit{Top panels}: Zoom-in on the MUSE color-composite image of UG103 highlighting the elongated background source, enclosed by the white dashed ellipse. The middle and right panels show flux maps integrated around each H$\alpha$ + [\ion{N}{ii}] complex. Physical scale bars are adapted according to the redshifts of the main and background galaxies. \textit{Bottom panel}: MUSE spectra integrated within the white dashed ellipse, zoomed on both H$\alpha$ + [\ion{N}{ii}] complexes. Shadowed regions indicate the wavelength ranges used to produce the middle and right images.}
    \label{fig:backsource}
\end{figure}

\section{Example of fitted spectra}
\label{appendix:spectra}

To illustrate the emission-line fitting procedure described in Sect.~\ref{subsec:linefitting}, in Fig.~\ref{fig:spectra_example} we present three fitted spectra, extracted from UG103. The top panel shows the spectrum at the galaxy center, with S/N~=~32 in the continuum, while the middle panel shows a spectrum near the center, with S/N~=~23. The bottom panel shows the spectrum from the brightest spaxel of a clump located far from the stellar disk, exhibiting S/N~=~1.5, therefore not included in the binning process. The first spectrum illustrates a case where the stellar continuum and ionized gas are well detected at the spaxel resolution, whereas the second case demonstrates the result of re-scaling the \ppxf solution to the spaxel resolution. The third case illustrates the treatment of spectra where S/N $\leqslant 2$ but with detected emission-line features. 

\begin{figure*}
    \centering
    \begin{tabular}{cc}
        \includegraphics[scale=0.5]{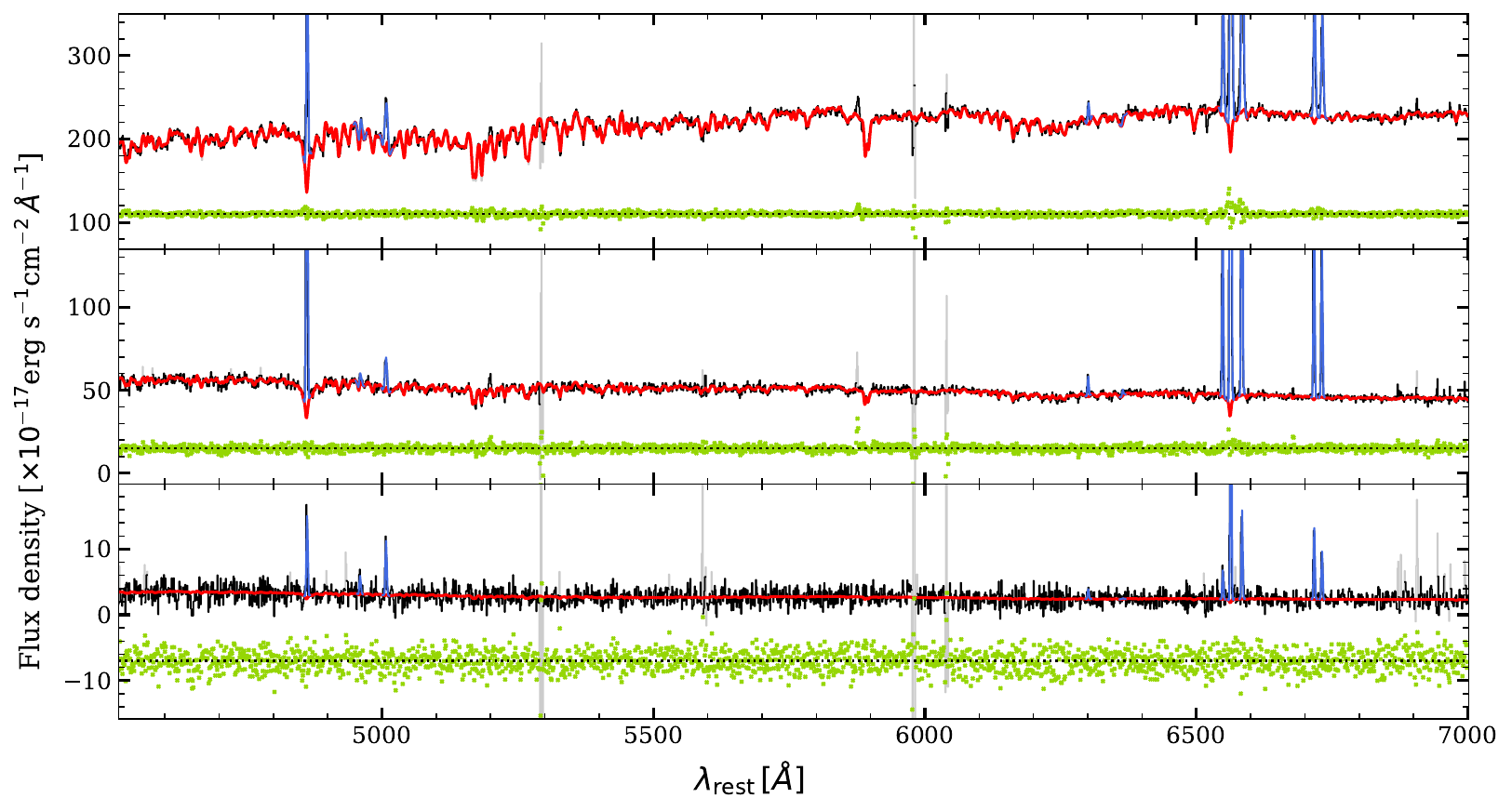} &
        \includegraphics[scale=0.5]{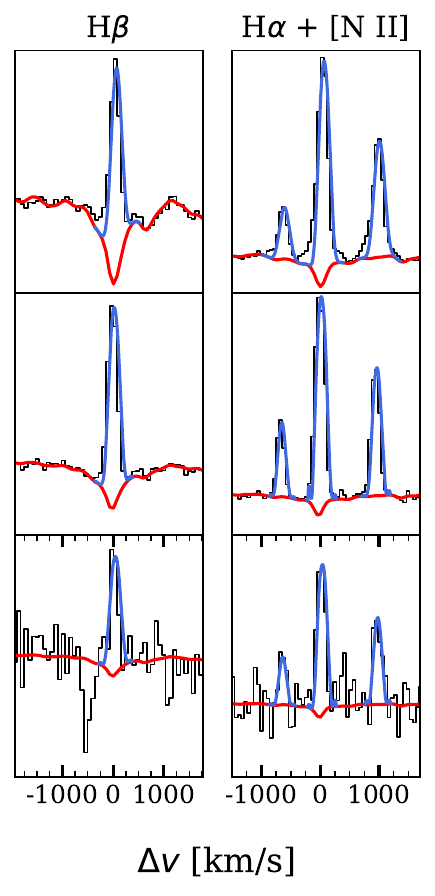}
    \end{tabular}        
    \caption{Example of three different spectra in UG103. \textit{Top panel}: Observed spectrum (black curve) at galaxy center. The red curve displays \ppxf best fit at the spaxel resolution, while the emission-line features modeled with \textsc{ifscube} are shown in blue. \textit{Middle panel}: Similar to the top panel, but shows a spectrum slightly offset from the galaxy center. The red curve corresponds to the \ppxf best-fit stellar emission re-scaled from the Voronoi bin to the spaxel resolution.
    \textit{Bottom panel}: Same as other panels, but now corresponding to the brightest spaxel of a SF clump distant from the galaxy center that does not meet the imposed S/N\,$\geqslant 2$ criterion in the continuum and is therefore excluded from the Voronoi binning scheme. In the absence of a \ppxf solution, the continuum is modeled using a moving median filter (red curve).
    In all panels green points show the fit residuals, scaled by the spectral variance. An arbitrary shift has been applied for clarity, and the dotted black line the position at which the modeled and observed spectra are identical. Light gray spectral regions flag the presence of sky residuals. The right panels show a zoom-in on H$\beta$ and H$\alpha$ + [\ion{N}{ii}] emission-line features.}
     \label{fig:spectra_example}
\end{figure*}

\section{Table with the pitch angles of spiral arms in UG101 and UG103}
\label{appendix:pitch_angles_table}

In Table~\ref{tab:pitch_angles} we list the pitch angles measured for the spiral arms identified in UG101 and UG103 (see Fig.~\ref{fig:pitch_angles}). The table includes the mean values of all pitch angles, as well as the mean considering only the inner ($< 2\,R_e$) and outer ($\geqslant 2\,R_e$) points. A detailed description of how these measurements were made can be found in Sect.~\ref{subsec:pitch_angles}.

\begin{table}[!t]
    \centering
    \caption{Pitch angles for the spiral arms in UG101 and UG103.}
    \label{tab:pitch_angles}
    \begin{tabular}{cccc}
\toprule
Galaxy/Arm ID & Global ($^{\circ}$) & Inner ($^{\circ}$) & Outer ($^{\circ}$)\\ 
\midrule
\midrule
UG101 / \#1   & 47.4 & 33.3 & 69.9  \\
UG101 / \#2   & 19.8 & 15.9 & 23.3  \\
UG101 / \#3   & 39.7 & 39.2 & 40.8  \\
UG101 / \#4   & 34.4 & 30.8 & 45.0  \\
UG101 / \#5   & 25.2 & 25.0 & 25.5  \\
UG101 / \#6   & 14.0 & 24.2 & 9.8   \\ \midrule
UG103 / \#1   & 37.9 & 49.7 & 28.4  \\
UG103 / \#2   & 33.4 & 32.3 & 33.9  \\
UG103 / \#3   & 32.1 & 24.0 & 47.5  \\
UG103 / \#4   & 22.5 & 17.3 & 43.3   \\
UG103 / \#5   & 32.8 & 25.5 & 46.6   \\
UG103 / \#6   & 48.6 & 27.5 & 67.0   \\
UG103 / \#7   & 25.0 & 25.0 & \ldots
\end{tabular}
    \tablefoot{Global pitch angles refer to the mean of all pitch angle measurements, while inner and outer consider the mean within and beyond $2\,R_e$, respectively. The angle is measured in the polar plane, increasing in the counter-clockwise direction.}
\end{table}

We note that in UG101 some spiral arms show similar inner and outer pitch angles (e.g., arms \#3 and \#5), whereas others exhibit a more abrupt increase (e.g., arms \#1 and \#4). This behavior contrasts with that of UG103, where most arms (arm \#2 is an exception) display a sharp change in pitch angles beyond $2\,R_e$. 
An interesting case is spiral arm \#6 in UG101, as it shows higher inner than outer pitch angles. This is evident in Fig.~\ref{fig:pitch_angles}, where the pitch angle decreases after $\phi \sim 300^{\circ}$. Similarly, arm \#1 of UG103 also exhibits slightly lower mean outer pitch angles. In this case, however, the pitch angle decreases mildly just beyond $2\,R_e$ but increases abruptly at $r \sim 22\,$kpc. Since there are fewer points at larger radii, the mean outer value is dominated by measurements at $r \lesssim 22\,$kpc. These arguments are illustrated by Fig.~\ref{fig:pangles_radprofile}, which shows how the pitch angle of each identified arm varies as function of galactocentric distance, comparing with the range of values derived in B21.

\begin{figure*}
    \centering
    \begin{tabular}{cc}
        \includegraphics[width=0.95\columnwidth]{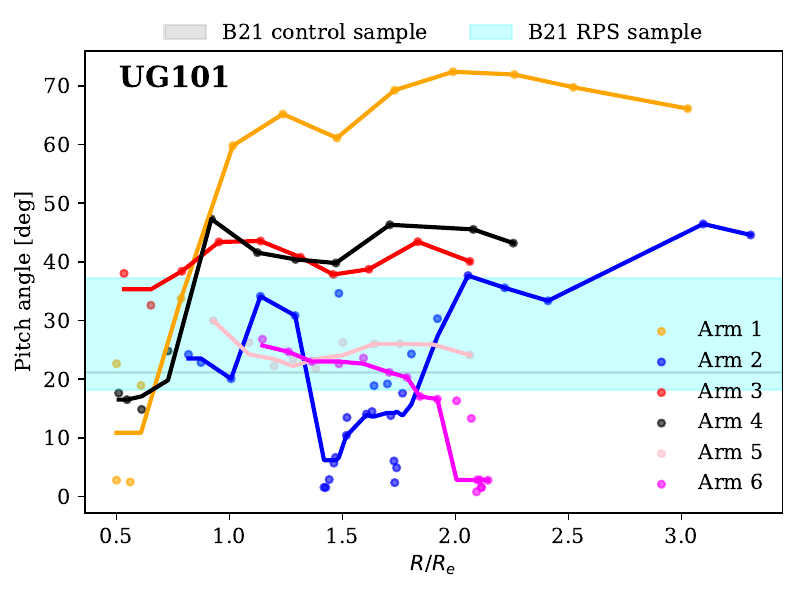} &
        \includegraphics[width=0.95\columnwidth]{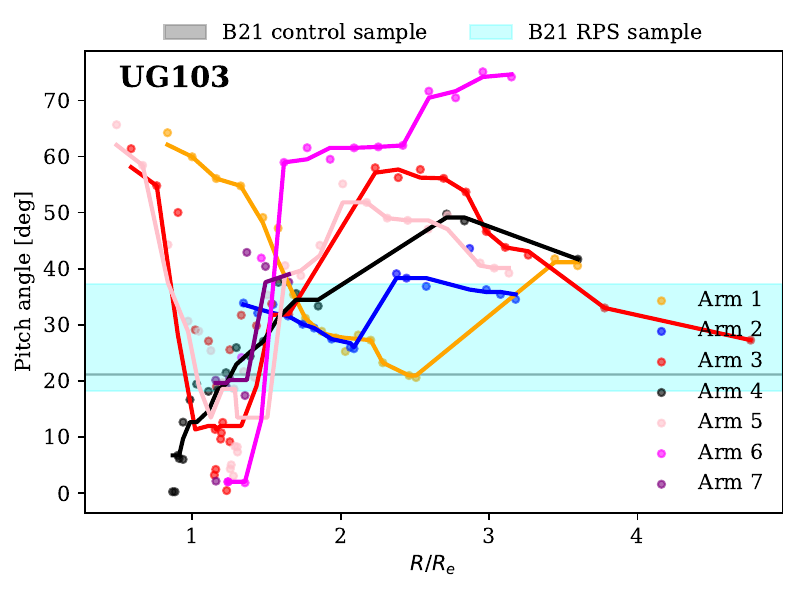}
    \end{tabular}
    \caption{Radial variation of the pitch angle for each spiral arm identified in UG101 (left) and UG103 (right). Points represent the measured pitch angles, while solid lines show the trends derived from a moving median. For reference, the shaded regions indicate the ranges spanned by the inner and outer spiral arms of undisturbed (gray) and RPS-driven galaxies, as reported by \citet{Bellhouse2021}.}
    \label{fig:pangles_radprofile}
\end{figure*}

\end{appendix}
\end{document}